# Broadband Multi-wavelength Properties of M87 during the 2017 Event Horizon Telescope Campaign

J. C. Algaba[1], J. Anczarski[2], K. Asada[3], M. Baloković[4,5], S. Chandra[6], Y.-Z. Cui[7,8], A. D. Falcone[9], M. Giroletti[10], C. Goddi[11,12], K. Hada[7,8], D. Haggard[13,14], S. Jorstad[15,16], A. Kaur[9], T. Kawashima[17], G. Keating[18], J.-Y. Kim[19,20], M. Kino[21,22], S. Komossa[20], E. V. Kravchenko[10,23,24], T. P. Krichbaum[20], S.-S. Lee[19], R.-S. Lu (路如森)[20,25,26], M. Lucchini[27], S. Markoff[27,28], J. Neilsen[2], M. A. Nowak[29], J. Park[30,31,244], G. Principe[10,32,33], V. Ramakrishnan[34], M. T. Reynolds[35], M. Sasada[21,36], S. S. Savchenko[37,38], K. E. Williamson[15]

(The Event Horizon Telescope Collaboration)

(The Fermi Large Area Telescope Collaboration)

(H.E.S.S. Collaboration)

(MAGIC Collaboration)

(VERITAS Collaboration)

(EAVN Collaboration)

(See the end matter for the full list of authors including corresponding authors for the partner facilities.)

*Received 2020 December 25; revised 2021 March 14; accepted 2021 March 16; published 2021 April 14*

## Abstract

In 2017, the Event Horizon Telescope (EHT) Collaboration succeeded in capturing the first direct image of the center of the M87 galaxy. The asymmetric ring morphology and size are consistent with theoretical expectations for a weakly accreting supermassive black hole of mass $\sim 6.5 \times 10^9 M_\odot$. The EHTC also partnered with several international facilities in space and on the ground, to arrange an extensive, quasi-simultaneous multi-wavelength campaign. This Letter presents the results and analysis of this campaign, as well as the multi-wavelength data as a legacy data repository. We captured M87 in a historically low state, and the core flux dominates over HST-1 at high energies, making it possible to combine core flux constraints with the more spatially precise very long baseline interferometry data. We present the most complete simultaneous multi-wavelength spectrum of the active nucleus to date, and discuss the complexity and caveats of combining data from different spatial scales into one broadband spectrum. We apply two heuristic, isotropic leptonic single-zone models to provide insight into the basic source properties, but conclude that a structured jet is necessary to explain M87's spectrum. We can exclude that the simultaneous γ-ray emission is produced via inverse Compton emission in the same region producing the EHT mm-band emission, and further conclude that the $\gamma$-rays can only be produced in the inner jets (inward of HST-1) if there are strongly particle-dominated regions. Direct synchrotron emission from accelerated protons and secondaries cannot yet be excluded.

---

[239] NASA Hubble Fellowship Program, Einstein Fellow.
[240] Now at University of Innsbruck.
[241] Also at Port d'Informació Científica (PIC) E-08193 Bellaterra (Barcelona), Spain.
[242] Also at Dipartimento di Fisica, Università di Trieste, I-34127 Trieste, Italy.
[243] Also at INAF Trieste and Dept. of Physics and Astronomy, University of Bologna.
[244] EACOA Fellow.
[245] UKRI Stephen Hawking Fellow.
[246] For questions concerning EHT results contact ehtcollaboration@gmail.com.
* H.E.S.S. corresponding author. For questions concerning H.E.S.S. results contact hess@hess-experiment.eu.
† MAGIC corresponding author. For questions concerning MAGIC results contact contact.magic@mpp.mpg.de.
‡ VERITAS corresponding author. For questions concerning VERITAS results contact wjin4@crimson.ua.edu, jmsantander@ua.edu.

## 1. Introduction

M87 is the most prominent elliptical galaxy within the Virgo Cluster, located just $16.8 \pm 0.8$ Mpc away (Blakeslee et al. 2009; Bird et al. 2010; Cantiello et al. 2018, and see also EHT Collaboration et al. 2019f). As one of our closest active galactic nuclei (AGNs), M87 also harbors the first example of an extragalactic jet to have been noticed by astronomers (Curtis 1918), well before these jets were understood to be a likely signature of black hole accretion. By now this famous one-sided jet has been well-studied in almost every wave band from radio (down to sub-parsec scales; e.g., Reid et al. 1989; Junor et al. 1999; Hada et al. 2011; Mertens et al. 2016; Kim et al. 2018b; Walker et al. 2018), optical (e.g., Biretta et al. 1999; Perlman et al. 2011), X-ray (e.g., Marshall et al. 2002; Snios et al. 2019), and $\gamma$-rays (e.g., Abdo et al. 2009a; Abramowski et al. 2012; MAGIC Collaboration et al. 2020). Extending over 60 kpc in length, the jet shows a system of multiple knot-like features, including an active feature HST-1 at a projected distance of $\sim$70 pc from the core, potentially marking the end of the black hole's sphere of gravitational influence (Asada & Nakamura 2012). In contrast, the weakly accreting supermassive black hole (SMBH) in the center of our own Milky Way galaxy, Sgr A$^*$, does not show obvious signs of extended jets, although theory predicts that such outflows should be formed (e.g., Dibi et al. 2012; Mościbrodzka et al. 2014; Davelaar et al. 2018). The conditions under which such jets are launched is one of the enduring questions in astrophysics today (e.g., Blandford & Znajek 1977; Blandford & Payne 1982; Sikora & Begelman 2013).

In 2019 April, the Event Horizon Telescope (EHT) Collaboration presented the first direct image of an SMBH "shadow" in the center of M87 (EHT Collaboration et al. 2019a, 2019b, 2019c, 2019d, 2019e, 2019f). The key result in these papers was the detection of an asymmetric ring (crescent) of light around a darker circle, due to the presence of an event horizon, along with detailed explanations of all the ingredients necessary to obtain, analyze, and interpret this rich data set. The ring itself stems from a convolution of the light produced near the last unstable photon orbit, as it travels through the geometry of the production region with radiative transfer in the surrounding plasma, and further experiences bending and redshifting due to the effects of general relativity (GR). Photons that orbit, sometimes multiple times before escaping, trace out a sharp feature revealing the shape of the spacetime metric, the so-called "photon ring". From the size of the measured, blurry ring and three different modeling approaches (see EHT Collaboration et al. 2019e), the EHT Collaboration (EHTC) calibrated for these multiple effects, to derive a mass for M87's SMBH of $(6.5 \pm 0.7) \times 10^9 M_\odot$.

One of the primary contributions to the $\sim$10% systematic error on this mass is due to uncertainties in the underlying accretion properties. As detailed in EHT Collaboration et al. (2019d) and Porth et al. (2019), the EHTC ran over 45 high-resolution simulations over a range of possible physical parameters in, e.g., spin, magnetic field configuration, and electron thermodynamics, using several different GR magnetohydrodynamic (GRMHD) codes. These outputs were then coupled to GR ray-tracing codes, to generate $\sim$60,000 images captured at different times during the simulation runs, which were verified in Gold et al. (2020). While the photon ring remains relatively robust to changes in spin, in part because of the small viewing angle (see, e.g., Johannsen & Psaltis 2010), the spreading of the light around this feature strongly depends on the plasma properties near the event horizon, introducing significant degeneracy. For instance, a smaller black hole produces a smaller photon ring, but in some emission models there is extended surrounding emission leading to a larger final blurry ring. Similarly, a larger black hole with significant emission produced along the line of sight will appear to have emission within the photon ring, and when convolved produces the appearance of a smaller blurry ring. The error in calibrating from a given image to a unique black hole mass is therefore a combination of image reconstruction limits, as well as our current level of uncertainty about the plasma properties and emission geometry very close to the black hole.

However, it is important to note that even in these first analyses, several of the models could already be ruled out using complementary information from observations with facilities at other wavelengths. For instance, the estimated minimum power in the jets, $P_{\rm jet} \geqslant 10^{42}$ erg s$^{-1}$, from prior and recent multi-wavelength studies (e.g., Reynolds et al. 1996; Stawarz et al. 2006; de Gasperin et al. 2012; Prieto et al. 2016), was already enough to rule out about half of the initial pool of models, including all models with zero spin. Furthermore, the X-ray fluxes from quasi-simultaneous observations with the Chandra X-ray Observatory and NuSTAR provided another benchmark that disfavored several models based on preliminary estimates of X-ray emission from the simulations. However, detailed fitting of these and other precision data sets were beyond the focus of the first round of papers and GRMHD model sophistication.

The current cutting edge in modeling accreting black holes, whether via GRMHD simulations or semi-analytical methods, focuses on introducing more physically self-consistent, reliable treatments of the radiating particles (electrons or electron-positron pairs). In particular, key questions remain about how the bulk plasma properties dictate the efficiency of heating, how many thermal particles are accelerated into a nonthermal population, and the dependence of nonthermal properties such as spectral index on plasma properties such as turbulence, magnetization, etc. (see, e.g., treatments in Howes 2010; Ressler et al. 2015; Mościbrodzka et al. 2016; Ball et al. 2018; Anantua et al. 2020).

To test the newest generation of models, it is important to have extensive, quasi-simultaneous or at least contemporaneous multi-wavelength monitoring of several AGNs, providing both spectral and imaging data (and ideally polarization where available) over a wide range of physical scales. These types of campaigns have been limited by the difficulty in obtaining time on multiple facilities and by scheduling challenges, so often data are combined from different epochs. However, the variability timescales for even SMBHs such as M87's are short enough (days to months) that combining data sets from different periods can skew modeling results for sensitive quantities such as radiating particle properties.

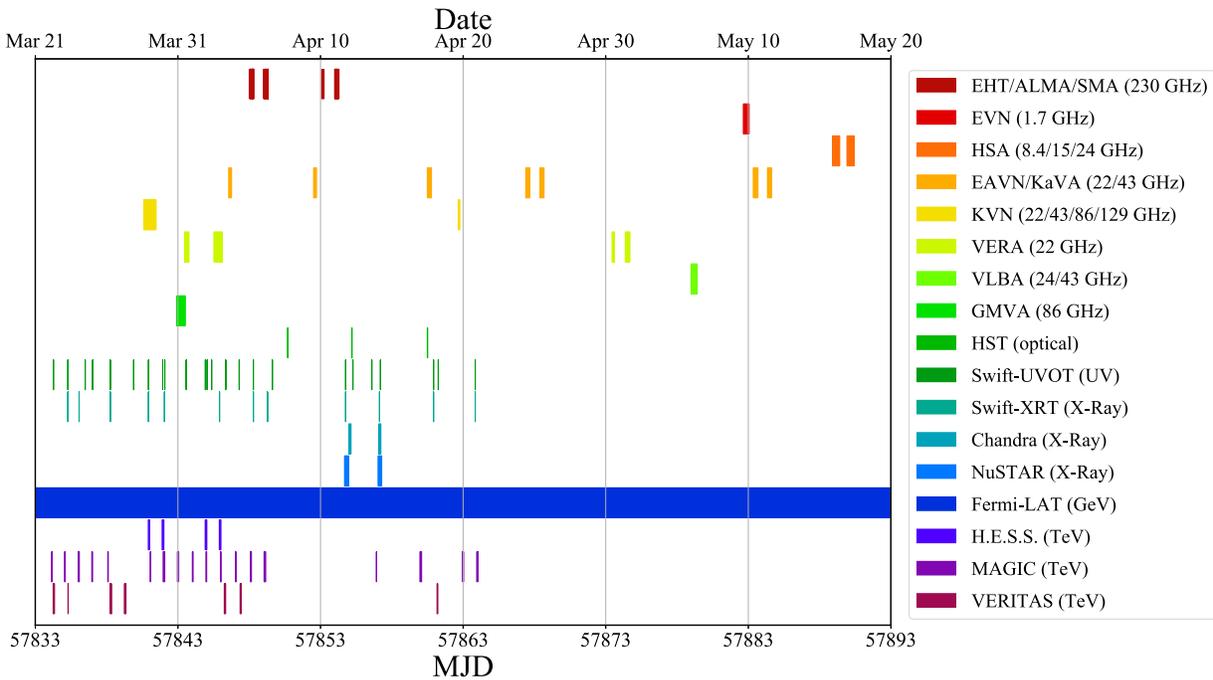

**Figure 1.** Instrument coverage summary of the 2017 M87 MWL campaign, covering MJD range 57833–57893. (Made with the MWLGen software by J. Farah.)

This Letter intends to be the first of a series, presenting the substantive multi-wavelength campaigns carried out by the EHT Multi-Wavelength Science Working Group (MWL WG), including EHTC members and partner facilities, for both our primary sources M87 and Sgr A*, as well as many other targets and calibrators such as 3C 279, 3C 273, OJ 287, Cen A, NGC 1052, NRAO 530, and J1924−2914. These legacy papers are meant to be companion papers to the EHT publications, and will be used for the detailed modeling efforts to come, both from the EHTC Theory and Simulations WG as well as from other groups. They serve as a resource for the entire community, to enable the best possible modeling outcomes and a benchmark for theory. Here we present the results from the 2017 EHT campaign on M87, combining very long baseline interferometry (VLBI) imaging and spectral index maps at longer wavelengths, with spectral data from submillimeter (submm) through TeV $\gamma$-rays (covering more than 17 decades in frequency).

In Section 2 we describe these observations, including images (a compilation of MWL images in one panel is shown in a later section), spectral energy distributions (SEDs) and, where relevant, lightcurves and comparisons to prior observations. In Section 3 we present a compiled SED together with a table of fluxes. We also fit this SED with a few phenomenological models and discuss the consequences for the emission geometry and high-energy properties. In Section 4 we give our conclusions. All data files and products are available for download, as described in Section 3.2.

## 2. Observations and Data Reduction

In Figure 1 we give a schematic overview of the 2017 MWL campaign coverage on M87. In the following subsections we provide detailed descriptions of the observations, data processing procedures, and band-specific analyses. To aid readability, all tabulated data are collected in Appendix A.

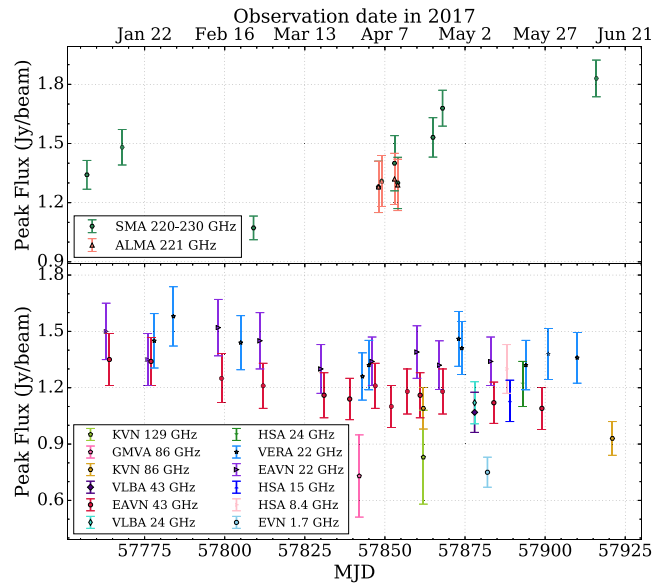

**Figure 2.** Radio lightcurves of the M87 core in 2017 at multiple frequencies. The top and bottom panels are for connected interferometers and VLBI, respectively. The corresponding beam sizes are indicated in Table A1. KVN data at 22 and 43 GHz are not shown here since KVN captures the data from the shortest baselines of EAVN.

### 2.1. Radio Data

In this subsection we describe the observations and data reduction of radio/mm data obtained with various VLBI facilities and connected interferometers. Especially regarding VLBI data, here we introduce the term "radio core" to represent the innermost part of the radio jet. A radio core in a VLBI jet image is conventionally defined as the most compact (often unresolved or partially resolved) feature seen at the apparent base of the radio jet in a given map (e.g., Lobanov 1998; Marscher 2008). For this reason, different angular resolutions by different VLBI instruments/frequencies, together with the frequency-dependent synchrotron optical depth (Marcaide & Shapiro 1984; Lobanov 1998), can make the identification of a radio core not exactly the same for each observation. See also Section 3.2 for related discussions.

### 2.1.1. EVN 1.7 GHz

M87 was observed with the European VLBI Network (EVN) at 1.7 GHz on 2017 May 9. The observations were made as part of a long-term EVN monitoring program of activity in HST-1, located at a projected distance of ∼70 pc from the core (Cheung et al. 2007; Chang et al. 2010; Giroletti et al. 2012; Hada et al. 2015). A total of eight stations joined a 10 hr long session with baselines ranging from 600 km to 10 200 km, yielding a maximum angular resolution of ∼3 mas at 1.7 GHz. The data were recorded at 1 Gbps with dual-polarization (a total bandwidth of 256 MHz, 16 MHz × 8 subbands for each polarization), and the correlation was performed at the Joint Institute for VLBI ERIC (JIVE). Automated data flagging and initial amplitude and phase calibration were also carried out at JIVE using dedicated pipeline scripts. This step was followed by frequency averaging within each spectral band (IF) and in time to 8 s. The final image was produced using the Difmap software (Shepherd 1997) after several iterations of phase and amplitude self-calibration (see Giroletti et al. 2012, for more detail). Here we provide the peak flux of the radio core (see Figure 2 and Table A1) and a large-scale jet image (presented in Section 3), while a dedicated analysis on the HST-1 kinematics will be presented in a separate paper.

### 2.1.2. HSA 8, 15, and 24 GHz

M87 was observed with the High Sensitivity Array (HSA) at 8.4, 15, and 24 GHz on 2017 May 15, 16 and 20, respectively, which are roughly a month after the EHT-2017 observations. Each session was made with a 12 hr long continuous track and the phased Very Large Array and Effelsberg 100 m antenna participated in the observations along with the 10 stations of the NRAO Very Long Baseline Array (VLBA). The data were recorded at 2 Gbps with dual-polarization (a total bandwidth of 512 MHz, 32 MHz × 8 subbands for each polarization), and the correlation was done with the VLBA correlator in Socorro (Deller et al. 2011). The initial data calibration was performed using the NRAO Astronomical Image Processing System (AIPS; Greisen 2003) based on the standard VLBI data reduction procedures (Crossley et al. 2012; Walker 2014). Similar to other VLBI data, images were created using the Difmap software with iterative phase/amplitude self-calibration.

A detailed study of the parsec-scale structure of the M87 jet from this HSA program will be discussed in a separate paper. Here we provide a core peak flux and VLBI-scale total flux at each frequency (Table A1 in Appendix A). We adopt 10% errors in flux estimate, which is typical for HSA.

### 2.1.3. VERA 22 GHz

The core of M87 was frequently monitored over the entire year of 2017 at 22 GHz with the VLBI Exploration of Radio Astrometry (VERA; Kobayashi et al. 2003), as part of a regular monitoring program of a sample of γ-ray bright AGNs (Nagai et al. 2013). A total of 17 epochs were obtained in 2017 (see Figure 2 and Table A1 in Appendix A). During each session, M87 was observed for 10–30 minutes with an allocated bandwidth of 16 MHz, sufficient to detect the bright core and create its light curves. All the data were analyzed in the standard VERA data reduction procedures (see Nagai et al. 2013; Hada et al. 2014, for more details). Note that VERA can recover only part of the extended jet emission due to the lack of short baselines, so the total fluxes of VERA listed in Table A1 in Appendix A significantly underestimate the actual total jet fluxes.

### 2.1.4. EAVN/KaVA 22 and 43 GHz

Since 2013 a joint network of the Korean VLBI Network (KVN) and VERA (KaVA; Niinuma et al. 2014) has regularly been monitoring M87 to trace the structural evolution of the pc-scale jet (Hada et al. 2017; Park et al. 2019). From 2017, the network was expanded to the East Asian VLBI Network (EAVN; Wajima et al. 2016; Asada et al. 2017; An et al. 2018) by adding more stations from East Asia, enhancing the array sensitivity and angular resolution. Between 2017 March and May, EAVN densely monitored M87 for a total of 14 epochs (five at 22 GHz, nine at 43 GHz; see Figure 2 and Table A1 in Appendix A). The default array configurations were KaVA+Tianma+Nanshan+Hitachi at 22 GHz and KaVA+Tianma at 43 GHz, respectively, while occasionally a few more stations (Sejong, Kashima, and Nobeyama) additionally joined if they were available. In addition, we also had four more sessions with KaVA alone (2+2 at 22/43 GHz) in 2017 January–February.

Each of the KaVA/EAVN sessions was made in a 5–7 hr continuous run at a data recording rate of 1 Gbps (2-bit sampling, a total bandwidth of 256 MHz was divided into 32 MHz × 8 IFs). All the data were correlated at the Daejeon hardware correlator installed at KASI. All the EAVN data were calibrated in the standard manner of VLBI data reduction procedures. We used the AIPS software package for the initial calibration of visibility amplitude, bandpass, and phase calibration. The imaging/CLEAN (Högbom 1974) and self-calibration were performed with the Difmap software. In Section 3 we present one of the 22 GHz EAVN images (taken in 2017 March 18) where the KaVA, Tianma—65 m, Nanshan—26 m, and Hitachi—32 m radio telescopes participated.

### 2.1.5. KVN 22, 43, 86, and 129 GHz

The KVN regularly observes M87 at frequencies of 22, 43, 86, and 129 GHz simultaneously via the interferometric Monitoring of γ-ray Bright Active galactic nuclei (iMOGABA) program, starting in 2012 December and lasting until 2020 January. The total bandwidth of the observations at each frequency band is 64 MHz and the typical beam sizes of the observations are $6.1 \times 3.1$ mas at 22 GHz, $2.8 \times 1.6$ mas at 43 GHz, $1.5 \times 0.8$ mas at 86 GHz, and $1.1 \times 0.5$ mas at 129 GHz. Details of the scheduling, observations, data reduction including frequency phase transfer technique, analysis (i.e., imaging and model-fitting), and early results for M87 are shown in Lee et al. (2016) and Kim et al. (2018a). Despite the limited coverage of baselines and capability of the array to image the extended jet structure in M87, the flux density of the compact core can be rather reliably measured (Kim et al. 2018a). We extract the core flux densities at the four frequencies and obtain light curves spanning observing periods between 2017 March and December at seven epochs. While typical flux density uncertainties at 22–86 GHz are of order of ∼10%, residual phase rotations and larger thermal noise at 129 GHz often lead to uncertainties of ∼30%. Accordingly, we adopt these uncertainties for all KVN observing epochs in this Letter.

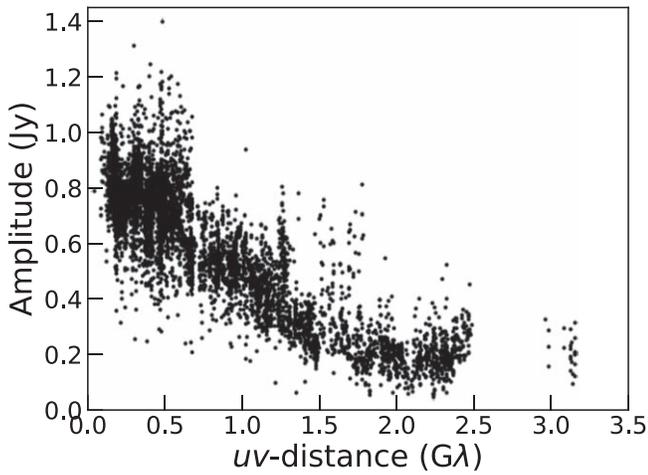

**Figure 3.** Visibility amplitude vs. *uv*-distance plot of GMVA data on 2017 March 30. In this display the visibilities are binned in 30 s intervals for clarity.

### 2.1.6. VLBA 24 and 43 GHz

M87 was observed with the VLBA at central frequencies of 24 and 43 GHz on 2017 May 5. These observations were carried out in the framework of the long-term monitoring program toward M87, which was initiated in 2006 (Walker et al. 2018) and lasted until 2020. For the 2017 session, the total on-source time amounts to about 1.7 hr at 24 GHz and 6 hr at 43 GHz. The sources OJ 287 and 3C 279 were observed to use as fringe finders and bandpass calibrators. In each band, eight 32 MHz wide frequency channels were recorded in both right and left circular polarization at a rate of 2 Gbps, and correlated with the VLBA software correlator in Socorro. The initial data reduction was conducted within AIPS following the standard calibration procedures for VLBI data. Deconvolution and self-calibration algorithms, implemented in Difmap, were used for phase and amplitude calibration and for constructing the final images. Amplitude calibration accuracy of 10% is adopted for both frequencies.

The resulting total intensity 43 GHz image of M87 is presented in Section 3, with the details given in Table A1 in Appendix A. The synthesized beam size amounts to $0.76 \times 0.40$ mas at the position angle (PA) of the major axis of $-8°$ at 24 GHz, and $0.41 \times 0.23$ mas at PA = $0°$ at 43 GHz. We note that these observations were used for the study of a linear polarization structure toward the M87 core, details of which can be found in a separate paper (Kravchenko et al. 2020).

### 2.1.7. GMVA 86 GHz

M87 was observed by the Global Millimeter-VLBI-Array (GMVA; Boccardi et al. 2017) on 2017 March 30 (project code MA 009). In total, 14 stations participated in the observations: VLBA (eight stations; without HN and SC), 100 m Green Bank Telescope, IRAM 30 m, Effelsberg 100 m, Yebes, Onsala, and Metsahovi. The observation was performed in full-track mode for a total of 14 hr. Nearby bright sources 3C 279 and 3C 273 were observed as calibrator targets. The raw data were correlated by using the DiFX correlator (Deller et al. 2011).[247] Further post-processing was performed using the AIPS software package, following typical VLBI data reduction procedures (see, e.g., Martí-Vidal et al. 2012). After the calibration, the data were frequency-averaged across the whole subchannels and IFs, and exported outside AIPS for imaging with the Difmap software. Within Difmap, the data were further time-averaged for 10 s, followed by flagging of outlying data points (e.g., scans with too low amplitudes due to pointing errors). Afterward, CLEAN and phase self-calibrations were iteratively performed near the peak of the intensity, but avoiding CLEANing of the faint counterjet side at the early stage. When no more significant flux remained for further CLEAN steps, a first amplitude self-calibration was performed using the entire time coverage as the solution interval, in order to find average station gain amplitude corrections. A similar procedure was repeated with progressively shorter self-calibration solution intervals, and the final image was exported outside Difmap when the shortest possible solution interval was reached and no more significant emission was visible in the dirty map compared to the off-source rms levels.

Calibrated visibilities are shown in Figure 3, and the CLEAN image is given in Section 3. We note that the final image has an rms noise level of $\sim 0.5$ mJy beam$^{-1}$. This noise level is a factor of a few higher than other 86 GHz images of M87 from previous observations, which were made with similar array configuration as the 2017 session (see Hada et al. 2016; Kim et al. 2018b). Therefore, we refer to this image as tentative, and it only reveals the compact core and faint base of the jet, mainly due to poor weather conditions during the observations. We consider an error budget of 30% for the flux estimate. The peak flux density on the resultant image amounts to $\sim 0.52$ Jy beam$^{-1}$ for the synthesized beam of $0.243 \times 0.066$ mas at PA = $-9°.3$ (see Table A1 in Appendix A for other details). We note that the peak flux density as well as the flux, integrated over the core region, are comparable to their historical values (Hada et al. 2016; Kim et al. 2018b), except for the 2009 May epoch, when the GMVA observations revealed about two times brighter core region in both intensity and integrated flux (Kim et al. 2018b).

### 2.1.8. Atacama Large Millimeter/submillimeter Array (ALMA) 221 GHz

The observations at Band 6 with phased-ALMA (Matthews et al. 2018) were conducted as part of the 2017 EHT campaign (Goddi et al. 2019). The VLBI observations were carried out while the array was in its most compact configurations (with longest baselines <0.5 km). The spectral setup at Band 6 includes four spectral windows (SPWs) of 1875 MHz, two in the lower and two in the upper sideband, correlated with 240 channels per SPW (corresponding to a spectral resolution of 7.8125 MHz). The central frequencies at this band are 213.1, 215.1, 227.1, and 229.1 GHz. Details about the ALMA observations and a full description of the data processing and calibration can be found in Goddi et al. (2019).

Imaging was performed with the Common Astronomy Software Applications (CASA; McMullin et al. 2007) package using the task tclean. Only phased antennas were used to produce the final images (with baselines <360 m), yielding synthesized beam sizes in the range 1″0–2″4 (depending on the observing band and date). We produced 256 × 256 pixel maps, with a cell size of 0″2 yielding a field of view of 51″ × 51″.

---
[247] We use the correlator at the Max-Planck-Institut für Radioastronomie (MPIfR) in Bonn, Germany.

The main observational and imaging parameters are summarized in Table A1 in Appendix A. For each data set and corresponding image, the table reports the flux-density values of the central compact core and the overall flux, including the extended jet. In order to isolate the core emission from the jet, we compute the sum of the central nine pixels of the model (CLEAN component) map (an area of 3 × 3 pixels);[248] the contribution from the jet is accounted for by also summing the clean components along the jet. The extended emission accounts for less than 20% of the total emission at 1.3 mm. The large-scale jet image is presented in Section 3. Details about the imaging and flux extraction methods can be found in Goddi et al. (2021).

### 2.1.9. Submillimeter Array (SMA) 230 GHz

The long term 1.3 mm band (230 GHz) flux density light-curve for M87 shown in Figure 2 was obtained at the SMA near the summit of Maunakea (Hawaii). M87 is included in an ongoing monitoring program at the SMA to determine flux densities for compact extragalactic radio sources that can be used as calibrators at mm wavelengths (Gurwell et al. 2007). Available potential calibrators are occasionally observed for 3–5 minutes, and the measured source signal strength calibrated against known standards, typically solar system objects (Titan, Uranus, Neptune, or Callisto). Data from this program are updated regularly and are available at the SMA Observer Center website (SMAOC[249]). Data were primarily obtained in a compact configuration (with baselines extending from 10 to 75 m) though a small number were obtained at longer baselines up to 210 m. The effective spatial resolution, therefore, was generally around 3″. The flux density was obtained by vector averaging of the calibrated visibilities from each observation.

Observations of M87 were additionally conducted as part of the 2017 EHT campaign, with the SMA running in phased-array mode, operating at 230 GHz. All observations were conducted while the array was in compact configuration, with the interferometer operating in dual-polarization mode. The SMA correlator produces four separate but contiguous 2 GHz spectral windows per sideband, resulting in frequency coverage of 208–216 and 224–232 GHz. Data were both bandpass and amplitude calibrated using 3C 279, with flux calibration performed using either Callisto, Ganymede, or Titan. Due in part to poor phase stability at the time of observations, phase calibration is done through self-calibration of the M87 data itself, assuming a point-source model. Data are then imaged and deconvolved using the CLEAN algorithm.

A summary of the measurements made from these data are shown in Table A1 in Appendix A, along with measurements taken within a month of these observations from the SMAOC data mentioned above. The reported core flux for M87 is the flux measured in the center of the cleaned map. Combined images of all data show the same jet-like structure seen in the ALMA image in Figure 13, although recovery of the flux for individual days through imaging is limited by a lack of (u, v)-coverage within individual tracks. Therefore, we estimate the total flux by measuring the mean flux density of all baselines within a (u, v)-angle of 110° ± 5°, as these baselines are not expected to resolve the jet and central region.

### 2.2. Optical and Ultraviolet (UV) Data

We performed optical and UV observations of M87 with the Neil Gehrels Swift Observatory during the EHT campaign, and have also analyzed contemporaneous archival observations from the Hubble Space Telescope (HST).

#### 2.2.1. UV/Optical Telescope (UVOT) Observations

The Neil Gehrels Swift Observatory (Gehrels et al. 2004) is equipped with UVOT (Roming et al. 2005), as well as with X-ray imaging optics (see Section 2.3.4 below). We retrieved UVOT optical and UV data from the NASA High-Energy Astrophysics Archive Research Center (HEASARC) and reduced them with v6.26.1 of the HEASOFT software[250] and CALDB v20170922. The observations were performed from 2017 March 22 to April 20 in six bands, $v$(5458 Å), $b$(4392 Å), $u$(3465 Å), $uvw1$(2600 Å), $uvm2$(2246 Å), and $uvw2$(1928 Å), with 24 measurements in each band, and effective filter wavelengths as given in Poole et al. (2008). The reduction of the data followed the standard prescriptions of the instrument team at the University of Leicester.[251] We checked the UVOT observations for small-scale sensitivity issues and found that none of our observations are affected by bad charge-coupled device (CCD) pixels. In 2020 November the Swift team released new UVOT calibration files, along with coefficients of the flux density correction as a function of time, for data reduced with the previous version of CALDB. For our observations the coefficients of the flux density correction (multiplicative factors) are very close to 1: $0.974 \pm 0.001$ (for the optical filters), $0.947 \pm 0.001$ ($uvw1$), $0.964 \pm 0.001$ ($uvm2$), and $0.958 \pm 0.001$ ($uvw2$). We have used these coefficients to correct our measurements for the UVOT sensitivity change.[252]

We performed aperture photometry for each individual observation using the tool UVOTSOURCE, with a circular aperture of a radius of 5″ centered on the sky coordinates of M87 and detection significance $\sigma \geqslant 5$. This aperture includes the M87 core, the knot HST-1, and some emission from the extended jet, which we cannot separate due to the size of the UVOT point-spread function (PSF; ~2″5). Since there is contamination from the bright host galaxy surrounding the core of M87, we measure the background level in three circular regions, each of 30″ radius in a source-free area located outside the host galaxy radius (see below). We used the count-rate to magnitude and flux density conversion provided by Breeveld et al. (2011) and retained only those measurements with magnitude errors of $\sigma_{mag} < 0.2$. This calibration of the UVOT broadband filters (Breeveld et al. 2011) includes additional calibration sources with a wider range of colors with respect to the one reported by Poole et al. (2008).

---

[248] An alternative method is provided by taking the integrated flux within the synthesized full width at half maximum (FWHM) size centered at the phase-center in each cleaned image, which provides consistent values with those obtained from the model cleaning components.
[249] http://sma1.sma.hawaii.edu/callist/callist.html
[250] NASA High-Energy Astrophysics Archive Research Center HEASOFT package https://heasarc.gsfc.nasa.gov/docs/software/heasoft/.
[251] https://www.swift.ac.uk/analysis/uvot/
[252] https://www.swift.ac.uk/analysis/uvot/index.php

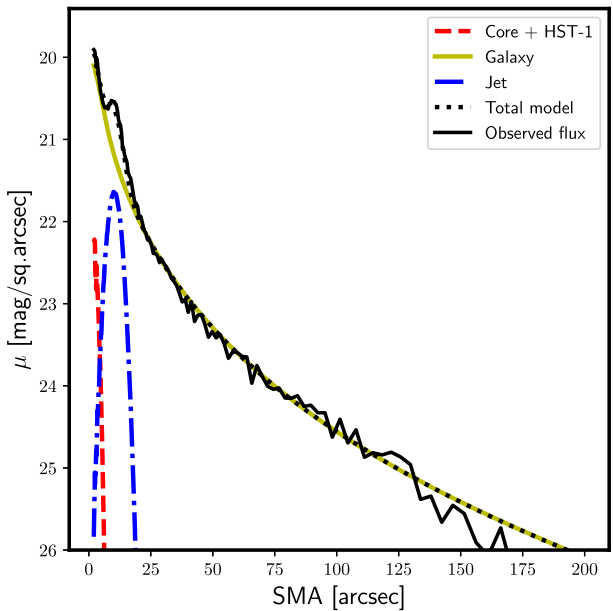

**Figure 4.** Mean surface brightness in *uvw1* band as a function of the semimajor axes of concentric ellipses with ellipticity and PA equal to those of the host galaxy; the red, blue, and green curves show the core + HST-1, jet, and host galaxy profiles, respectively; the solid black curve represents the observed flux, and the black dotted curve gives the total flux of the model; all of the curves show the average profiles over all images in this band.

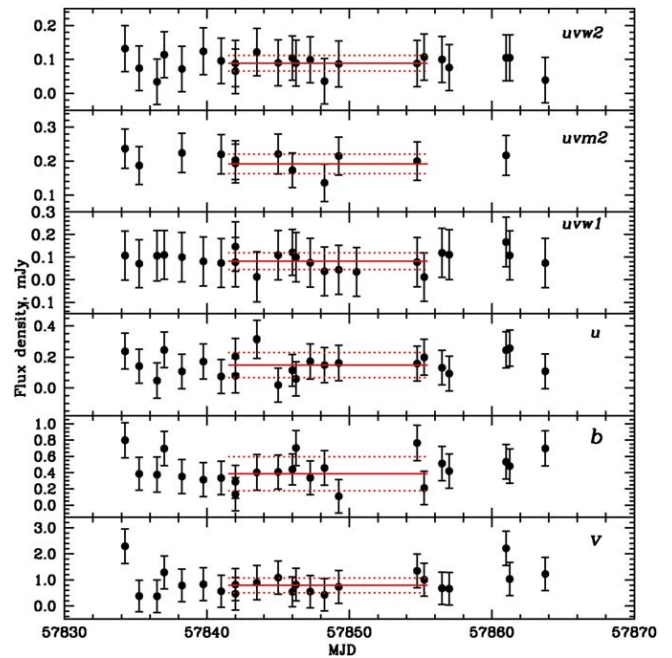

**Figure 5.** Optical and UV light curves of the core region of M87 in all six UVOT filters; the red lines show the average flux density in each filter during the EHT campaign and their standard deviations (red dotted lines) along with the time and duration of the campaign.

To estimate the contributions of the host galaxy to the derived flux densities, we modeled the UVOT images. First we combined individual images using the tool `UVOTIMSUM` to produce a stacked image in each filter. Ordinarily, this would have allowed us to obtain the best signal-to-noise ratio (S/N) for modeling. However, since the source was not centered in the different images at the same CCD position, the area of intersection between the images was too small. Hence, even in the outer regions the flux from the galaxy was still significant. Therefore, we made a set of models using individual images and then found a mean model for each band after a mask of background objects was used to exclude all objects (except for M87 itself). Since a unified mask was used, the images have the same pixels included and excluded from the analysis, so no bias is introduced.

After this procedure, the decomposition process was run for each image using a model consisting of three components: the core region, jet, and host galaxy. The core region includes the core and knot HST-1, which was modeled by a point source, while the jet was fitted by a highly elliptical Gaussian. The galaxy was modeled by a Sérsic function: $I(R) = I_e \exp(-b_n[(R/R_e)^{1/n} - 1])$, where $n$ is the Sérsic parameter, $b_n = 2n - 1/3$, $R_e$ is the half-light radius, and $I_e$ is the intensity at $R_e$. All three structural features can be seen in the combined image in the *uvw1* band presented in Figure 13. A point source model for the core + HST-1 region was convolved with the PSF determined for each filter using several isolated non-saturated stars in a number of images. The parameters of each PSF were averaged over stars and images. Before the decomposition, the images were normalized to a one-second exposure to make them uniform. The program `imfit` (Erwin 2015) was employed to perform decomposition for each image. Each model parameter for a given filter corresponds to the median value averaged over all images using the best-fit image parameters (according to $\chi^2$), with $>2\sigma$ outliers removed (number of outliers for each filter $\leqslant 4$).

Table A2 in Appendix A lists the derived median values of the Sérsic model $n$-parameter and its uncertainty for different bands. The uncertainty corresponds to a scatter of models among the individual images. The derived values of the Sérsic $n$-parameter show a dependence on wavelength that is caused by a difference in the stellar population and, probably, the scattered light from the core, which is blue, leading to higher $n$-values for blue filters. Note that there is significant scatter among values of the $n$-parameter reported in the literature, from $n = 2.4$ (Vika et al. 2012) to $n = 3.0$ (D'Onofrio 2001), $n = 6.1$ (Ferrarese et al. 2006), $n = 6.9$ (Graham & Driver 2007), and $n = 11.8$ (Kormendy et al. 2009).

Table A2 in Appendix A also gives the derived values of the effective radius of the galaxy in different bands, $R_e$, ellipticity, $\epsilon$, and PA, $\Phi$, of the major axis of the galaxy calculated counter clockwise from north. Figure 4 shows an example of a comparison between the mean observed surface brightness profile along the semimajor axis of the host galaxy and the results of the modeling. The mean observed profile is obtained by azimuthally averaging along a set of concentric ellipses with the ellipticity and PA equal to those of the host galaxy.

According to our X-ray data analysis (see Section 2.3.3), the hydrogen column density corresponding to absorption in both our Galaxy and near M87 is equal to $N_H = 0.050^{+0.003}_{-0.002} \times 10^{22}$ cm$^{-2}$, while there is evidence from our X-ray data for additional X-ray absorption within the central 1″ around M87 with a column density of $0.12^{+0.05}_{-0.04} \times 10^{22}$ cm$^{-2}$. However, this latter value is most likely variable since much lower values of $N_H$ for M87 were detected previously, e.g., $N_H \approx 0.01 \times 10^{22}$ cm$^{-2}$ (Sabra et al. 2003) based on UV spectroscopy. Therefore, as discussed in Prieto et al. (2016) we assume that the additional X-ray absorber is dust-free and employ the extinction curve ($R_V = 3.1$) and the extinction value ($E_{B-V} = 0.022$) given by Schlegel et al. (1998) along with the formalism provided by Cardelli et al. (1989) to derive the extinction values in different bands, which are listed in Table A2 in Appendix A.

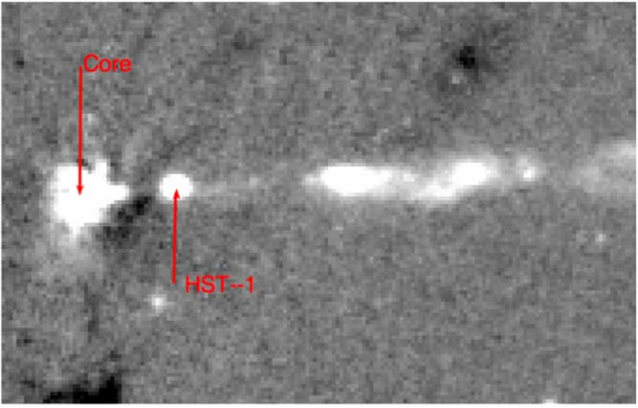

**Figure 6.** HST image of M87 in F606W filter with host galaxy subtracted; the core and HST-1 are designated, the distance between the features is $0\rlap{.}''86 \pm 0\rlap{.}''04$.

Tables A3 and A4 in Appendix A give flux densities of the core region of M87 corrected for the host galaxy contamination and extinction in optical and UV bands, respectively. Figure 5 shows light curves of the core region during the campaign. According to Table A2 in Appendix A and as can be seen from Figure 5 the standard deviation of $F_{core}$ averaged over the EHT campaign is significantly less than the uncertainty of an individual measurement in all filters. The uncertainty is dominated by that of the host galaxy decomposition, which is added (in quadrature) to the photometric uncertainty of each measurement. Therefore, based on the UVOT observations we cannot detect variability in the core region of M87 during the EHT campaign, which is consistent with apparently a low activity state of both the core and HST-1 knot. Based on these results we have calculated the UVOT spectral index of the core region during the EHT campaign using the average values of the flux density in the UVOT bands as $S \propto \nu^{-\alpha}$, which results in $\alpha = 1.88 \pm 0.55$. Although $\alpha$ has a significant error, most likely connected with large uncertainties in the host galaxy decomposition in different bands, the spectral index is consistent with the optical/UV spectral index of the core of M87 reported by Perlman et al. (2011). This indicates that the core dominates the innermost region of M87 at UV/optical wavelengths during the campaign.

### 2.2.2. HST Observations

We have downloaded HST images of M87 from the HST archive obtained on 2017 April 7, 12, and 17 with the Wide Field Camera 3 (WFC3) camera in two wide bands, F275W and F606W (ID5o30010, ID5o30020, ID5o31010, ID5o31020, ID5o32010, and ID5o32020). We used fully calibrated and dither combined images. The image obtained on 2017 April 12 in the F606W filter is a part of a composite of M87 multi-wavelength images presented in Section 3. The decomposition of the HST images was made using the same imfit package as for the images obtained with the UVOT (Section 2.2.1). The substantially higher spatial resolution of the HST images (1 pixel is $0\rlap{.}''04$), however, led to a somewhat different approach. First, we masked out the jet on the images: the HST images reveal its very complex structure, which cannot be approximated with a simple analytical model, as was the case for the low-resolution UVOT images. We also masked out knot HST-1 and the core, which are clearly resolved in the HST images (Figure 6). Second, we used a more complex model for the host galaxy: a sum of two Sérsic models (instead of one as in the case of the UVOT images), which gave us significantly better residual (observations—model) maps. Finally, we used the HST library of PSF images provided at the HST website[253] to model the PSF in each filter. Figure 6 shows the resulting map of M87 on 2017 April 11 in F606W filter, with the host galaxy subtracted.

The decomposition analysis was limited to the central region of the images ($16'' \times 16''$), since the main goal of the process was to subtract the host galaxy flux before the aperture photometry of the core and HST-1. Fitting the full image of the galaxy would have required a more complex model to have comparable quality of the fit for the central regions (see, e.g., Huang et al. 2013). After subtracting the best-fit model from the images, we performed the aperture photometry with a radius of $0\rlap{.}''4$ to find the flux densities from the core and HST-1 in each band. imfit allows one to perform a bootstrap method to derive estimates of uncertainties of the decomposition parameters. For each image, the program was run 100 times, each with same number of random pixels involved in the decomposition process. For each decomposition, a residual FITS file was constructed containing only the core and HST-1 (100 cases for each band and date). For each such residual file, we performed the aperture photometry with a radius of $0\rlap{.}''4$ and calculated the standard deviation of flux density measurements of the core and HST-1 over different decompositions. This standard deviation was added in quadrature to the uncertainty of the photometry from the decomposition of the original image in each band and date. Table A5 in Appendix A gives flux densities of the core and HST-1, measured in two different bands and at epochs contemporaneous with the EHT observations. The flux densities are corrected for the extinction in the same manner as described in Section 2.2.1.

According to Table A5 in Appendix A the core shows a slight increase in flux density over 10 days, while knot HST-1 has a constant flux density. We have estimated optical/UV spectral indices of the core and HST-1, which are $1.44 \pm 0.09$ and $0.60 \pm 0.02$, respectively (the spectral index is defined in the same way as in Section 2.2.1). These are in good agreement with those given by Perlman et al. (2011), ~1.5 for the core and ~0.5 for HST-1, which confirm a flatter spectral index of HST-1 with respect to that of the core at UV/optical wavelengths.

To determine the activity state of M87 during the 2017 campaign, we have constructed light curves of the core and HST-1 in two bands, F275W (from 1999 to 2017) and F606W (from 2002 to 2017). During the period 1999 to 2010 different instruments were used at HST: UV observations were performed with STIS/F250QTZ $\lambda_{eff} = 2365$ Å, ACS/F220W $\lambda_{eff} = 2255.5$ Å, and ACS/F250W $\lambda_{eff} = 2716$ Å (e.g., Madrid 2009). We have used the UV measurements presented in Madrid (2009) and translated them into WFC3/F275W using spectral indices reported by Perlman et al. (2011), who observed M87 in four UV/optical bands during the same period. In addition, Madrid (2009) performed photometry with an aperture of radius $0\rlap{.}''25$, so that to construct a uniform UV lightcurve, we have recalculated our measurements in F275W band using the same aperture. For the optical lightcurve in F606W band, we have used measurements provided by

---

[253] https://www.stsci.edu/hst/instrumentation/wfc3/data-analysis/psf

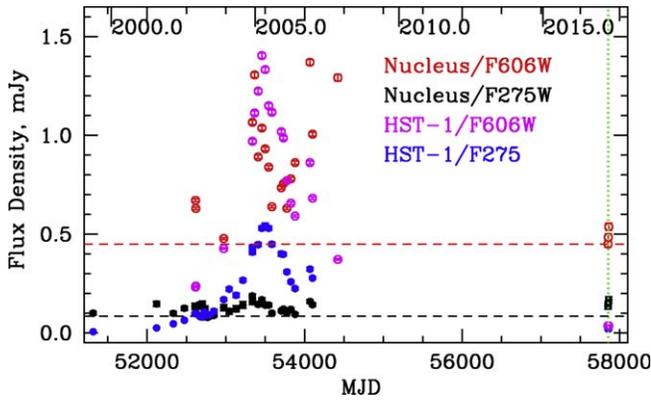

**Figure 7.** Historical optical and UV light curves of M87's core and the HST-1 knot (see also Madrid 2009; Perlman et al. 2011). Horizontal dashed lines indicate the lowest flux density level of the core in the F606W (red) and F275W (black) filters, the vertical green dotted line marks the time of the EHT campaign.

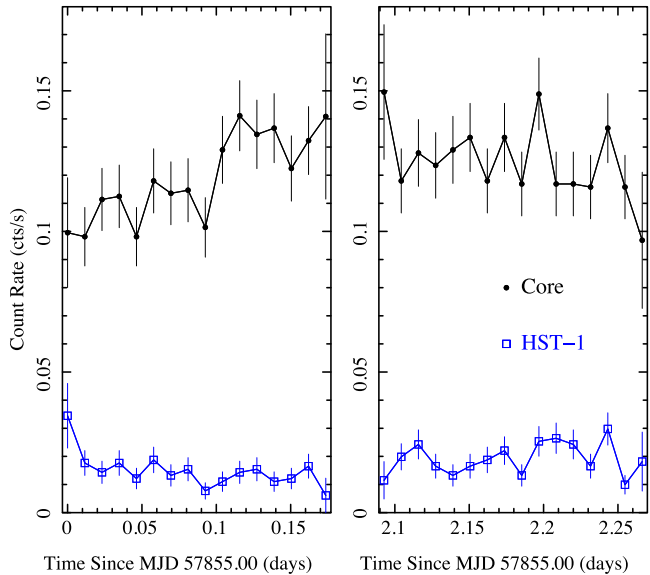

**Figure 8.** Chandra X-ray lightcurve of the core of M87 (black) and HST-1 (blue) in 2017 April, showing a small amount of variability between observations 20034 (left panel) and 20035 (right panel).

Perlman et al. (2011) obtained with Advanced Camera for Surveys (ACS)/F606W and Wide Field Camera 2 (WFC2)/F606W. Since $\lambda_{eff}$ of these instruments in F606W band is the same as for WFC3/F606W and photometry was performed with the same aperture of radius $0\rlap.{''}4$, no corrections to the measurements have been applied. Figure 7 shows the historical UV/optical light curves of M87. Independent of the filter used, HST-1 was in its lowest brightness state ever observed during the EHT campaign. Although the core was in its lowest brightness state at optical wavelengths in 2017, the better-sampled UV lightcurve suggests that the core was in an average quiescent state.

### 2.3. X-Ray Observations

#### 2.3.1. Chandra

We requested Director's Discretionary Time (DDT) observations of M87 with the Chandra X-ray Observatory to coordinate with the EHT campaign. The source was observed with Advanced CCD Imaging Spectrometer (ACIS)-S for 13.1 ks starting on 2017 April 11 23:46:58 UT (ObsID 20034) and again starting on 2017 April 14 02:00:28 UT (ObsID 20035). In order to perform spatially resolved spectral and variability analysis (see Section 2.3.3), we also analyzed several archival Chandra observations. Following Wong et al. (2017), for constraints on the intra-cluster medium (ICM), we included ObsIDs 352, 3717, and 2707, which were acquired on 2000 July 29, 2002 July 5, and 2004 July 6 and have good exposures of 37.7 ks, 98.7 ks, and 20.6 ks, respectively.

The Chandra observations were processed using standard data reduction procedures in CIAO v4.9.[254] We focused on extracting spectra from the core, the knot HST-1, and the outer jet, along with instrumental response files for spectral analysis. We took positions for the core and HST-1 from Perlman & Wilson (2005). For the core, we used a circular source extraction region with a radius of $0\rlap.{''}4$ centered on M87 (the approximate FWHM of $0\rlap.{''}8$ is quoted in Table A8 in Appendix A). The core background region is a half annulus centered on M87 with inner and outer radii of $2''$ and $3\rlap.{''}5$, respectively; we excluded the half of the annulus that is on the same side of the core as the extended X-ray jet. For HST-1, we used a similar circular source region, but for the background annulus we excluded ~90° wedges containing the core on one side and the extended jet on the other. For the jet itself, we used a $19\rlap.{''}5 \times 3''$ rectangular source region centered on the jet, with $19\rlap.{''}5 \times 1\rlap.{''}5$ rectangular regions on either side. To illustrate the relative brightness and variability of the core and HST-1, we show their lightcurves (1 ks bins) in Figure 8. Sun et al. (2018) presented a Chandra study spanning ~8 yr from 2002 to 2010 with coverage of the core and HST-1 in low and high states (see their Figure 3). In their study, the core flux drops as low as $\sim 10^{-12}$ erg s$^{-1}$ cm$^{-2}$ (the average is $\sim 4 \times 10^{-12}$ erg s$^{-1}$ cm$^{-2}$). In the 2017 Chandra observations, the unabsorbed core flux in the 0.3–7 keV band is $3 \times 10^{-12}$ erg s$^{-1}$ cm$^{-2}$, and the absorbed flux is $2 \times 10^{-12}$ erg s$^{-1}$ cm$^{-2}$—hence our observations show the core below the historical mean.

Because the ICM contributes significantly to the NuSTAR background, we used a single set of extraction regions to produce the Chandra ICM spectrum and the NuSTAR spectra (circular regions of radius $45''$, see Section 2.3.2 for more details). In extracting the Chandra ICM spectrum, we excluded the source regions for the core, HST-1, and the extended jet, all of which fit well within the NuSTAR PSF (Section 2.3.2).

#### 2.3.2. NuSTAR

We also requested two DDT observations of M87 with NuSTAR (Harrison et al. 2013) to coordinate with the EHT campaign in 2017 April. These observations are contemporaneous with the Chandra observations described in Section 2.3.1, and they are summarized in Table A6 in Appendix A. NuSTAR observations of M87 in 2017 February and April have been presented in Wong et al. (2017).

Raw data from NuSTAR observations were processed using standard procedures outlined in the NuSTAR data analysis software guide (Perri et al. 2017). We used data analysis software (NuSTARDAS, version 1.8.0), distributed by HEASARC/HEASOFT, version 6.23. Instrumental responses were calculated based on HEASARC CALDB version 20180312.

---
[254] https://cxc.cfa.harvard.edu/ciao/threads/index.html

We cleaned and filtered raw event data for South Atlantic Anomaly (SAA) passages using the `nupipeline` script; both minimal and maximal filtering levels were considered, with no substantial differences in the results. We used a source extraction region that is a circle with 45″ radius, while the background region was chosen as an identical region well separated from the peak of the hard X-ray emission (which is a point source above 12 keV, per Wong et al. 2017). Note that this does not include all of the low-energy extended emission resolved with Chandra (see Section 2.3.1), nor does it subtract it out. However, we modeled the low-energy ($\lesssim$3 keV) X-ray and high-energy ($\gtrsim$8 keV) X-ray emission jointly, as described in Section 2.3.3. Source and background spectra were computed from the calibrated and cleaned event files using the `nuproducts` tool.

*2.3.3. Joint Chandra and NuSTAR Spectral Analysis*

Given the spatial complexity of the underlying X-ray emission, any detailed analysis of the X-ray spectrum of M87 must involve a joint treatment of Chandra and NuSTAR data. Here we discuss our strategy for this analysis.

To properly recover the intrinsic spectrum of the core in M87 up to 40–50 keV, it was necessary to either subtract or model the bright emission of the ICM. Given our interest in the core and HST-1, which are moderately bright point sources, we must correct our spectra for pileup (see Section 2.3.1). Since this correction depends on the total count rate per pixel per frame time (Davis 2001), it must be applied without background subtraction. Thus, we chose to model the ICM spectrum.

We also needed to account for variations between different detectors and epochs. It is common practice to include a cross-normalization constant between NuSTAR's FPMA and FPMB detectors. We opted to take the same approach when jointly modeling NuSTAR and Chandra data: we allowed additional cross-normalization constants between the Chandra spectrum and the NuSTAR spectrum. Because the archival Chandra observations we used to constrain the ICM spectrum are nearly 20 yr old (and because of Chandra's changing effective area at soft X-ray energies due to contaminant buildup[255]), we allowed those data to have a different cross-normalization constant than the 2017 data.

Schematically, then, we can represent the model for the NuSTAR spectra of M87 as follows:

$$F_{\text{NuSTAR,FPMA}} = A \times F_{\text{Chandra,ICM}} \\ + B \times (F_{\text{Chandra,core}} + F_{\text{Chandra,HST}-1} + F_{\text{Chandra,jet}})$$

and

$$F_{\text{NuSTAR,FPMB}} = C \times F_{\text{Chandra,ICM}} \\ + D \times (F_{\text{Chandra,core}} + F_{\text{Chandra,HST}-1} + F_{\text{Chandra,jet}}),$$

where $A$, $B$, $C$, and $D$ are cross-normalization constants, $F_{\text{Chandra,ICM}}$ is the model for the Chandra ICM emission, and similar notation holds for the Chandra spectra of the core, the nearby knot HST-1, and the rest of the jet. Again, the purpose of this procedure is to use Chandra to account for all of the X-ray emission inside the NuSTAR extraction region, while allowing for time-dependent normalization constants between missions and detectors.

[255] https://cxc.cfa.harvard.edu/ciao/why/acisqecontamN0013.html

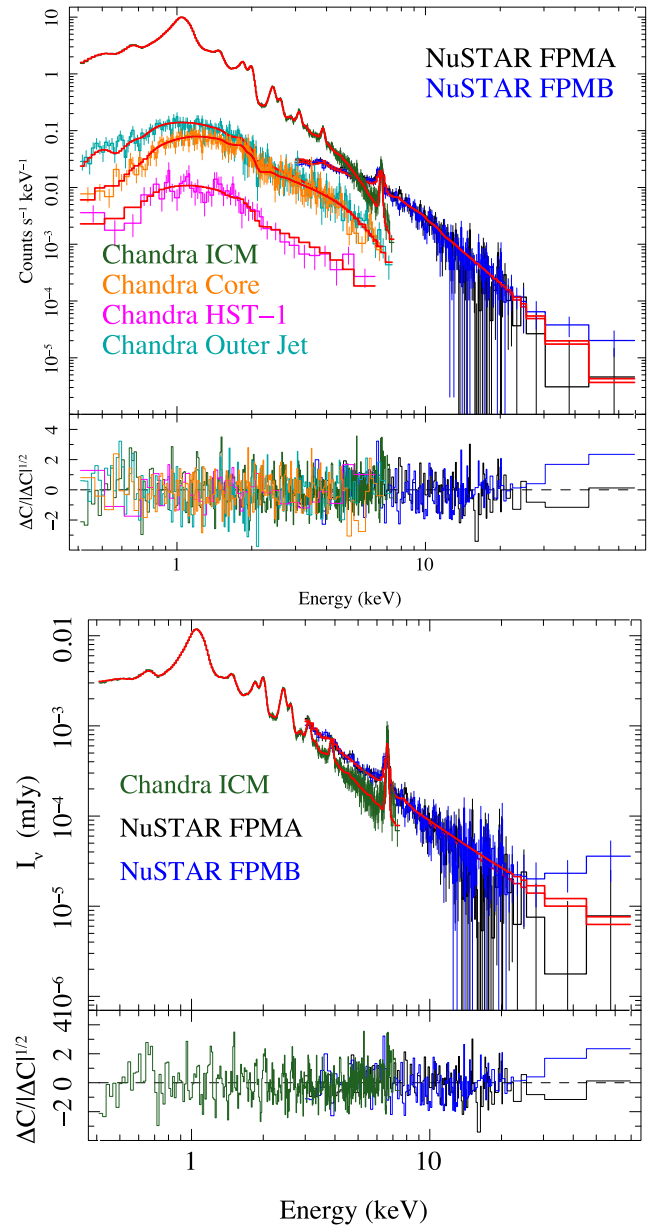

**Figure 9.** Chandra and NuSTAR spectra of M87. Top panel: count rate spectra from both observatories from 2017 April. Spectra have been rebinned for plotting purposes. In both panels, the Chandra spectrum of the underlying ICM is shown in green, while the NuSTAR FPMA and FPMB are shown in black and blue, respectively; the red curve is the total model spectrum for each data set. In the top panel, the spectrum of the core, HST-1, and the outer jet are shown in orange, magenta, and cyan, respectively. Bottom panel: unfolded spectra from Chandra and NuSTAR. Because the Chandra core/HST-1/jet spectra have the pileup kernel applied (Section 2.3.1), they are excluded from the flux plot.

With the structure of our model defined, we proceeded to select models for each of the components. The ICM has a complex physical structure and three-dimensional temperature profile. Our purpose here was not to model the cluster gas physics but to reconstruct the ICM spectrum with sufficient accuracy to model the X-ray continuum from M87 itself. Following Wong et al. (2017), we adopted a two-temperature Astrophysical Plasma Emission Code (APEC) model (Smith et al. 2001) with variable abundances (vvapec). We allowed the Ne, Na, Mg, Al, Si, S, Ar, Ca, and Fe abundances to vary and included three Gaussian emission lines; note that this treatment differs slightly from Wong et al. (2017), who used solar metallicities and 5 Gaussians to account for their residuals relative to the APEC models. We also included a cutoff power-law to model the low-mass X-ray binary (LMXB) emission in the extraction region with photon index $\Gamma = 0.5$ and cutoff energy $E_c = 3$ keV (Revnivtsev et al. 2014). This model was modified by interstellar absorption; we use the model tbnew (Wilms et al. 2000) with atomic cross-sections from Verner & Yakovlev (1995).

For the core, HST-1, and the remainder of the jet, we modeled their continuum emission using power laws (modified by the same interstellar absorption component as the ICM emission). However, inspection of the spectra in the top panel of Figure 9 revealed a steeper rollover in the soft X-ray spectrum of the core (orange) than in HST-1 or the outer jet. This is typical of ISM absorption, so we included a second tbnew component in the model for the spectrum of the core (see Wilson & Yang 2002; Perlman & Wilson 2005, but also Di Matteo et al. 2003). As noted above, these spectra required a pileup correction; the "grade migration" parameter $\alpha$ is tied between the two contemporaneous Chandra spectra of each spatial component.

While the Chandra data effectively disentangle the spatial components of M87, the NuSTAR data are superior when it comes to constraining the photon index, given their wider energy range. But because the count rates are low, we allowed the model X-ray flux to vary between observations while assuming the photon index was constant.

For fitting, the Chandra spectra were binned to a minimum S/N of 3 between 0.4 and 8 keV. To maximize the useful energy range of the NuSTAR data, we grouped the spectrum from each FPM to a minimum S/N of only 1.1; we also rebinned the spectra by factors of 3, 4, 5, 6, and 7 over the energy ranges 3–51 keV, 51–59 keV, 59–67 keV, 67–74.9 keV, and 74.9–79 keV, respectively (this reduces oversampling of the energy resolution; J. Steiner 2021, private communication). Because very low count rates are involved for some energy bins, we used Cash statistics (Cash 1979). Our best fit gave a reduced Cash statistic of 1.26 (a Cash statistic of 1392 for 1139 data bins and 38 free parameters). Our power-law model does not rule out possible curvature in the high-energy X-ray spectrum.

Given the possibility of complex correlations between parameters in this model, we opted to determine confidence intervals for our parameters using Markov Chain Monte Carlo (MCMC) methods. We used an implementation of the affine-invariant sampler emcee, after Goodman & Weare (2010) and Foreman-Mackey et al. (2013), with 10 walkers for each of the 38 free parameters and let the sampler run for 20,000 steps.

At present, we focus on the spectral index, flux, and ISM absorption for each of our spatial components; additional parameters shall be described below as necessary. Unless otherwise noted, quoted uncertainties represent 90% credible intervals; these generally align well with our 90% confidence intervals from direct fitting.

Our results confirm our inspection of the Chandra spectrum of the core of M87: it is more highly absorbed than the rest of the jet. All components required a small absorbing column density $N_H = 0.050^{+0.003}_{-0.002} \times 10^{22}$ cm$^{-2}$, though this is larger than the Galactic value from H I studies ($0.025 \times 10^{22}$ cm$^{-2}$; Dickey & Lockman 1990, $0.025 \times 10^{22}$ cm$^{-2}$; HI4PI Collaboration et al. 2016). The core, however, required an additional $N_{H,excess} = 0.12^{+0.05}_{-0.04} \times 10^{22}$ cm$^{-2}$ of intervening gas, over and above the column toward the rest of the jet. The origin of this excess absorption is not known. It was not detected by Di Matteo et al. (2003), but was mentioned by Wilson & Yang (2002) and Perlman & Wilson (2005). If it is real, it is most likely absorption in the galaxy M87 itself, though it is difficult to speculate on its properties without knowing more about its variability, which will be discussed in future work (J. Neilsen et al. 2021, in preparation). An alternative interpretation is that there is measurable curvature in the core spectrum at these energies.

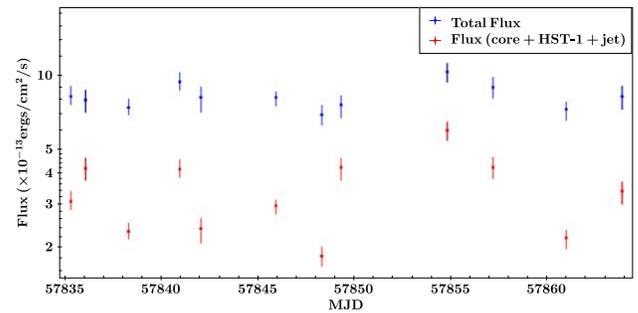

**Figure 10.** Variability in the Swift-XRT unabsorbed flux (2–10 keV) from 2017 March 27 to April 20. The Total Flux refers to the overall unabsorbed flux derived from the spectral fitting to the composite model. The net flux from only the three components, i.e., core, HST-1 and outer jet is referred to as the Flux(core+HST-1+jet), which was calculated from their respective spectral fitted parameters.

The flux itself is not a parameter of our fits, but it is easily calculated from the normalization of the power law. We drew 1000 samples from our chain, excluding the first 4000 steps to account for the burn-in period. For each sample we used the power-law photon index and normalization for each spectral component to calculate the 2–10 keV luminosity, assuming a distance of 16.8 Mpc as in EHT Collaboration et al. (2019a). For 2017 April, we found an X-ray luminosity of $(4.8 \pm 0.2) \times 10^{40}$ erg s$^{-1}$, with a photon index $\Gamma = 2.06^{+0.10}_{-0.07}$.

### 2.3.4. Swift-X-Ray Telescope (XRT)

M87 was observed using Swift-XRT from 2017 March 27 to April 20. A total of 24 observations in the photon counting mode were conducted, out of which 12 were discarded due to the fact that M87 was centered on the bad columns, which resulted from a micrometeorite impact in 2005. For the purpose of this work, we analyzed data for the remaining 12 epochs. First, all the cleaned level 3 event files were generated using xrtpipeline version 0.13.5. For spectral extraction, a circular source region with a diameter of 15 pixels (35″) and an annular background region were chosen. If the source counts exceeded 0.5 cts s$^{-1}$, a pile-up correction was performed. In this case, an appropriate annular region was selected as the source region for the final spectrum extraction, ensuring that the piled-up region is removed from the final extraction process. The exposure maps created by the xrtpipeline were utilized to create the Ancillary Response File (arf) for each spectrum employing the xrtmkarf command. The Response Matrix File (rmf) used in this process was later used to group all these spectral files for each epoch, using the grppha command.

The spectral resolution of Chandra provided well-constrained model parameters for M87 data obtained during the same time as Swift-XRT. The composite model contained a complex background model, which is a sum of the ISM, ICM, and any remaining background. All the details for this model, in addition to the details for the power-law models for the core, HST-1 and the jet, respectively, are described in the previous section. These Chandra derived parameters were used as a guide to fit the spectra obtained from Swift-XRT, freezing all parameters other than (1) the overall normalization of the power-law continuum components (spectral slopes and *relative* normalizations remained fixed) and (2) the normalization of the background component, vvapec. The overall variability seen in Figure 10 contains contributions from both the power-law continuum and background component variations. The variability of the background component may be an artifact of our assumptions about the cross-normalization of the Swift-XRT models relative to Chandra and NuSTAR. It could also indicate some residual systematic uncertainty in the spectral decomposition (background versus core, jet, and HST-1) that is not accounted for in our error bars. In any case, the non-ICM fluxes represent an upper limit on the core X-ray emission for our SED modeling, and so this precise decomposition does not have a significant impact on our conclusions about the SED.

The overall evolution of the lightcurve for M87 in 2017 is displayed in Figure 10. We derived an overall total unabsorbed flux using the spectral fits for each epoch. We also calculated the net flux resulting from the core, HST-1, and outer jet only from these fits by removing the background model from the final fit, which helped estimate the combined flux only from these three components of M87. Both these flux values are reported in Figure 10. A day to day variability of the order of about ∼5%–20% is seen during this time. However, an overall variation in the flux by about a factor of 3 was seen during 2017. Note that the spatial location of the variability can not be provided due to the lower angular resolution of Swift-XRT.

### 2.4. γ-Ray Data

#### 2.4.1. Fermi-Large Area Telescope (LAT) Observations

The γ-ray emission at GeV energies from M87 was first reported in 2009, using the first 10 months of observations with Fermi-LAT (Abdo et al. 2009a). Fermi-LAT (Atwood et al. 2009) mainly operates in survey mode, observing the whole sky every three hours. Connected to the EHT campaign, during the period 2017 March 22 to April 20, pointing mode observations[256] of M87 have been specially performed providing additional 500 ks of data on this source, allowing the source to be effectively observed >56% of the time (w.r.t. the standard ∼48%) during the EHT campaign.

We performed a dedicated analysis of the Fermi-LAT data of M87 using 11 yr of LAT observations taken between 2008 August 4 and 2019 August 4. Similarly, we repeated the analysis using three-month time bins centered on the EHT observation period (i.e., 2017 March 1–May 31). We selected P8R3 Source class events (Bruel et al. 2018), in the energy range between 100 MeV and 1 TeV, in a region of interest (ROI) of 20° radius centered on the M87 position. The low-energy threshold is motivated by the large uncertainties in the arrival directions of the photons below 100 MeV, leading to a possible confusion between point-like sources and the Galactic diffuse component. See Principe et al. (2018, 2019) for a different analysis implementation to solve this and other issues at low energies with Fermi-LAT.

The analysis (which consists of model optimization, and localization, spectrum, and variability study) was performed with Fermipy[257] (Wood et al. 2017), a Python package that facilitates analysis of LAT data with the Fermi Science Tools, of which the version 11-07-00 was used. The counts maps were created with a pixel size of 0°.1. All γ-rays with zenith angle larger than 95° were excluded in order to limit the contamination from secondary γ-rays from the Earth's limb (Abdo et al. 2009b). We made a harder cut at low energies by reducing the maximum zenith angle and by selecting event types with the best PSFs.[258] For energies below 300 MeV we excluded events with zenith angle larger than 85°, as well as photons from PSF0 event type, while above 300 MeV we use all event type. The P8R3_Source_V2 instrument response functions (IRFs) are used. The model used to describe the sky includes all point-like and extended LAT sources, located at a distance <25° from the source position, listed in the Fourth Fermi-LAT Source Catalog (4FGL; Abdollahi et al. 2020), as well as the Galactic diffuse and isotropic emission. For these two latter contributions, we made use of the same templates[259] adopted to compile the 4FGL. For the analysis we first optimized the model for the ROI (fermipy.optimize), then we searched for the possible presence of new sources (fermipy.find_sources) and finally we re-localized the source (fermipy.localize). We investigated the possible presence of additional faint sources, not in 4FGL, by generating test statistic[260] (TS) maps and we found two candidate new sources, detected at $5\sigma$ level, that we added into our model. The best-fit positions of these new sources are R.A., decl. = (184°.85, 5°.82) and (191°.99, 7°.20), with 95% confidence-level uncertainty $R_{95} = 0°.07$. We left free to vary the diffuse background and the spectral parameters of the sources within 5° of our target. For the sources at a distance between 5° and 10° only the normalization was fitted, while we fixed the parameters of all the sources within the ROI at larger angular distances from our target. The spectral fit was performed over the energy range from 100 MeV to 1 TeV. To perform a study of the γ-ray emission variability of M87 we divided the Fermi-LAT data into time intervals of 4 months. For the lightcurve analysis we have fixed the photon index to the value obtained for 11 yr and left only the normalization free to vary. The 95% upper limit has been reported in each time interval with TS < 10.

The results on the 11 yr present a significant excess, TS = 1781 corresponding to a significance $>40\sigma$, centered on the position (R.A., decl.) = (187°.73 ± 0°.03, 12°.36 ± 0°.04), compatible with the position of 4FGL J1230.8 + 1223 and associated with M87. The averaged flux for the entire period is $\mathrm{Flux}_{11\,\mathrm{yr}} = (1.72 \pm 0.12) \times 10^{-8}\,\mathrm{ph\,cm^{-2}\,s^{-1}}$. We model the spectrum of the source with a power-law function ($\frac{dN}{dE} = N_0 \times \left(\frac{E}{E_0}\right)^{-\Gamma}$). The spectral best-fit results for 11 yr of LAT data of M87 are $\Gamma = 2.03 \pm 0.03$ and $N_0 = (1.64 \pm 0.07) \times 10^{-12}$ (MeV cm$^{-2}$ s$^{-1}$) for $E_0 = 1$ GeV, which are in agreement with the 4FGL results for this source ($\Gamma_{4FGL} = 2.05 \pm 0.04$). For the three-month EHT observation period specifically (i.e., 2017 March 1–May 31), the source is detected with a significance of $\simeq 8\sigma$ (TS = 72). The spectrum obtained for the three-month period centered on the EHT observation presents a slightly harder photon index with respect to the one obtained using 11 yr, with a power-law index of $\Gamma = 1.84 \pm 0.18$ and $N_0 = (1.50 \pm 0.44) \times 10^{-12}$ MeV cm$^{-2}$ s$^{-1}$. During the EHT observational period, M87 reveals a brightness level comparable to the average state of the source, with a flux ($F_{2017} = 1.58 \pm 0.50 \times 10^{-8}$ ph cm$^{-2}$ s$^{-1}$) compatible with the averaged flux estimated for the entire period.

---

[256] Fermi ToO: https://fermi.gsfc.nasa.gov/ssc/observations/timeline/too/090606-1-1.html.

[257] http://fermipy.readthedocs.io/en/latest/

[258] A measure of the quality of the direction reconstruction is used to assign events to four quartiles. The γ-rays in Pass 8 data can be separated into four PSF event types: 0, 1, 2, 3, where PSF0 has the largest PSF and PSF3 has the best.

[259] https://fermi.gsfc.nasa.gov/ssc/data/access/lat/BackgroundModels.html

[260] The test statistic is the logarithmic ratio of the likelihood of a source being at a given position in a grid to the likelihood of the model without the source, $\mathrm{TS} = 2\log(\frac{\mathrm{likelihood_{src}}}{\mathrm{likelihood_{null}}})$ (Mattox et al. 1996).

### 2.4.2. H.E.S.S., MAGIC, VERITAS Observations

In 1998 the first strong hint of very-high energy (VHE; $E > 100$ GeV) $\gamma$-ray emission from M87 was measured by the High Energy Gamma Ray Astronomy (HEGRA) Collaboration (Aharonian et al. 2003). Since 2004, the source has been frequently monitored (Acciari et al. 2009; Abramowski et al. 2012) in the VHE band with the High Energy Stereoscopic System (H.E.S.S.; Aharonian et al. 2006a), the Major Atmospheric Gamma Imaging Cherenkov (MAGIC; Albert et al. 2008; Aleksić et al. 2012; MAGIC Collaboration et al. 2020) telescopes, and the Very Energetic Radiation Imaging Telescope Array System (VERITAS; Acciari et al. 2008, 2010; Aliu et al. 2012; Beilicke & VERITAS Collaboration 2012).

Over the course of its monitoring, M87 exhibited high VHE emission states in 2005 (Aharonian et al. 2006a), 2008 (Albert et al. 2008), and 2010 (Abramowski et al. 2012) that lasted one to a few days. A weak two-month VHE enhancement was also reported for 2012 (Beilicke & VERITAS Collaboration 2012), but since 2010, no major outburst has been detected in VHE $\gamma$-rays. During the observed high states the flux typically increased by factors of two to 10, while no significant spectral hardening was detected. In the following we describe the participating instruments, observations, and data analyses of the coordinated 2017 MWL campaign.

The H.E.S.S. observations presented here were performed using stereoscopic observations with the four 12 m CT1-4 telescopes (see Table A7). These observations of a total of 7.9 hr of live time were taken in so-called "wobble" mode with an offset of 0°.5 from the position of M87, allowing simultaneous background estimation. The reconstruction of the Cherenkov shower properties was done using the ImPACT maximum likelihood-based technique (Parsons & Hinton 2014; Parsons et al. 2015) and hadron events were rejected with a boosted decision tree classification (Ohm et al. 2009). The background was estimated using the so-called "ring-background model" for the signal estimation and the "reflected-background model" for the calculation of the flux and the spectrum (Aharonian et al. 2006b). For the 7.9 hr of data a total statistical significance of $3.7\sigma$ was calculated. A source with 1% of the Crab Nebula's flux can be detected in $\sim$10 hr. Based on the estimated systematic uncertainties following Aharonian et al. (2006b), the systematic uncertainty of the flux has been adopted to be 20%.

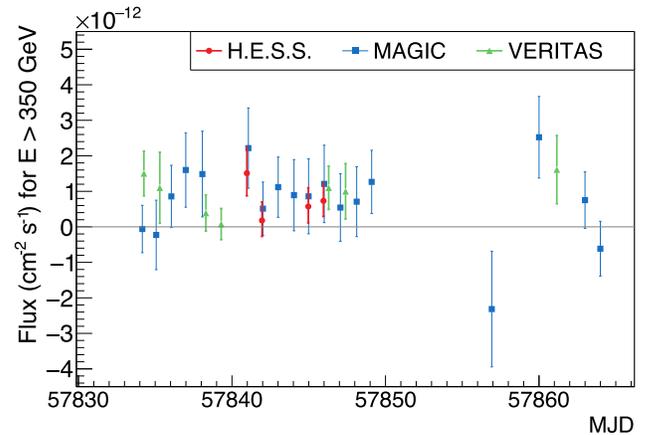

**Figure 11.** Flux measurements of M87 above 350 GeV with 1$\sigma$ uncertainties obtained with H.E.S.S., MAGIC, and VERITAS during the coordinated MWL campaign in 2017. Upper limits for flux points with a significance below 2$\sigma$ are provided in Table A7 in Appendix A.

For this study MAGIC observed M87 for a total of 27.2 h after quality cuts with the standard wobble offset of 0°.4. The data were analyzed with the standard MAGIC software framework MAGIC Analysis and Reconstruction Software (MARS; Zanin et al. 2013; Aleksić et al. 2016). We used the package SkyPrism, a spatial likelihood analysis, to determine the PSF in true energy (Vovk et al. 2018). 6.7 hr of these data were collected in the presence of moonlight and analyzed according to Ahnen et al. (2017). All MAGIC observations together yielded a total significance of $4.6\sigma$. MAGIC can detect a source with 1% of the Crab Nebula's flux in $\sim$26 hr above an energy threshold of 290 GeV. A detailed description of the various sources and estimates of systematic uncertainties in the MAGIC telescopes and analysis can be found in Aleksić et al. (2016). From there we estimate a systematic flux normalization uncertainty of 11%, a systematic uncertainty on the energy scale of 15% and a systematic uncertainty of $\pm 0.15$ on the reconstructed spectral slope for the MAGIC observations, which sums up to a total of $\sim$30% integral flux uncertainty.

VERITAS collected 15 hr of quality-selected observations of M87 during the 2017 EHT campaign window using a 0°.5 wobble. Data were analyzed using standard analysis tools (Cogan 2007; Daniel 2008; Maier & Holder 2017) with background-rejection cuts optimized for sources with spectral photon indices in the 2.5–3.5 range. The analysis yields an overall statistical significance of $3.8\sigma$. In its current configuration, VERITAS can detect a source with a flux of 1% of the $\gamma$-ray flux of the Crab Nebula within 25 hr of observation (Park 2015). We estimate a systematic uncertainty on the quoted photon flux of 30%.

Separate differential spectra and differential upper limits were produced assuming a power law of the form $dN/dE \propto E^{-\Gamma}$ for each instrument. For the different instruments the $\gamma$-ray flux is measured in five independent bins per energy decade, respectively, with flux points being quoted for bins with a significance larger than 2$\sigma$. A differential upper limit at the 95% confidence level following the method described in Rolke et al. (2005) is quoted otherwise. The bin edges vary among the different Imaging Atmospheric Cherenkov Telescope (IACT) measurements due to differences in instrumental and observational conditions. Night-wise integrated flux points and upper limits were calculated for the light curves above an energy threshold of 350 GeV assuming a differential spectrum that follows a simple power law with a spectral index of $\Gamma = 2.3$. Due to technological limits, and ultimately limited by fluctuations in the development of air showers, IACTs are not able to spatially resolve M87 and so the measurements are compatible with a point source (Hofmann 2006).

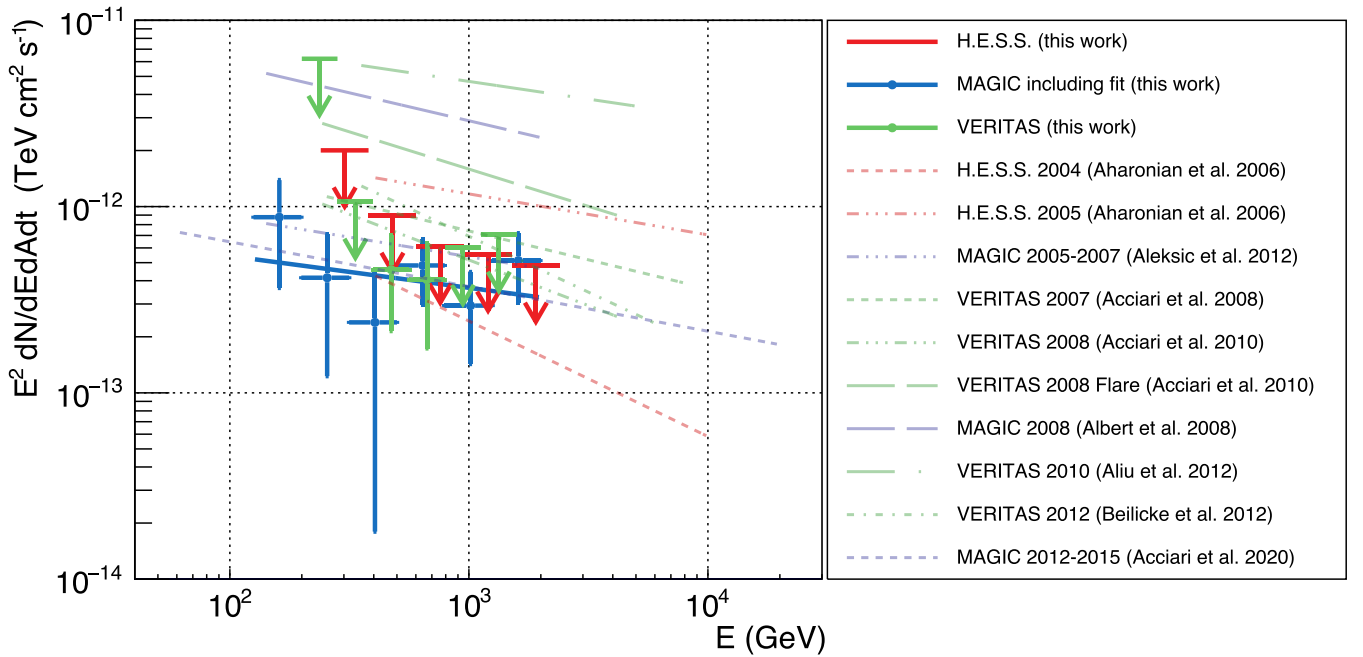

**Figure 12.** VHE SED measured with H.E.S.S., MAGIC, and VERITAS during the 2017 MWL campaign. Upper limits are represented by arrows taking only statistical uncertainties into account. The error bars are equivalent to $1\sigma$ confidence level and upper limits of the flux are given at the $2\sigma$ confidence level. Historical spectral shapes are taken from Aharonian et al. (2006a, 2006b), Aleksić et al. (2012), Acciari et al. (2008, 2010), Albert et al. (2008), Aliu et al. (2012), Beilicke & VERITAS Collaboration (2012), and MAGIC Collaboration et al. (2020).

A summary of the individual observation nights can be found in Table A7 in Appendix A. Figure 11 shows the measured fluxes for each observation night with $1\sigma$ statistical uncertainty while the flux upper limits for nights with $<2\sigma$ are given in Table A7 in Appendix A. All measurements are compatible with constant emission within uncertainties at an overall low state during the 2017 campaign (MAGIC Collaboration et al. 2020 and references therein). The resulting SEDs assuming that the intrinsic flux is constant for each instrument are shown in Figure 12 together with historical measured spectra. The index of a power-law fit to the MAGIC data is $\Gamma = 2.17 \pm 0.46$. This is in good agreement with previous results obtained during low-flux states MAGIC Collaboration et al. (2020). The data from H.E.S.S. and VERITAS do not allow one to perform a measurement of the VHE $\gamma$-ray spectral shape of M87, but yield upper limits and energy flux measurements that are consistent with the spectrum determined with the MAGIC data.

## 3. Results and Discussion

### 3.1. Multi-wavelength Images and Core Shift

In Figure 13 we show a composite of M87 multi-wavelength images obtained by various instruments during the 2017 campaign, including the EHT image. The M87 jet is imaged at all scales from ∼1 kpc down to a few Schwarzschild radii, summarizing a contemporaneous MWL view of M87 during the 2017 EHT campaign.

The large-scale jet structure is well imaged with ALMA, HST, and Chandra. The observed kpc-scale jet morphology in 2017 is overall consistent with that known in the literature: a straight, highly collimated jet continues along PA ∼ 290° with several discrete knots (e.g., Owen et al. 1989; Biretta et al. 1999; Snios et al. 2019). The near-core knot HST-1 is well isolated from the nucleus by EVN, HST, and Chandra. The EVN at 1.7 GHz further resolves the inner jet at ∼10–100 pc scales, revealing a continuous, straight jet morphology.

At parsec scales, high-resolution VLBI observations reveal the detailed transverse structure of the jet, resolving a limb-brightened structure that follows a parabolic collimation profile (Asada & Nakamura 2012; Hada et al. 2013; Nakamura & Asada 2013; Kim et al. 2018b). Closer to the black hole, the jet morphology may change more rapidly than at large scales (e.g., Britzen et al. 2017; Walker et al. 2018). Walker et al. (2018) reported a significant oscillation of the inner-jet position angle in their VLBA 43 GHz images on timescales of several years. During the 2017 campaign, our VLBA and EAVN 43 GHz images (see also Figure 14) reveal a jet direction within 2 mas to be close to east–west (PA ∼ 270°), significantly offset from the large-scale jet direction (PA ∼ 290°). This inner-jet direction observed in 2017 seems to follow the long-term periodic oscillation found by Walker et al. (2018). A consistent jet direction can also be seen in the GMVA 86 GHz image obtained in 2017 March, although the image quality is poorer than that of VLBA at 43 GHz. If the jet is conically precessing, an apparent change of ∼20° in PA implies a variation of jet inclination by $\Delta\theta \sim PA \times \sin\theta \sim 6°$ (for $\theta \sim 17°$) on timescales of several years. Interestingly, where the jet follows the east–west direction (see Figure 14), the southern jet limb tends to be brighter than the northern limb, which is similar to the north–south asymmetric brightness pattern seen in the EHT-2017 image (EHT Collaboration et al. 2019a). We do not see any clear signature of prominent component ejection from the core during our campaign, although some motion in the underlying flow may exist near the core (e.g., Mertens et al. 2016; Walker et al. 2018; Park et al. 2019).

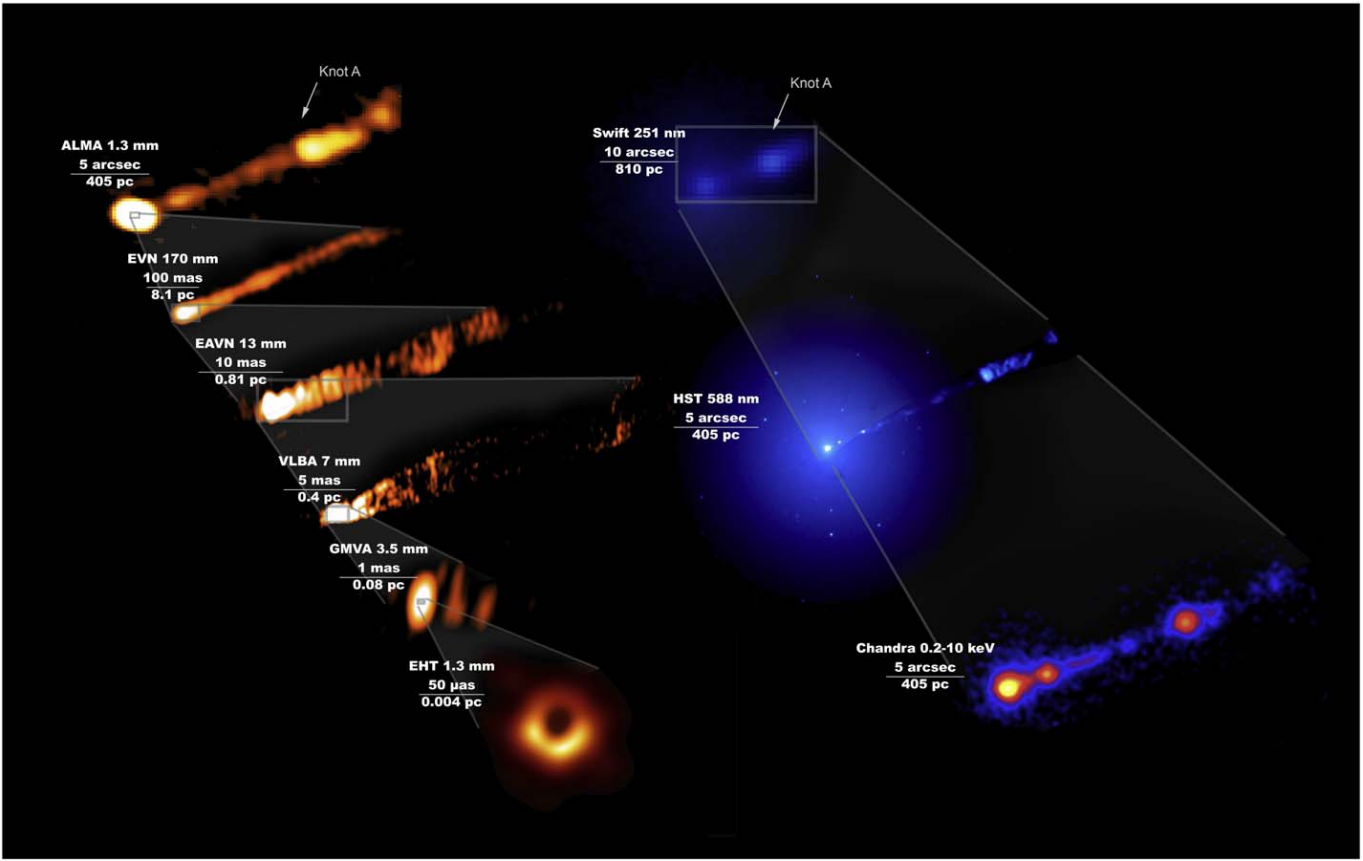

**Figure 13.** Compilation of the quasi-simultaneous M87 jet images at various scales during the 2017 campaign. The instrument, observing wavelength, and scale are shown on the top-left side of each image. Note that the color scale has been chosen to highlight the observed features for each scale, and should not be used for rms or flux density calculation purposes. Location of the Knot A (far beyond the core and HST-1) is shown in the top figures for visual aid.

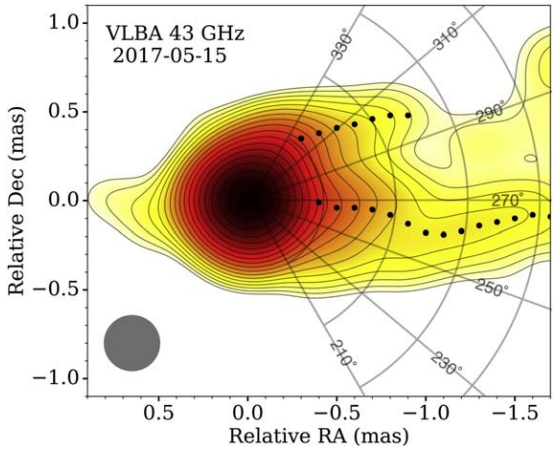

**Figure 14.** Innermost region of the M87 jet observed with VLBA at 43 GHz. To better describe the transverse jet structure, the image is convolved with a circular beam of 0.31 mas (shown in the bottom-left corner of the figure), which is slightly superresolved along the north–south direction. Black dots indicate the ridges of the northern and southern limbs beyond 0.2 mas from the core. For reference, a polar axis is overlaid to show corresponding PA. Contour levels are scaled as $(1, 1.4, 2, 2.8...) \times 3$ mJy beam$^{-1}$.

In Figure 15, we show two-frequency spectral index distribution maps within the mas-scale region of the M87 jet. For the analysis, the $(u, v)$-coverage was matched for the corresponding pair of frequencies and images. The 22–43 GHz map was obtained by stacking EAVN images of the three pairs of quasi-simultaneous epochs around the dates of the EHT campaign (2017 March 18–April 18) at 22 and 43 GHz. The total intensity images at different frequencies were aligned by using the two-dimensional (2D) cross-correlation analysis (Walker et al. 2000) considering the optically thin regions of the jet. The 22–43 GHz spectral index map shows a flat-spectrum radio core, while the extended jet regions become progressively optically thin, which is typical for relativistic jets in AGN (e.g., Hovatta et al. 2014). A comparison of the spectral index maps produced from stacked 22–43 GHz EAVN and 24–43 GHz VLBA observations, which are separated in time by roughly a month, shows no principal difference.

In the bottom panel of Figure 15 we also show a spectral index map zoomed into the central ~1 mas of the radio core, which was produced using the 43 GHz VLBA and 86 GHz GMVA images, aligned at the position of the peak flux densities on corresponding images. The core shows a flat spectrum, which together with the core shift indicates that the inner jets are stratified and synchrotron self-absorbed at least up to 86 GHz, as predicted by (e.g., Blandford & Königl 1979). Spectra of the jet details downstream of the core may give unreliable results owing to the sparseness of the GMVA data and non-simultaneity of GMVA and VLBA observations (more than a month), thus we omit their discussion here.

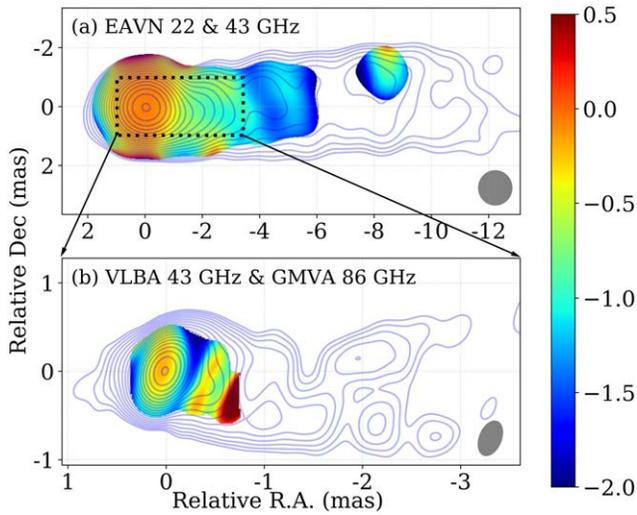

**Figure 15.** (a) Spectral-index map of the stacked EAVN image including three pairs of quasi-simultaneous epochs at 22 and 43 GHz. A common beam size of 1.2 mas × 1.2 mas, corresponding to the 22 GHz beam, is used for both frequencies. The images at different frequencies were aligned by doing the 2D cross-correlation analysis using optically thin regions of the jet. The dashed box indicates the region of the bottom panel. (b) Spectral-index map from VLBA 43 GHz on 2017 May 5 and GMVA 86 GHz on 2017 March 30. A common beam size of 0.2 mas × 0.4 mas at PA = 1°.2, corresponding to the 43 GHz beam, is used for both frequencies. We note that the small patch of high opacity is likely due to poor quality and large uncertainty in the flux of the 86 GHz data (see also Section 2.1.7 for details). Both of these two images are rotated clockwise by 18°. Contours in (a) indicate the EAVN 22 GHz total intensity map, while those in (b) are the VLBA 43 GHz total intensity map. Contour levels for both (a) and (b) are scaled as (1, 1.4, 2, 2.8...) × 1.9 mJy beam$^{-1}$. A restoring beam is indicated in the bottom-right corner of each panel. The clipping levels of both images are 10 mJy beam$^{-1}$, which corresponds to over 10 times the image rms.

### 3.2. Multi-wavelength SED of the M87 Core and High-energy Jet Emission

In Figures 2, 5, 7, 8, 10, and 11 we show the lightcurves from the radio to the $\gamma$-ray bands, taken contemporaneously with the 2017 April 5–11 EHT observations. Decades of long-term flux-monitoring programs and other previous work on M87 allow us to place these multi-wavelength fluxes in a historical context. From this analysis we conclude that during the 2017 EHT campaign, the emission from the innermost regions of M87, to the extent to which they can be resolved with our multi-wavelength observations, is consistent with being in a historically low/quiescent state. Specifically, HST-1 remained in a typical low state at radio (15 mJy at 1.7 GHz). In the optical and X-rays, HST-1 seems to be in the lowest state observed so far, following the continued flux decaying trend over the last decade (e.g., Abramowski et al. 2012; Sun et al. 2018). While the GeV/TeV facilities cannot resolve the source of the emission, they also find M87 to be in a typical low state in 2017. The fact that HST-1 is so sub-dominant to the core in the higher energy bands lends confidence to the association of the total fluxes from NuSTAR and Swift with the core emission, while the localization of the VHE emission is less constrained.

As described in Section 2, many different facilities carried out observations of M87 during or around the EHT observing campaign. Although these observations were not all strictly simultaneous, within the context of the typical variability timescale of the core of several days to weeks, they effectively provide a snapshot of the broadband SED of the M87 core in a quiescent, low-flux state. In addition to near simultaneity of the multi-wavelength observations, our data processing included careful analysis of non-core emission components in order to isolate the core emission as much as possible. Therefore, this legacy data set comprises the definitive target constraints for any model seeking to explain the source behavior at the time of the EHT 2017 image acquisition.

The list of fluxes over a broad range between frequency of $\simeq$1 GHz and photon energy $\simeq$1 TeV is given in Table A8 in Appendix A. We provide both $f_\nu$ and $\nu f_\nu$ values, although the differing data analysis procedures typically yield only one of these values. The other one is calculated by multiplication or division with the representative frequency. In the high-energy regime (>0.1 keV), we adopt the geometrical mean of the band's limits as appropriate even for relatively wide bands sampling power-law spectra. As described in Section 2, all flux points are based on statistically significant detections (>3$\sigma$ for all instruments except >2$\sigma$ for IACTs) and provided with uncertainties equivalent to the 1$\sigma$ confidence level. Upper limits on flux are given at the 2$\sigma$ confidence level.

Table A8 in Appendix A also contains band labels typical for the given energy or frequency range, as well as representative angular scales and the time ranges over which the data were acquired. While most observations were performed within a few days of the EHT observations, some of the data included in our broadband SED were acquired in single observations up to one month later, e.g., EVN, HSA, VLBA. For the majority of instruments we have multiple observations; in those cases we average the fluxes over observations taken within a week of the EHT campaign. In the case of Fermi-LAT, the fluxes provided for the SED are integrated over a period of three months, chosen as a compromise between simultaneity and photon-limited data quality.

It is very important to note that the emission measured by the instruments contributing to this broadband SED spans a factor of $\gtrsim 10^8$ in angular scale: from the effective spatial resolution of the EHT, $\simeq$20 $\mu$as, to the resolving power of Fermi-LAT in low-energy $\gamma$-rays, $\simeq$2°. In the radio bands, we list unresolved peak fluxes to represent the core. One exception is the EHT, for which we list the estimated total flux within a 60 × 60 $\mu$as$^2$ region, as described in EHT Collaboration et al. (2019f, their Appendix B). For elliptical beams, we list the average of the axes as a representative angular scale. At higher energies, the scale typically corresponds to the FWHM of the point-spread function within a given band or the diameter of the region used for data extraction. For Chandra and NuSTAR data analyzed jointly, the scale is tied to Chandra's greater resolving power, even though NuSTAR can only separate out the core emission spectroscopically (see Section 2.3.3). Because Swift-XRT cannot resolve the core from other components detected by Chandra, we treat its unresolved flux as an upper limit.

The data listed in Table A8 in Appendix A are plotted in Figure 16 with labels highlighting different observatories and instruments. For clarity, we omitted every other point in the X-ray spectrum and one high upper limit from the figure. In the figure we also highlight the upper limits on emission region size for the VLBI measurements. The machine-readable form of the table is provided to the community via the EHT Collaboration Data Webpage[261] along with associated documentation and all data files needed for spectral modeling of the Chandra, NuSTAR, and Swift-XRT data. The list of all supplementary material published along with this paper is provided in Appendix B.

---

[261] https://eventhorizontelescope.org/for-astronomers/data (alternatively, https://doi.org/10.25739/mhh2-cw46).

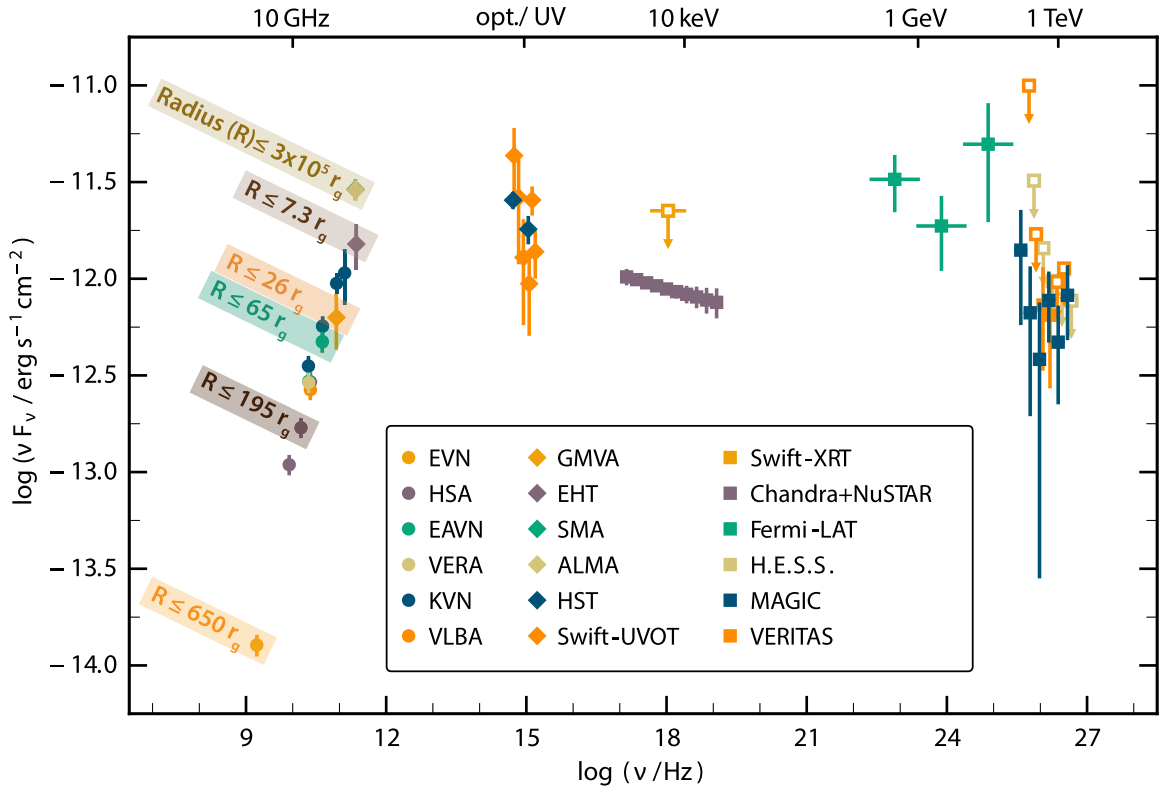

**Figure 16.** Observed broadband SED of M87 quasi-simultaneous with the EHT campaign in 2017 April (see Table A8 in Appendix A) with fluxes measured by various instruments highlighted with different colors and markers. Note that only every other point in the X-ray spectrum is plotted here and that one Fermi-LAT upper limit is missing, off to the upper right of the figure. For the mm-radio VLBI, the upper limits on emission size for several representative frequencies are labeled to clarify the constraints used in Section 3.3. An illustration of the resolved flux differences depending on spatial resolution is shown by the comparison of the differing EHT and ALMA-only 230 GHz fluxes and size limits.

### 3.3. Single-zone Heuristic Modeling and Interpretation

As is the case for many nearby low-luminosity AGNs where broadband SEDs are available, M87's radio core and SED indicate a multi-scale, stratified, and self-absorbed jet up to at least 86 GHz (see Section 3.1 and Figure 16). As this requires detailed dynamical modeling at a level that goes well beyond the goal of this Letter, we have opted to use the simpler single-zone approach common to many AGN blazar papers, in order to highlight some baseline properties of the peak emission regions. Furthermore, this approach will enable *relative* comparisons to other M87 epochs, as well as to the other AGNs observed by the EHT, which we plan to present in future papers.

A structured, multi-zone jet can to first order be considered as a summation of many single-zones. We consider two different single-zones, one representing the launchpoint of the jets and the EHT-detected emission region, and one roughly 100 times larger and thus probing a region further down in the inner jets. For each of these size scales we also explore a different approach. The first seeks to maximize the contribution to the entire broadband SED from the most compact regions imaged by EHT, while the second focuses on statistical fitting of the X-rays, allowing an exploration of the degeneracies inherent to such modeling. Both classes of model share a spherical geometry, assume isotropy, and calculate synchrotron and synchrotron self-Compton (SSC) emission, but otherwise have somewhat different parameters and approaches. However, in all cases the predicted radio emission must be at or below the flux measured for the associated size scale, to avoid violating the radio constraints. We have labeled these constraints for clarity in Figure 16 as well as provide a quick-reference in Table 1. It is important to emphasize that any single-zone model that dominates over a large range of radio/mm frequency will be in conflict with the core size constraints.

**Table 1**
Spatial Extent (Diameter of a Circle) within which Compact Radio Fluxes were Measured with VLBI Observations at Each Frequency

| Frequency (GHz) | Spatial Extent (Diameter) | |
|---|---|---|
| | (mas) | ($r_g$) |
| 230 | 0.06 | 15.5 |
| 129 | 1.0 | 260 |
| 86 | 0.2 | 52 |
| 43 | 0.5 | 130 |
| 15–24 | 1.0 | 260 |
| 8.4 | 1.5 | 390 |
| 1.7 | 5.0 | 1300 |

For all models, we take the mass to be $M_{\rm BH} = 6.5 \times 10^9 M_\odot$, the distance to be 16.8 Mpc, and a source inclination (viewing angle of the emitting region with respect to the line of sight) of $17°$, as adopted by EHT Collaboration et al. (2019a).

Table 2
Single-zone Model Parameters

| Model | $L_j$ ($L_{\rm Edd}$)[a] | $\delta$ | $R$ ($r_g$)[b] | $n_e'$ (cm$^{-3}$)[c] | $B'$ (mG) | $\gamma_{\min}$ | $\gamma_{\rm br}$ ($10^4$) | $\gamma_{\max}$ ($10^6$) | $p_1$ | $p_2$ | $U_e/U_B$ |
|---|---|---|---|---|---|---|---|---|---|---|---|
| 1a | $6 \times 10^{-3}$ | $1$ | $5.6$ | $3.6 \times 10^5$ | $4700$ | $1$ | $\cdots$ | $3.5$ | $2.2$ | $\cdots$ | $2.3$ |
| 1b | $3 \times 10^{-3}$ | $1$ | $5.2$ | $5.0 \times 10^5$ | $5000$ | $1$ | $0.5$ | $10$ | $2.6$ | $3.6$ | $1.1$ |
| 2 | $2.8^{+2.0}_{-1.4} \times 10^{-5}$ | $3.3$ | $626^{+256}_{-301}$ | $9.5^{+7.5}_{-7.8} \times 10^{-3}$ | $1.5^{+1.6}_{-0.9}$ | $4100^{+2100}_{-1500}$ | $\cdots$ | $64^{+26}_{-36}$[d] | $\cdots$ | $3.03^{+0.03}_{-0.05}$ | $635^{+465}_{-288}$ |

**Notes.** All values above in *italics* are not fitted parameters, they are either fixed before modeling, or derived for comparison between models (see the text). For model 2, the errors for $n_e'$ and $B'$ are calculated from the $1\sigma$ range for all the model parameters, using Equations (2) and (3).
[a] The total power for model 1a/1b is calculated in the same way as for model 2 assuming one proton per electron, but as protons are unconstrained this is solely for comparison purposes.
[b] For M87 $r_g = 9.8 \times 10^{14}$ cm = 65.51 Astronomical Units = 9.08 lt-hr.
[c] The number density of total nonthermal electrons or electron/positron pairs.
[d] Parameter pegged to the upper limit of the allowed interval.

### 3.3.1. Model 1: EHT-oriented Model

First, we consider a single-zone model that aims to provide a straightforward description of the flux and compact emission region size measured by EHT and other VLBI facilities. Even in the framework of a single-zone model, the predicted spectra depend strongly on how nonthermal electron (and positron in principle) distribution functions (eDF) are prescribed. In order to take into account such uncertainties, we therefore explore two different model scenarios with different eDF treatments. The first model, hereafter referred to as model 1a, includes the effects of radiative cooling on the initial single power-law eDF (see, e.g., Kino et al. 2002, and references therein). We also apply a broken power-law model to allow additional degrees of freedom in the eDF shape, but without directly calculating radiative cooling (RAIKOU code, see Kawashima et al. 2019, and also T. Kawashima et al. 2021, in preparation), hereafter referred to as model 1b. In this way we explore a large range of eDF parameter space for the single-zone approach.

Models 1a/1b share the following parameters determining the macroscopic characteristics of the spherical emission region: the radius ($R$), the Doppler factor of the bulk motion ($\delta$), and the global magnetic field strength ($B'$). The power law of nonthermal electrons injected in the emission region is characterized by the following quantities: the total (energy-integrated) number density of non-thermal electrons ($n_e$), the spectral indices of the injected power law $N(\gamma) \sim \gamma^{-p}$, and the minimum/maximum Lorentz factors ($\gamma_{\min}$ and $\gamma_{\max}$). Model 1a includes the feedback of radiative cooling on the eDF, thus it self-consistently calculates the break Lorentz factor ($\gamma_{\rm br}$), which is not included in Table 2 as a free parameter. On the other hand, model 1b treats the $\gamma_{\rm br}$ as a free parameter and thus it needs the second power-law index $p_2$ above $\gamma_{\rm br}$. Thus, model 1b includes additional degrees of freedom in the eDF.

Because models 1a/1b focus on describing an emission region with size and flux as determined from the EHT observations in 2017 April (EHT Collaboration et al. 2019a), we fix the emission region radius to around what is expected theoretically for the source of the observed blurred ring image in M87 (see Table 2). Since the bulk motion of the emission region has likely not yet reached a relativistic speed, we fix the Doppler factor to $\delta = 1$ and this is consistent with the brightness temperature measured at 230 GHz to be between $10^9$ and $10^{10}$ K (EHT Collaboration et al. 2019a). We also set the minimum Lorentz factor of the injected nonthermal electrons to $\gamma_{\min} = 1$–2, which has little impact on the resultant spectra but will affect the derived total power for the single-zone. To avoid over-producing the UV flux density, we fix the injected power-law spectral index of nonthermal electrons to be steeper than $p_1 = 2$.

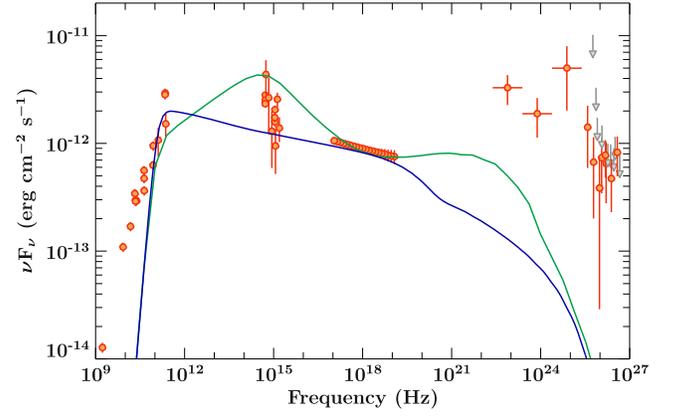

**Figure 17.** SED fit focusing on the EHT data. Blue and green lines display the resulting SEDs for models 1a and 1b, respectively. At $\gamma$-ray energy bands, one can see the SSC emission components although they underestimate the observed $\gamma$-ray flux density.

Once $R$ and $\delta$ are fixed to the above explained values (see Table 2), the only properties that can change the normalization of the synchrotron flux at 230 GHz are the magnetic field strength $B'$ and the eDF (particularly the particle distribution normalization) and the synchrotron-self-absorption (SSA) turnover frequency ($\nu_{\rm SSA}$; the SSA frequency for the most compact region that we are modeling). As discussed above, the radio core must be optically thick up to at least 86 GHz, and we are "tuning" this component to account for the EHT flux at 230 GHz. In order to not underpredict this flux we find that $\nu_{\rm SSA}$ must in fact be rather close to 230 GHz.

Then, together with the standard theory of the synchrotron self-absorption process, one can obtain an order of magnitude estimation of the magnetic field strength as follows:

$$B' \approx 12 \left(\frac{\theta_{\rm ring}}{40\,\mu{\rm as}}\right)^4 \left(\frac{\nu_{\rm SSA}}{190\,{\rm GHz}}\right)^5 \left(\frac{S_{\nu_{\rm SSA}}}{1\,{\rm Jy}}\right)^{-2} {\rm G}, \quad (1)$$

where the numerical factor (conventionally denoted as $b(p)$) related to synchrotron absorption has been described in the literature (Marscher 1983; Hirotani 2005; Kino et al. 2014) in detail and the slight variation in $b(p)$ in those literature values does not affect the estimation of the magnetic field strength shown here. It is thus clear that too small a value for $B'$ ($\sim$mG) cannot consistently reproduce the observed EHT synchrotron flux and optical depth constraints.

With this approximate scale for $B'$, we adjust the eDF to optimize the model, in order to have the SSC component also contribute to the X-ray flux. For model 1a, the resultant eDF steepens compared to its injected value of $p_1 = 2.2$, due to the rapid cooling of the injected electrons caused by the large $B'$. We then tune the model parameters in order to maximize the $\gamma$-ray flux, primarily by increasing $n_e$ to the maximum value that is still consistent with the VLBI observations. For model 1b, taking advantage of the flexibility of the double power law, we have tried to choose as large a $p_1$ value as possible within the range consistent with the UVOT and HST flux densities.

Our resulting best fits are shown in Figure 17. Both models 1a/1b show that the radio core emission peaking in the submm, as measured by VLBI, can be well explained by the optically thick part of the nonthermal synchrotron radiation. The model fitting gives the SSA turnover frequency $\sim$230 GHz which is consistent with the detection of the core shift constraints above, and in Hada et al. (2011).

The key difference between models 1a/1b can be seen by the spectra in the infrared (IR) band. The spectral break seen in model 1b is the result of tuning $\gamma_{br}$ in order to increase the $\gamma$-ray flux. The optical/UV data impose the strongest limits on the synchrotron model spectra. Both models avoid over-producing the Swift-UVOT flux in the optically thin part of the synchrotron spectrum, although model 1b is higher than the HST points, which may argue for even less $\gamma$-ray production than in this maximal case.

For both models, we find that synchrotron emission dominates over the SSC contribution in the X-rays. A key result is that neither EHT-oriented model can produce sufficient GeV and TeV $\gamma$-ray emission. We attempted to maximize the $\gamma$-ray flux, so these results indicate the upper limit possible without violating the low-energy constraints. Therefore, our results support the idea that the $\gamma$-rays detected in 2017 likely originate from a region that is more extended than the EHT-observed region ($R \lesssim 10\,r_g$).

### 3.3.2. Model 2: High-energy Oriented Model

As it is clear that the $\gamma$-ray data cannot be explained simultaneously with the high-frequency/mm-radio emission via a single-zone model, we focus here on whether the non-VLBI (O/UV, X-ray and $\gamma$-ray) data could be consistent with originating in a single emission region. Due to the difficulties of disentangling the various jet and ICM components in the X-ray observations, because of their spatial resolution particularly in the NuSTAR band, it is important to model the data in detector space. We do this by importing a single-zone model into ISIS. We model all components except the core X-ray emission with the same models and parameters as described in Section 2.3.3, except swap in the single-zone model for the core power law.

The model is similar to models 1a/1b overall, but with somewhat different choices in fixed and fitted parameters. We assume that the emitting region is a sphere of radius $R$ moving with a bulk Lorentz factor $\Gamma_j$ and corresponding speed $\beta c$, carrying a power $L_j = \pi R^2 \Gamma_j^2 \beta c (U'_e + U'_B + U'_p)$ in the comoving frame, divided between relativistic electrons, non-relativistic protons and a global magnetic field. We assume one cold proton per electron (no positrons), and that the electron eDF is described by a power law with slope $p_2$ between a minimum and maximum Lorentz factors $\gamma_{\min}$ and $\gamma_{\max}$. We assign the index to $p_2$ to allow better comparison to model 1b, as our steeper distribution is likely in the radiatively cooled regime, although we do not calculate $\gamma_{br}$ explicitly.

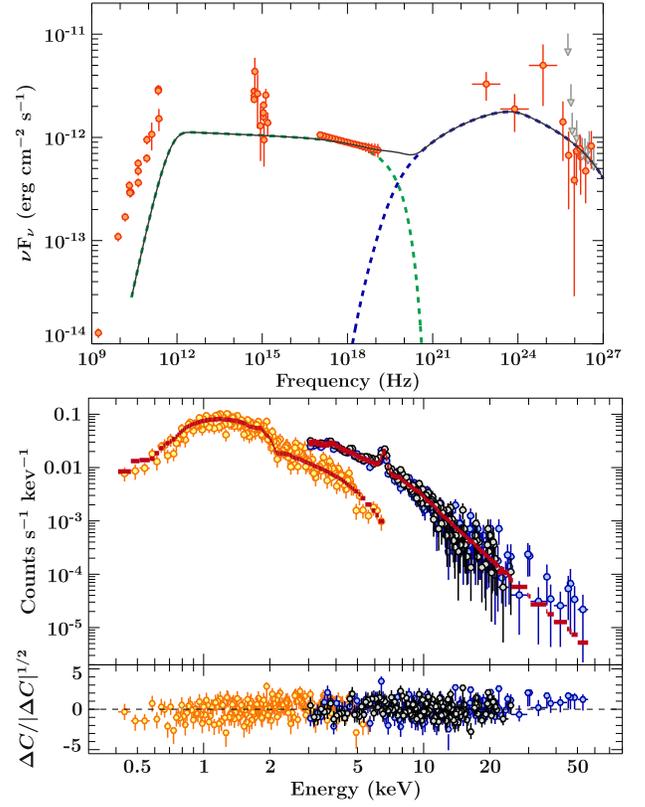

**Figure 18.** Top panel: same data as Figure 17, showing the model 2 fit focusing on the higher energy data. Bottom panel: statistical model 2 fit to the X-ray data. The orange points are the Chandra data, the blue and black points the NuSTAR data, and the red line shows the total model (which for NuSTAR includes the contribution of the ICM, HST-1 and kpc-scale jet). Note that when performing this fit, we modeled the X-ray spectra in detector space, simultaneously with the multiwavelength data shown in the top panel.

The ratio between radiating particle and magnetic energy density is another fitted parameter, $U'_e/U'_B$, where $U'_e = \langle\gamma\rangle n'_e m_e c^2$, where $\langle\gamma\rangle$ is the average Lorenz factor of the electrons, and $U'_B = B'^2/8\pi$. We calculate the particle number density $n'_e$ and magnetic field strength $B'$ from $U'_e/U'_B$ and $L_j$:

$$n'_e = \frac{L_j}{\pi R^2 \Gamma_j^2 \beta c^3 \langle\gamma\rangle m_e (1 + m_p/(m_e\langle\gamma\rangle) + U'_B/U'_e)} \quad (2)$$

$$B' = \sqrt{\frac{8\pi n'_e \langle\gamma\rangle m_e c^2}{U'_e/U'_B}}. \quad (3)$$

We take $\Gamma_j = 3$, which for our assumed viewing angle of 17° results in a Doppler factor $\delta \approx 3.4$, consistent with being inwards of the region constrained to $\Gamma \lesssim 6$ (Snios et al. 2019).

We fit the model to the data via the spectral modeling tool ISIS (Houck & Denicola 2000), using its implementation of the emcee algorithm (Foreman-Mackey et al. 2013) in order to thoroughly explore the parameter space, as well as evaluate model degeneracies and parameter uncertainties, using Cash statistics.[262] We run the algorithm with 20 walkers, and for 20,000 loops, conservatively taking the initial 14,000 iterations as the chain "burn-in" period in order to ensure convergence. We define the best-fitting parameters as the median of the posterior distribution in the final 6000 loops, and the $1\sigma$ error bars as the intervals that contain 68% of the walkers.

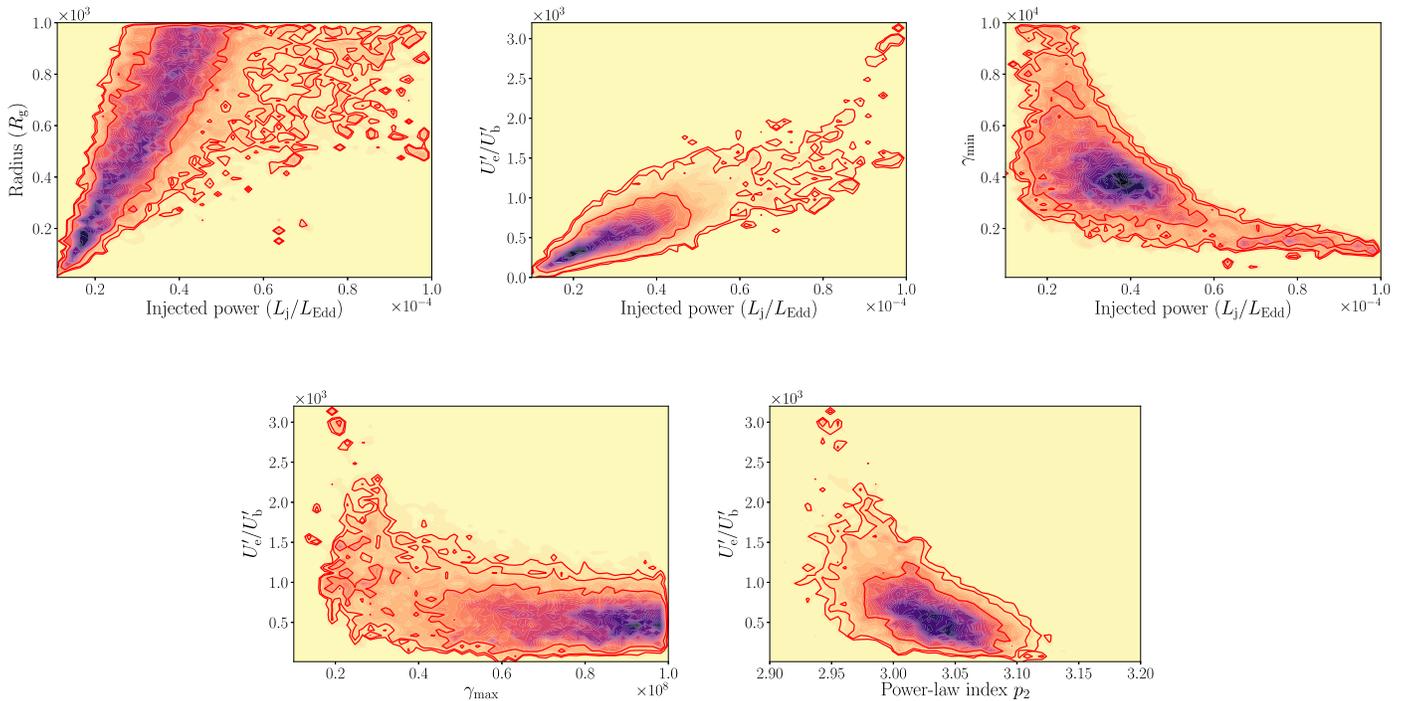

**Figure 19.** 2D posterior distributions for the parameters that show degeneracies in the high-energy single-zone SSC model. Left to right, top row: injected power $L_j$ (always given in units of $L_{Edd}$) vs. emitting region radius $r$, injected power $L_j$ vs. $U'_e/U'_B$, injected power $L_j$ vs. minimum electron Lorentz factor $\gamma_{min}$; left to right, bottom row: maximum electron Lorentz factor $\gamma_{max}$ vs. $U'_e/U'_B$, electron power-law index $s$ vs. $U'_e/U'_B$. The red lines represent the location of 68, 95 and 98 percent of the walkers.

The best-fitting parameters are shown in Table 2, and fits to the broadband SED and X-ray spectra of the core components are shown in Figure 18. Without a strong prior constraint on the size as in models 1a/1b, we found that most model parameters are highly degenerate and cannot be constrained satisfactorily by our fit. This behavior is shown in Figure 19, which shows the 2D posterior distributions where this degeneracy is most noticeable. This degeneracy is slightly reduced for models 1a/1b because of the additional constraints from optical depth effects and because $\gamma_{min}$ is free in model 2 and frozen at 1 in models 1a/1b.

### 3.3.3. Conclusions from Single-zone Fitting

As expected from the VLBI constraints, a stratified outflow model is necessary in order to capture the primary features of the entire broadband M87 SED. A compact, isotropic, single-zone model that is mildly magnetically dominated can be "maximally tuned" to capture both the EHT-mm flux as well as the X-ray power, but falls significantly below the $\gamma$-rays. This model does not, however, account for the fact that the synchrotron cooling time for the X-ray producing electrons is ~30 s in a ~5 G magnetic field, which is in tension with the modest X-ray variability. A larger single-zone model that is significantly particle-dominated can describe the X-rays well statistically, but cannot provide a compelling fit to the radio-mm VLBI core nor the VHE emission particularly in the GeV range, and even then requires rather extreme parameter values (discussed further below). These two approaches effectively approximate regions within either the base and inner jet of an expanding outflow (longitudinally structured), or an inner spine and outer sheath in a radially structured outflow. Therefore, we conclude that a structured jet is necessary to explain M87's observational properties, and that an additional emission component other than the EHT-core seems necessary to explain the $\gamma$-ray emission observed in 2017.

The physical conditions in the jet base of M87 are critically important for understanding the jet power and jet-launching mechanism (see the discussion below regarding the jet Lorentz factor). Recently there have been divergent views based on more sophisticated modeling approaches applied to non-simultaneous data sets, with Kino et al. (2015) claiming magnetic dominance of $8 \times 10^{-7} \lesssim U'_e/U'_B \lesssim 1 \times 10^{-4}$. In contrast, MAGIC Collaboration et al. (2020), focusing on the MAGIC/VHE bands, prefer the opposite extreme of particle dominance with $U'_e/U'_B \sim 2 \times 10^4$. Interestingly, our results fall between these two extremes, consistent with mildly magnetically dominated conditions in the jet launching region probed by EHT, evolving to moderate particle domination either further out along the jet axis, or radially (sheath).

This picture is somewhat consistent with the MAGIC Collaboration et al. (2020) results as the jet base in their model is already over 400 $r_g$, and thus similar to our model 2 scenario. This larger size scale, however, is in conflict with the VLBI size constraints above ∼5 GHz, and the core shift/optical depth requirements described above. However, we note that their predicted mm flux is a factor of a few lower than that measured by EHT, and thus emission from this component could in principle dominate the VHE emission while contributing only slightly to the radio/mm core.

---
[262] For the bands outside of the X-ray, we use artificial diagonal response matrices with a combination of area/exposure times so that we fit a "counts" in each frequency bin with a Cash statistic in its $\chi^2$ limit, with the resulting error corresponding to the Gaussian error of that frequency bin. Thus the $1\sigma$ Gaussian error corresponds to the square root of these "counts," which are in the $\chi^2$ limit of the Cash statistic. Only the X-ray band is treated in the "low counts" regime where the Cash statistics deviate significantly from $\chi^2$ statistics.

The bigger issue pinpointed by MAGIC Collaboration et al. (2020) and common to many VHE-focused models is the need for extreme particle domination ($U'_e/U'_B \gg 10^3$) on scales that correspond to M87's inner jets ($z \lesssim 10^4 r_g$). The extreme values obtained by this and other models may originate in the non-simultaneity of earlier data sets, and/or the non-inclusion of the VLBI constraints on geometry. VLBI images at cm wavelengths (Asada & Nakamura 2012; Hada et al. 2013; Nakamura & Asada 2013) show a roughly parabolic radial jet profile on these scales, usually interpreted as a magnetically accelerated bulk flow being collimated by the external medium (e.g., Nakamura et al. 2018). The fact that GRMHD simulations of large-scale jets can self-consistently obtain such a profile further supports this argument (e.g., Chatterjee et al. 2019, Figure 13). Upcoming EHTC results on the polarization properties of M87 will provide even more stringent constraints (EHT Collaboration 2021a, 2021b).

Furthermore, recent results constraining the innermost knot HST-1 to be moving superluminally outwards at an apparent speed requiring a Lorentz factor of ∼6 (Snios et al. 2019, and references therein) indicate a moderately relativistic bulk flow velocity. Even disregarding the above arguments, such a speed would be challenging to obtain for extreme particle domination in the regions probed by EHT. In the ideal MHD scenario, the bulk flow velocity at large scales reflects the magnetization (the ratio of magnetic to gas enthalpy) at the jet base (see, e.g., Komissarov et al. 2007; Tchekhovskoy et al. 2009), which is now also reproduced in large scale numerical GRMHD simulations (Chatterjee et al. 2019). If the base of M87's jets are extremely magnetically dominated, then instabilities leading to mass entrainment may be required very close in, to reduce the final bulk flow velocity.

Interestingly, despite being in a historically low state for the high-energy emission, our conclusions regarding M87's mm/radio core properties and power are very similar to those presented in Reynolds et al. (1996). Based on earlier 5 GHz VLBI measurements, the authors concluded that the jets are more likely to be pair-dominated because of power requirements. Similarly, when we estimate the power assuming (somewhat arbitrarily) an electron-proton plasma with charge neutrality, we find models 1a/b to be in tension with estimates of the total jet power of $L_j \sim 10^{-4} L_{\rm Edd}$, based on the pressure arguments for the knots. As Reynolds et al. also discuss, this power issue can be mitigated by raising $\gamma_{\rm min}$, as we also found for model 2. This overall consistency between power and inferred properties for the inner jets over decades may provide further support for the scenario where the changes seen at high-energy originate further out in the jets.

Constraining the jet composition near the launch point will have key consequences for not only the power requirements, but potentially the high-energy emission. If hadrons are indeed present and accelerated, a population of >100 TeV protons in the jet base could potentially produce γ-rays via secondary neutral $\pi_0$ decay or via inverse Comptonization from secondary leptons injected at similar energies. As a consequence, the models as presented would need to be renormalized in order to avoid overproducing the lower frequency emission, in turn requiring even more particle domination over the magnetic fields. As described above, this energy partition would further exacerbate the problems explaining the jet acceleration and the core shift, as well as increasing the jet power, particularly for accelerated proton power laws with index $p < 2$. Another possibility is direct proton synchrotron from >EeV protons, as well as secondary muon synchrotron, given the potential for strong magnetic fields in the EHT-emission region (see, e.g., Reimer et al. 2004). We cannot rule this scenario out without more detailed modeling, however it may be difficult to reconcile with the accompanying low-energy emission if leptons are assumed accelerated at the same rate. Regardless of mechanism, all models seeking to produce the observed ∼$10^{42}$ erg s$^{-1}$ in γ-rays within the innermost $10r_g$ of M87 would have to find away around the extreme attenuation expected due to pair production off the core optical/UV photons.

In conclusion, while we cannot constrain the composition of the jets, power arguments favor lighter jets at least in the jet base. Our results also favor moderate magnetic domination at the launch region of M87's jets, at least where the EHT and possibly some X-ray emission are produced, while a larger and more particle-dominated region than the "EHT-zone" is likely responsible for the VHE emission. Either this region is not in the accelerating part of the jet flow, or it could be interacting with the surrounding ICM and may require an additional source of seed photons to the jet emission alone. The VHE emission could also be coming from the extended jet and knots outward of HST-1, not included in our radio through X-ray images but still unresolved in the γ-rays. Correlated multi-wavelength variability in future campaigns will be the ultimate test of these scenarios. We will return to these questions with a more time-dependent analysis in the EHT 2018 multi-wavelength campaign paper.

## 4. Summary

We present the most extensive, quasi-simultaneous, broad-band spectrum of M87 taken yet, together with the highest ever resolution mm-VLBI images using the Event Horizon Telescope from its 2017 April campaign. The primary result of this Letter is the presentation of this legacy data set, for which all data and some analysis scripts are available to the community via a Cyverse repository; see Appendix B. From the observational side, our main conclusions are that the M87 core was in a relatively low state compared to historical observations, but clearly still dominating over the nearest knot HST-1, which was seemingly at its lowest historical brightness state. This provides the ideal observing conditions for a multi-wavelength campaign combining data over a large range of spatial resolution, because M87's core was dominating the total flux in radio through X-ray bands.

While we defer detailed modeling to further work by ourselves and the broader community, we can already make some baseline conclusions based on simple single-zone models. The first conclusion is that M87's complex, broadband spectral energy distribution cannot be modeled by a single zone. The stratified nature of the radio through millimeter bands is clearly indicated by the high dynamic-range images that place significant size constraints on the emitting geometry per frequency band. However by selecting two representative regions for the EHT-observed region and the inner jets, respectively, we can already come to a few solid conclusions. First, it is not yet clear where the VHE γ-rays originate, but we can robustly rule out that they coincide with the EHT region for leptonic processes. The energetics and small size of this region, together with the VLBI constraints on optical depth and thus strong magnetic fields, cannot be reconciled with significant $\gamma$-ray production via SSC emission. A $\sim$100 times larger region can provide VHE $\gamma$-ray fluxes but requires an uncomfortably high particle domination compared to magnetic fields, which may be inconsistent with the observed jet velocity and parabolic geometrical profile. Direct proton and muon synchrotron emission from the EHT-emission region contributing to the GeV/TeV range cannot be ruled out at this time, but can be tested in the future using the upcoming improved constraints on the magnetic fields expected from EHT polarization results. We emphasize in any case that a structured jet model including time-dependence will be important for any detailed interpretation. Additional information will be presented in the upcoming paper on the 2018 campaign, where there will be new constraints on the multi-wavelength variability properties.

This "quiescent active" core spectrum for M87 is also important for comparison with the upcoming analysis on the Galactic center supermassive black hole Sgr A$^*$. Sgr A$^*$ has also been in an extended, much weaker quiescent period since (at least) its first identification as a high-energy source (Baganoff et al. 2001, 2003), punctuated only by moderate flaring (e.g., Neilsen et al. 2015; Witzel et al. 2018; Do et al. 2019; Haggard et al. 2019). M87's low state thus provides the ideal comparison spectrum to understand how sources evolve from quiescence into low-luminosity AGN and launch more extended jet outflows.

With results from the 2017 EHT campaign for 3C 279 already published (Kim et al. 2020), and other 2017 papers on M87's polarization (EHT Collaboration et al. 2021, in preparation), Sgr A$^*$ (EHT Collaboration et al. 2021, in preparation), Cen A (M. Janssen et al. 2021, in preparation), OJ 287 (J. L. Gómez et al. 2021, in preparation), and many more in the pipeline, we expect the context and impact of the multi-wavelength results to significantly broaden in the coming years. The precision EHT mm-VLBI images together with broadband SED information will be essential ingredients for developing a fuller understanding of the AGN phenomenon. Beyond simple accretion rate changes, this new body of work will allow us to directly study the interplay between strong gravity and magnetic fields and to trace those effects to the larger jet structures and their feedback onto the external environment. With our legacy sample, enhanced by multi-year campaigns with the steadily enhanced EHT array, as well as eventually with the future next-generation EHT, we aim to provide a rich, public data set to the community for decades of projects to come.

J.-C.A acknowledges support from the Malaysian Fundamental Research Grant Scheme (FRGS) FRGS/1/2019/STG02/UM/02/6.

M.B. acknowledges support from the YCAA Prize Postdoctoral Fellowship and the Black Hole Initiative at Harvard University, which is funded in part by the Gordon and Betty Moore Foundation (grant GBMF8273) and in part by the John Templeton Foundation.

S.C. is supported by the South African Research Chairs Initiative (grant No. 64789) of the Department of Science and Innovation and the National Research Foundation of South Africa.

K.H. acknowledges support from JSPS KAKENHI grant Nos. JJP18H03721, JP19H01943, JP18KK0090, and support from the Mitsubishi Foundation grant No. 201911019.

D.H. acknowledges support from the Natural Sciences and Engineering Research Council of Canada (NSERC) Discovery Grant and the Canada Research Chairs program.

T.K. acknowledges support from JSPS KAKENHI grant Nos. JP18K13594, JP19H01908, JP19H01906, MEXT as "Program for Promoting Researches on the Supercomputer Fugaku" (Toward a unified view of the universe: from large scale structures to planets), MEXT as "Priority Issue on post-K computer" (Elucidation of the Fundamental Laws and Evolution of the Universe) and JICFuS. A part of the calculations were carried out on the XC50 at the Center for Computational Astrophysics, National Astronomical Observatory of Japan.

R.-S.L. is supported by the Max Planck Partner Group of the MPG and the CAS and acknowledges support from the Key Program of the National Natural Science Foundation of China (grant No. 11933007) and the Research Program of Fundamental and Frontier Sciences, CAS (grant No. ZDBS-LY-SLH011).

M.L. and S.M. are thankful for support from an NWO (Netherlands Organisation for Scientific Research) VICI award, grant No. 639.043.513.

J.N. acknowledges support from SAO award DD7-18089X and NASA award 80NSSC20K0645.

J.P. acknowledges financial support from the Korean National Research Foundation (NRF) via Global PhD Fellowship grant 2014H1A2A1018695 and support through the EACOA Fellowship awarded by the East Asia Core Observatories Association, which consists of the Academia Sinica Institute of Astronomy and Astrophysics, the National Astronomical Observatory of Japan, Center for Astronomical Mega-Science, Chinese Academy of Sciences, and the Korea Astronomy and Space Science Institute.

M.S. acknowledges support from JSPS KAKENHI grant Nos. JP19K14761.

The authors of the present Letter thank the following organizations and programs: the Academy of Finland (projects 274477, 284495, 312496, 315721); Agencia Nacional de Investigación y Desarrollo (ANID), Chile via NCN19_058 (TITANs) and FONDECYT 3190878; the Alexander von Humboldt Stiftung; the Black Hole Initiative at Harvard University, through a grant (60477) from the John Templeton Foundation; the China Scholarship Council; Consejo Nacional de Ciencia y Tecnología (CONACYT, Mexico, projects 104497, 275201, 279006, 281692); the Delaney Family via the Delaney Family John A. Wheeler Chair at Perimeter Institute; Dirección General de Asuntos del Personal Académico—Universidad Nacional Autónoma de México (DGAPA—UNAM, project IN112417); the European Research Council Synergy Grant "BlackHoleCam: Imaging the Event Horizon of Black Holes" (grant 610058); the Generalitat Valenciana postdoctoral grant APOSTD/2018/177 and GenT Program (project CIDEGENT/2018/021); the Gordon and Betty Moore Foundation (grants GBMF-3561, GBMF-5278); the Istituto Nazionale di Fisica Nucleare (INFN) sezione di Napoli, iniziative specifiche TEONGRAV; the International Max Planck Research School for Astronomy and Astrophysics at the Universities of Bonn and Cologne; the Japanese Government (Monbukagakusho: MEXT) Scholarship; the Japan Society for the Promotion of Science (JSPS) Grant-in-Aid for JSPS Research Fellowship (JP17J08829); the Key Research Program of Frontier Sciences, Chinese Academy of Sciences (CAS, grants QYZDJ-SSW-SLH057, QYZDJ-SSW-SYS008); the Leverhulme Trust Early Career Research Fellowship; the Max-Planck-Gesellschaft (MPG); the Max Planck Partner Group of the MPG and the CAS; the MEXT/

JSPS KAKENHI (grants 18KK0090, JP18K13594, JP18K03656, JP18H03721, 18K03709, 18H01245, 25120007); the MIT International Science and Technology Initiatives (MISTI) Funds; the Ministry of Science and Technology (MOST) of Taiwan (105-2112-M-001-025-MY3, 106-2112-M-001-011, 106-2119-M-001-027, 107-2119-M-001-017, 107-2119-M-001-020, 107-2119-M-110-005, 108-2112-M-001-048 and 109-2124-M-001-005); the National Aeronautics and Space Administration (NASA, Fermi Guest Investigator grant 80NSSC20K1567 and Hubble Fellowship grant HST-HF2-51431.001-A awarded by the Space Telescope Science Institute, which is operated by the Association of Universities for Research in Astronomy, Inc., for NASA, under contract NAS5-26555, NuSTAR award 80NSSC20K0645, Chandra award DD7-18089X); the National Institute of Natural Sciences (NINS) of Japan; the National Key Research and Development Program of China (grant 2016YFA0400704, 2016YFA0400702); the National Science Foundation (NSF, grants AST-0096454, AST-0352953, AST-0521233, AST-0705062, AST-0905844, AST-0922984, AST-1126433, AST-1140030, DGE-1144085, AST-1207704, AST-1207730, AST-1207752, MRI-1228509, OPP-1248097, AST-1310896, AST-1337663, AST-1440254, AST-1555365, AST-1715061, AST-1615796, AST-1716327, OISE-1743747, AST-1816420, AST-2034306); the Natural Science Foundation of China (grants 11573051, 11633006, 11650110427, 10625314, 11721303, 11725312, 11991053); the Natural Sciences and Engineering Research Council of Canada (NSERC, including a Discovery Grant and the NSERC Alexander Graham Bell Canada Graduate Scholarships-Doctoral Program); the National Research Foundation of Korea (the Global PhD Fellowship Grant: grants NRF-2015H1A2A1033752, 2015-R1D1A1A01056807, the Korea Research Fellowship Program: NRF-2015H1D3A1066561); the Netherlands Organization for Scientific Research (NWO) VICI award (grant 639.043.513) and Spinoza Prize SPI 78-409; the New Scientific Frontiers with Precision Radio Interferometry Fellowship awarded by the South African Radio Astronomy Observatory (SARAO), which is a facility of the National Research Foundation (NRF), an agency of the Department of Science and Innovation (DSI) of South Africa; the South African Research Chairs Initiative of the Department of Science and Innovation and National Research Foundation; the Onsala Space Observatory (OSO) national infrastructure, for the provisioning of its facilities/observational support (OSO receives funding through the Swedish Research Council under grant 2017-00648) the Perimeter Institute for Theoretical Physics (research at Perimeter Institute is supported by the Government of Canada through the Department of Innovation, Science and Economic Development and by the Province of Ontario through the Ministry of Research, Innovation and Science); the Spanish Ministerio de Economía y Competitividad (grants PGC2018-098915-B-C21, AYA2016-80889-P); the State Agency for Research of the Spanish MCIU through the "Center of Excellence Severo Ochoa" award for the Instituto de Astrofísica de Andalucía (SEV-2017-0709); the Toray Science Foundation; the US Department of Energy (USDOE) through the Los Alamos National Laboratory (operated by Triad National Security, LLC, for the National Nuclear Security Administration of the USDOE (Contract 89233218CNA 000001)); the European Union's Horizon 2020 research and innovation program under grant agreement No. 730562 RadioNet; ALMA North America Development Fund; the Academia Sinica; Chandra TM6-17006X; the GenT Program (Generalitat Valenciana) Project CIDEGENT/2018/021. This work used the Extreme Science and Engineering Discovery Environment (XSEDE), supported by NSF grant ACI-1548562, and CyVerse, supported by NSF grants DBI-0735191, DBI-1265383, and DBI-1743442. XSEDE Stampede2 resource at TACC was allocated through TG-AST170024 and TG-AST080026N. XSEDE JetStream resource at PTI and TACC was allocated through AST170028. The simulations were performed in part on the SuperMUC cluster at the LRZ in Garching, on the LOEWE cluster in CSC in Frankfurt, and on the HazelHen cluster at the HLRS in Stuttgart. This research was enabled in part by support provided by Compute Ontario (http://computeontario.ca), Calcul Quebec (http://www.calculquebec.ca) and Compute Canada (http://www.computecanada.ca). We thank the staff at the participating observatories, correlation centers, and institutions for their enthusiastic support. This Letter makes use of the following ALMA data: ADS/JAO.ALMA#2016.1.01154.V. ALMA is a partnership of the European Southern Observatory (ESO; Europe, representing its member states), NSF, and National Institutes of Natural Sciences of Japan, together with National Research Council (Canada), Ministry of Science and Technology (MOST; Taiwan), Academia Sinica Institute of Astronomy and Astrophysics (ASIAA; Taiwan), and Korea Astronomy and Space Science Institute (KASI; Republic of Korea), in cooperation with the Republic of Chile. The Joint ALMA Observatory is operated by ESO, Associated Universities, Inc. (AUI)/NRAO, and the National Astronomical Observatory of Japan (NAOJ). The NRAO is a facility of the NSF operated under cooperative agreement by AUI. APEX is a collaboration between the Max-Planck-Institut für Radioastronomie (Germany), ESO, and the Onsala Space Observatory (Sweden). The SMA is a joint project between the SAO and ASIAA and is funded by the Smithsonian Institution and the Academia Sinica. The JCMT is operated by the East Asian Observatory on behalf of the NAOJ, ASIAA, and KASI, as well as the Ministry of Finance of China, Chinese Academy of Sciences, and the National Key R&D Program (No. 2017YFA0402700) of China. Additional funding support for the JCMT is provided by the Science and Technologies Facility Council (UK) and participating universities in the UK and Canada. he LMT is a project operated by the Instituto Nacional de Astrofísica, Óptica, y Electrónica (Mexico) and the University of Massachusetts at Amherst (USA), with financial support from the Consejo Nacional de Ciencia y Tecnología and the National Science Foundation. The IRAM 30-m telescope on Pico Veleta, Spain is operated by IRAM and supported by CNRS (Centre National de la Recherche Scientifique, France), MPG (Max-Planck-Gesellschaft, Germany) and IGN (Instituto Geográfico Nacional, Spain). The SMT is operated by the Arizona Radio Observatory, a part of the Steward Observatory of the University of Arizona, with financial support of operations from the State of Arizona and financial support for instrumentation development from the NSF. The SPT is supported by the National Science Foundation through grant PLR- 1248097. Partial support is also provided by the NSF Physics Frontier Center grant PHY-1125897 to the Kavli Institute of Cosmological Physics at the University of Chicago, the Kavli Foundation and the Gordon and Betty Moore Foundation grant GBMF 947. The SPT hydrogen maser was provided on loan from the GLT, courtesy of ASIAA. The EHTC has received generous donations of FPGA chips from Xilinx Inc., under the Xilinx University Program. The EHTC has benefited from technology shared under open-source license by the Collaboration for Astronomy Signal Processing and Electronics Research (CASPER). The EHT project is grateful to T4Science and Microsemi for their assistance with Hydrogen Masers. This research has made use of NASA's Astrophysics Data System. We gratefully acknowledge the support provided by the extended staff of the ALMA, both from the inception of the ALMA Phasing Project through the observational campaigns of 2017 and 2018. We would like to thank A. Deller and W. Brisken for EHT-specific support with the use of DiFX. We acknowledge the significance that Maunakea, where the SMA and JCMT EHT stations are located, has for the indigenous Hawaiian people.

The European VLBI Network is a joint facility of independent European, African, Asian, and North American radio astronomy institutes. Scientific results from data presented in this publication are derived from the following EVN project code: EH033.

This research has made use of data obtained with 12 radio telescopes from the East Asian VLBI Network (EAVN): 2 stations from the Chinese VLBI Network (the Tianma radio telescope operated by Shanghai Astronomical Observatory of Chinese Academy of Sciences (CAS) and the Nanshan radio telescope operated by Xinjiang Astronomical Observatory of CAS), 2 stations from the Japanese VLBI Network (the Hitachi station operated by Ibaraki University and the Kashima station operated by the National Institute of Information and Communications Technology), the Korean VLBI Network operated by the Korea Astronomy and Space Science Institute (KASI), the VLBI Exploration of Radio Astrometry (VERA) operated by the National Astronomical Observatory of Japan (NAOJ), and the Nobeyama 45 m radio telescope operated by NAOJ. We thank VERA staff members who helped the operation. We are grateful to KVN staff who helped to operate the array and to correlate the data for quasi-simultaneous multi-wavelength observations of M87. The KVN is a facility operated by KASI. The KVN operations are supported by KREONET (Korea Research Environment Open NETwork), which is managed and operated by KISTI (Korea Institute of Science and Technology Information)

The VLBA is an instrument of the National Radio Astronomy Observatory. The National Radio Astronomy Observatory is a facility of the National Science Foundation operated by Associated Universities, Inc.

This study has been supported in part by the Russian Science Foundation: project 20-62-46021.

The Fermi LAT Collaboration acknowledges generous ongoing support from a number of agencies and institutes that have supported both the development and the operation of the LAT as well as scientific data analysis. These include the National Aeronautics and Space Administration and the Department of Energy in the United States, the Commissariat à l'Energie Atomique and the Centre National de la Recherche Scientifique/Institut National de Physique Nucléaire et de Physique des Particules in France, the Agenzia Spaziale Italiana and the Istituto Nazionale di Fisica Nucleare in Italy, the Ministry of Education, Culture, Sports, Science and Technology (MEXT), High Energy Accelerator Research Organization (KEK) and Japan Aerospace Exploration Agency (JAXA) in Japan, and the K. A. Wallenberg Foundation, the Swedish Research Council and the Swedish National Space Board in Sweden. Additional support for science analysis during the operations phase is gratefully acknowledged from the Istituto Nazionale di Astrofisica in Italy and the Centre National d'Etudes Spatiales in France. This work is performed in part under DOE Contract DE-AC02-76SF00515.

The support of the Namibian authorities and of the University of Namibia in facilitating the construction and operation of H.E.S.S. is gratefully acknowledged, as is the support by the German Ministry for Education and Research (BMBF), the Max Planck Society, the German Research Foundation (DFG), the Helmholtz Association, the Alexander von Humboldt Foundation, the French Ministry of Higher Education, Research and Innovation, the Centre National de la Recherche Scientifique (CNRS/IN2P3 and CNRS/INSU), the Commissariat à l'énergie atomique et aux énergies alternatives (CEA), the U.K. Science and Technology Facilities Council (STFC), the Knut and Alice Wallenberg Foundation, the National Science Centre, Poland grant No. 2016/22/M/ST9/00382, the South African Department of Science and Technology and National Research Foundation, the University of Namibia, the National Commission on Research, Science & Technology of Namibia (NCRST), the Austrian Federal Ministry of Education, Science and Research and the Austrian Science Fund (FWF), the Australian Research Council (ARC), the Japan Society for the Promotion of Science and by the University of Amsterdam. We appreciate the excellent work of the technical support staff in Berlin, Zeuthen, Heidelberg, Palaiseau, Paris, Saclay, Tübingen, and in Namibia in the construction and operation of the equipment. This work benefited from services provided by the H.E.S.S. Virtual Organisation, supported by the national resource providers of the EGI Federation.

We would like to thank the Instituto de Astrofisica de Canarias for the excellent working conditions at the Observatorio del Roque de los Muchachos in La Palma. The financial support of the German BMBF, MPG and HGF; the Italian INFN and INAF; the Swiss National Fund SNF; the ERDF under the Spanish Ministerio de Ciencia e Innovacion (MICINN) (FPA2017-87859-P, FPA2017-85668-P, FPA2017-82729-C6-5-R, FPA2017-90566-REDC, PID2019-104114RB-C31, PID2019-104114RB-C32, PID2019-105510GB-C31,PID 2019-107847RB-C41, PID2019-107847RB-C42, PID2019-107988GB-C22); the Indian Department of Atomic Energy; the Japanese ICRR, the University of Tokyo, JSPS, and MEXT; the Bulgarian Ministry of Education and Science, National RI Roadmap Project DO1-268/16.12.2019 and the Academy of Finland grant No. 320045 is gratefully acknowledged. This work was also supported by the Spanish Centro de Excelencia "Severo Ochoa" SEV-2016-0588 and CEX2019-000920-S, and "Maria de Maeztu" CEX2019-000918-M, the Unidad de Excelencia "Maria de Maeztu" MDM-2015-0509-18-2 and the "la Caixa" Foundation (fellowship LCF/BQ/PI18/11630012) and by the CERCA program of the Generalitat de Catalunya; by the Croatian Science Foundation (HrZZ) Project IP-2016-06-9782 and the University of Rijeka Project 13.12.1.3.02; by the DFG Collaborative Research Centers SFB823/C4 and SFB876/C3; the Polish National Research Centre grant UMO-2016/22/M/ST9/00382; and by the Brazilian MCTIC, CNPq, and FAPERJ.

This research is supported by grants from the U.S. Department of Energy Office of Science, the U.S. National Science Foundation and the Smithsonian Institution, by NSERC in Canada, and by the Helmholtz Association in Germany. This research used resources provided by the Open Science Grid, which is supported by the National Science Foundation and the U.S. Department of Energy's Office of Science, and resources of the National Energy Research

Scientific Computing Center (NERSC), a U.S. Department of Energy Office of Science User Facility operated under Contract No. DE-AC02-05CH11231. We acknowledge the excellent work of the technical support staff at the Fred Lawrence Whipple Observatory and at the collaborating institutions in the construction and operation of the instrument.

*Facilities:* ALMA, Chandra X-ray Observatory, EAVN, EHT, EVN, Fermi, GMVA, H.E.S.S., HSA, HST, KVN, MAGIC, Neil Gehrels Swift Observatory, NuSTAR, SMA, VERA, VERITAS, VLBA.

## Appendix A
## Tabulated Data

This section contains all tabulated data mentioned in Sections 2 and 3.2, as follows.

1. Summary of radio observations in Table A1,
2. UV/optical host galaxy and core region parameters in Table A2,
3. Optical flux densities from Swift-UVOT observations in Table A3,

Table A1
Summary of Radio Observations

| Obs. Code | Obs. Date (yyyy-mm-dd) | Obs. Date (MJD) | Frequency (GHz) | Peak Brightness[a] (Jy beam$^{-1}$) | Total Flux Density[b] (Jy) |
|---|---|---|---|---|---|
| | | EVN (beam: 5.0 mas circ.) | | | |
| eh033 | 2017-05-09 | 57882 | 1.7 | $0.75 \pm 0.08$ | $2.85 \pm 0.29$ |
| | | HSA (beam: 1.5/1.0/1.0 mas circ. at 8.4/15/24 GHz) | | | |
| bh221a | 2017-05-15 | 57888 | 8.4 | $1.30 \pm 0.13$ | $2.88 \pm 0.29$ |
| bh221b | 2017-05-16 | 57889 | 15.4 | $1.13 \pm 0.11$ | $2.12 \pm 0.21$ |
| bh221c | 2017-05-20 | 57893 | 23.8 | $1.22 \pm 0.12$ | $1.94 \pm 0.19$ |
| | | VERA (beam: 1.0 mas circ.) | | | |
| r17025c | 2017-01-25 | 57778 | 22.2 | $1.45 \pm 0.15$ | $1.75 \pm 0.18$ |
| r17031c | 2017-01-31 | 57784 | 22.2 | $1.58 \pm 0.16$ | $1.89 \pm 0.19$ |
| r17052c | 2017-02-21 | 57805 | 22.2 | $1.44 \pm 0.15$ | $1.69 \pm 0.17$ |
| r17090c | 2017-03-31 | 57843 | 22.2 | $1.26 \pm 0.13$ | $1.45 \pm 0.15$ |
| r17092c | 2017-04-02 | 57845 | 22.2 | $1.32 \pm 0.13$ | $1.60 \pm 0.16$ |
| r17120c | 2017-04-30 | 57873 | 22.2 | $1.46 \pm 0.15$ | $1.74 \pm 0.17$ |
| r17121c | 2017-05-01 | 57874 | 22.2 | $1.41 \pm 0.14$ | $1.68 \pm 0.17$ |
| r17141c | 2017-05-21 | 57894 | 22.2 | $1.32 \pm 0.13$ | $1.57 \pm 0.16$ |
| r17148c | 2017-05-28 | 57901 | 22.2 | $1.38 \pm 0.14$ | $1.65 \pm 0.17$ |
| r17157c | 2017-06-06 | 57910 | 22.2 | $1.36 \pm 0.14$ | $1.61 \pm 0.16$ |
| r17254c | 2017-09-11 | 58007 | 22.2 | $1.20 \pm 0.12$ | $1.62 \pm 0.16$ |
| r17276c | 2017-10-03 | 58029 | 22.2 | $1.20 \pm 0.12$ | $1.40 \pm 0.14$ |
| r17300c | 2017-10-27 | 58053 | 22.2 | $1.38 \pm 0.14$ | $1.66 \pm 0.17$ |
| r17324c | 2017-11-20 | 58077 | 22.2 | $1.09 \pm 0.11$ | $1.61 \pm 0.16$ |
| r17343c | 2017-12-09 | 58096 | 22.2 | $1.37 \pm 0.14$ | $1.58 \pm 0.16$ |
| r17352c | 2017-12-18 | 58105 | 22.2 | $1.37 \pm 0.14$ | $1.55 \pm 0.16$ |
| | | EAVN/KaVA (beam: 1.0/0.5 mas circ. at 22/43 GHz) | | | |
| r17010b | 2017-01-10 | 57763 | 22.2 | $1.50 \pm 0.15$ | $2.11 \pm 0.21$ |
| r17011b | 2017-01-11 | 57764 | 43.1 | $1.35 \pm 0.14$ | $1.89 \pm 0.19$ |
| r17023b | 2017-01-23 | 57776 | 22.2 | $1.41 \pm 0.14$ | $1.96 \pm 0.20$ |
| r17024b | 2017-01-24 | 57777 | 43.1 | $1.34 \pm 0.13$ | $1.93 \pm 0.19$ |
| r17045b | 2017-02-14 | 57798 | 22.2 | $1.52 \pm 0.15$ | $2.16 \pm 0.22$ |
| r17046b | 2017-02-15 | 57799 | 43.1 | $1.25 \pm 0.13$ | $1.87 \pm 0.19$ |
| r17058b | 2017-02-27 | 57811 | 22.2 | $1.45 \pm 0.15$ | $2.05 \pm 0.21$ |
| r17059b | 2017-02-28 | 57812 | 43.1 | $1.21 \pm 0.12$ | $1.80 \pm 0.18$ |
| a17077a | 2017-03-18 | 57830 | 22.2 | $1.30 \pm 0.13$ | $1.97 \pm 0.20$ |
| a17078a | 2017-03-19 | 57831 | 43.1 | $1.16 \pm 0.12$ | $1.54 \pm 0.15$ |
| a17086a | 2017-03-27 | 57839 | 43.1 | $1.14 \pm 0.11$ | $1.60 \pm 0.16$ |
| a17093a | 2017-04-03 | 57846 | 22.2 | $1.34 \pm 0.13$ | $1.98 \pm 0.20$ |
| a17094a | 2017-04-04 | 57847 | 43.1 | $1.21 \pm 0.12$ | $1.64 \pm 0.16$ |
| a17099a | 2017-04-09 | 57852 | 43.1 | $1.10 \pm 0.11$ | $1.55 \pm 0.16$ |
| a17104a | 2017-04-14 | 57857 | 43.1 | $1.18 \pm 0.12$ | $1.57 \pm 0.16$ |
| a17107a | 2017-04-17 | 57860 | 22.2 | $1.39 \pm 0.14$ | $1.96 \pm 0.20$ |
| a17108a | 2017-04-18 | 57861 | 43.1 | $1.16 \pm 0.12$ | $1.55 \pm 0.16$ |
| a17114a | 2017-04-24 | 57867 | 22.2 | $1.32 \pm 0.13$ | $1.81 \pm 0.18$ |
| a17115a | 2017-04-25 | 57868 | 43.1 | $1.18 \pm 0.12$ | $1.45 \pm 0.15$ |
| a17130a | 2017-05-10 | 57883 | 22.2 | $1.34 \pm 0.13$ | $1.88 \pm 0.19$ |
| a17131a | 2017-05-11 | 57884 | 43.1 | $1.12 \pm 0.11$ | $1.52 \pm 0.15$ |
| a17146a | 2017-05-26 | 57899 | 43.1 | $1.09 \pm 0.11$ | $1.43 \pm 0.14$ |
| | | KVN (beam: 1.0 mas circ. at 22/43/86/129 GHz) | | | |
| p17sl01c | 2017-03-28 | 57840 | 43.4 | $1.25 \pm 0.13$ | $1.29 \pm 0.13$ |
| p17sl01d | 2017-04-19 | 57862 | 21.7 | $1.63 \pm 0.16$ | $1.73 \pm 0.17$ |
| p17sl01d | 2017-04-19 | 57862 | 43.4 | $1.31 \pm 0.13$ | $1.41 \pm 0.14$ |

Table A1
(Continued)

| Obs. Code | Obs. Date (yyyy-mm-dd) | Obs. Date (MJD) | Frequency (GHz) | Peak Brightness[a] (Jy beam$^{-1}$) | Total Flux Density[b] (Jy) |
|---|---|---|---|---|---|
| p17sl01d | 2017-04-19 | 57862 | 86.8 | 1.09 ± 0.11 | 1.26 ± 0.13 |
| p17sl01d | 2017-04-19 | 57862 | 129.3 | 0.83 ± 0.25 | 0.91 ± 0.27 |
| p17sl01e | 2017-05-21 | 57894 | 21.7 | 1.65 ± 0.17 | 1.71 ± 0.17 |
| p17sl01f | 2017-06-17 | 57921 | 21.7 | 1.67 ± 0.17 | 1.73 ± 0.17 |
| p17sl01f | 2017-06-17 | 57921 | 43.4 | 1.28 ± 0.13 | 1.32 ± 0.13 |
| p17sl01f | 2017-06-17 | 57921 | 86.8 | 0.93 ± 0.09 | 0.95 ± 0.10 |
| p17sl01g | 2017-09-19 | 58015 | 21.7 | 1.59 ± 0.16 | 1.64 ± 0.16 |
| p17sl01g | 2017-09-19 | 58015 | 43.4 | 1.12 ± 0.11 | 1.16 ± 0.12 |
| p17sl01g | 2017-09-19 | 58015 | 86.8 | 0.80 ± 0.08 | 0.82 ± 0.08 |
| p17sl01i | 2017-11-04 | 58061 | 86.8 | 1.17 ± 0.12 | 1.19 ± 0.12 |
| p17sl01k | 2017-12-06 | 58093 | 21.7 | 1.63 ± 0.16 | 1.71 ± 0.17 |
| p17sl01k | 2017-12-06 | 58093 | 86.8 | 0.89 ± 0.09 | 0.97 ± 0.10 |
| p17sl01k | 2017-12-06 | 58093 | 129.3 | 0.64 ± 0.19 | 0.69 ± 0.21 |
| VLBA (beam: 1.0/0.5 mas circ. at 24/43 GHz) | | | | | |
| bg251a | 2017-05-05 | 57878 | 23.8 | 1.12 ± 0.11 | 1.75 ± 0.18 |
| bg251a | 2017-05-05 | 57878 | 43.1 | 1.09 ± 0.11 | 1.63 ± 0.16 |
| GMVA (beam: 0.2 mas circ.) | | | | | |
| MA009 | 2017-03-30 | 57842 | 86.3 | 0.73 ± 0.22 | 1.02 ± 0.30 |
| ALMA | | | | | |
| 2016.1.01154.V | 2017-04-05 | 57848 | 221 | 1.28 ± 0.13 | 1.54 ± 0.15 |
| 2016.1.01154.V | 2017-04-06 | 57849 | 221 | 1.31 ± 0.13 | 1.58 ± 0.16 |
| 2016.1.01154.V | 2017-04-10 | 57853 | 221 | 1.33 ± 0.13 | 1.58 ± 0.16 |
| 2016.1.01154.V | 2017-04-11 | 57854 | 221 | 1.34 ± 0.13 | 1.56 ± 0.16 |
| SMA | | | | | |
| 2016B-S079 | 2017-04-05 | 57848 | 220 | 1.28 ± 0.13 | 1.60 ± 0.16 |
| 2016B-S079 | 2017-04-06 | 57849 | 220 | 1.31 ± 0.13 | 1.64 ± 0.16 |
| 2016B-S079 | 2017-04-10 | 57853 | 220 | 1.40 ± 0.14 | 1.57 ± 0.16 |
| 2016B-S079 | 2017-04-11 | 57854 | 220 | 1.30 ± 0.13 | ... |
| N/A | 2017-04-22 | 57865 | 225 | 1.53 ± 0.10 | ... |
| 2016B-A012 | 2017-04-25 | 57868 | 221 | 1.68 ± 0.09 | ... |

**Notes.**
[a] Peak brightness represents the unresolved flux of the radio core (the most compact and brightest feature at the jet base) within the beam size.
[b] Total flux density is measured by integrating the radio fluxes over the entire jet structure within the field of view of each observation.

Table A2
UV/Optical Host Galaxy and Core Region Parameters

| Filter | $A_\lambda$ (mag) | $n$ | $R_e$ (″) | $\epsilon$ | $\Phi$ (deg) | $F_{hg}$ (mJy) | $F_{corereg}$ (mJy) |
|---|---|---|---|---|---|---|---|
| (1) | (2) | (3) | (4) | (5) | (6) | (7) | (8) |
| v | 0.068 | 2.1 ± 0.1 | 85 ± 3 | 0.10 ± 0.01 | −22 ± 2 | 11.2 ± 0.5 | 0.77 ± 0.27 |
| b | 0.091 | 2.1 ± 0.1 | 110 ± 4 | 0.11 ± 0.01 | −23 ± 3.5 | 3.8 ± 0.2 | 0.38 ± 0.20 |
| u | 0.109 | 2.8 ± 0.1 | 92 ± 8 | 0.09 ± 0.01 | −23 ± 5 | 1.67 ± 0.10 | 0.15 ± 0.08 |
| uvw1 | 0.147 | 3.5 ± 0.1 | 113 ± 12 | 0.06 ± 0.02 | −33 ± 14 | 0.63 ± 0.10 | 0.08 ± 0.04 |
| uvm2 | 0.207 | 3.8 ± 0.3 | 74 ± 27 | 0.17 ± 0.03 | −43 ± 6.5 | 0.23 ± 0.05 | 0.18 ± 0.03 |
| uvw2 | 0.182 | 4.0 ± 0.2 | 102 ± 13 | 0.15 ± 0.03 | −39 ± 7 | 0.33 ± 0.05 | 0.09 ± 0.02 |

**Note.** Sérsic function is $I(R) = I_e \exp(-b_n[(R/R_e)^{1/n} - 1])$, with $b_n = 2n - 1/3$, $R_e$—half-light radius, and $I_e$—intensity at $R_e$. Columns are: (1)—UVOT band; (2)—total extinction; (3)—value of Sérsic parameter $n$; (4)—half-light radius of host galaxy; (5)—its ellipticity; (6)—PA of the major axis of host galaxy; (7)—host galaxy flux density in a circle aperture of a radius of 5″; (8)—flux density of the core in a circle aperture of a radius of 5″, averaged over the period of the EHT campaign and its standard deviation. Flux densities are corrected for the extinction.

Table A3
Optical Flux Densities from Swift-UVOT Observations

| MJD (days) (1) | $F_v$ (mJy) (2) | MJD (days) (3) | $F_b$ (mJy) (4) | MJD (days) (5) | $F_u$ (mJy) (6) |
|---|---|---|---|---|---|
| 57834.2809 | 2.23 ± 0.62 | 57834.2759 | 0.78 ± 0.21 | 57834.2749 | 0.23 ± 0.11 |
| 57835.2827 | 0.37 ± 0.57 | 57835.2770 | 0.37 ± 0.19 | 57835.2759 | 0.14 ± 0.11 |
| 57836.5034 | 0.37 ± 0.59 | 57836.5013 | 0.37 ± 0.21 | 57836.5008 | 0.05 ± 0.11 |
| 57837.0201 | 1.25 ± 0.60 | 57837.0148 | 0.68 ± 0.20 | 57837.0137 | 0.24 ± 0.11 |
| 57838.2706 | 0.77 ± 0.59 | 57838.2657 | 0.34 ± 0.20 | 57838.2647 | 0.10 ± 0.11 |
| 57839.8949 | 0.81 ± 0.60 | 57839.8924 | 0.31 ± 0.20 | 57839.8919 | 0.16 ± 0.11 |
| 57840.9337 | 0.55 ± 0.58 | 57840.9286 | 0.33 ± 0.20 | 57840.9275 | 0.07 ± 0.11 |
| 57841.9369 | 0.46 ± 0.59 | 57841.9322 | 0.13 ± 0.20 | 57841.9312 | 0.08 ± 0.11 |
| 57842.0836 | 0.80 ± 0.58 | 57842.0808 | 0.28 ± 0.20 | 57842.0802 | 0.20 ± 0.11 |
| 57843.5795 | 0.87 ± 0.62 | 57843.5756 | 0.39 ± 0.21 | 57843.5748 | 0.30 ± 0.12 |
| 57844.9235 | 1.06 ± 0.60 | 57844.9186 | 0.40 ± 0.20 | 57844.9176 | 0.02 ± 0.11 |
| 57846.0459 | 0.53 ± 0.54 | 57846.0414 | 0.43 ± 0.18 | 57846.0404 | 0.11 ± 0.10 |
| 57846.3732 | 0.80 ± 0.60 | 57846.3684 | 0.68 ± 0.21 | 57846.3674 | 0.06 ± 0.11 |
| 57847.3696 | 0.54 ± 0.59 | 57847.3647 | 0.33 ± 0.20 | 57847.3637 | 0.17 ± 0.11 |
| 57848.2995 | 0.42 ± 0.59 | 57848.2949 | 0.45 ± 0.20 | 57848.2939 | 0.14 ± 0.11 |
| 57849.2965 | 0.72 ± 0.60 | 57849.2919 | 0.11 ± 0.20 | 57849.2909 | 0.16 ± 0.11 |
| 57854.7593 | 1.31 ± 0.61 | 57854.7545 | 0.75 ± 0.21 | 57854.7535 | 0.15 ± 0.11 |
| 57855.2771 | 0.98 ± 0.60 | 57855.2722 | 0.21 ± 0.20 | 57855.2712 | 0.19 ± 0.11 |
| 57856.6054 | 0.66 ± 0.59 | 57856.6004 | 0.50 ± 0.20 | 57856.5994 | 0.13 ± 0.11 |
| 57857.1997 | 0.64 ± 0.59 | 57857.1973 | 0.41 ± 0.20 | 57857.1968 | 0.09 ± 0.11 |
| 57860.9346 | 2.15 ± 0.63 | 57860.9296 | 0.52 ± 0.20 | 57860.9286 | 0.24 ± 0.11 |
| 57861.2591 | 1.01 ± 0.60 | 57861.2542 | 0.47 ± 0.20 | 57861.2532 | 0.25 ± 0.11 |
| 57863.8557 | 1.20 ± 0.60 | 57863.8508 | 0.68 ± 0.21 | 57863.8498 | 0.10 ± 0.11 |

**Note.** $F_v$, $F_b$, $F_u$—flux densities within a radius of 5″ in bands $v$, $b$, and $u$, respectively, corrected for the extinction and host galaxy contamination.

Table A4
UV Flux Densities from Swift-UVOT Observations

| MJD (days) (1) | $F_{uvw1}$ (mJy) (2) | MJD (days) (3) | $F_{uvm2}$ (mJy) (4) | MJD (days) (5) | $F_{uvw2}$ (mJy) (6) |
|---|---|---|---|---|---|
| 57834.2733 | 0.10 ± 0.10 | 57834.2828 | 0.23 ± 0.06 | 57834.2784 | 0.13 ± 0.07 |
| 57835.2741 | 0.07 ± 0.10 | 57835.2847 | 0.18 ± 0.05 | 57835.2799 | 0.07 ± 0.06 |
| 57836.5002 | 0.10 ± 0.11 | … | … | 57836.5024 | 0.03 ± 0.06 |
| 57837.0121 | 0.10 ± 0.10 | … | … | 57837.0175 | 0.11 ± 0.07 |
| 57838.2633 | 0.09 ± 0.10 | 57838.2724 | 0.22 ± 0.06 | 57838.2682 | 0.07 ± 0.06 |
| 57839.8912 | 0.08 ± 0.10 | … | … | 57839.8937 | 0.12 ± 0.07 |
| 57840.9260 | 0.07 ± 0.10 | 57840.9356 | 0.21 ± 0.06 | 57840.9312 | 0.09 ± 0.06 |
| 57841.9298 | 0.07 ± 0.10 | 57841.9389 | 0.20 ± 0.05 | 57841.9346 | 0.08 ± 0.07 |
| 57842.0794 | 0.14 ± 0.10 | 57842.0845 | 0.19 ± 0.05 | 57842.0823 | 0.06 ± 0.06 |
| 57843.5736 | 0.01 ± 0.11 | … | … | 57843.5775 | 0.12 ± 0.07 |
| 57844.9161 | 0.10 ± 0.10 | 57844.9252 | 0.21 ± 0.06 | 57844.9211 | 0.09 ± 0.07 |
| 57846.0390 | 0.11 ± 0.10 | 57846.0476 | 0.17 ± 0.05 | 57846.0435 | 0.10 ± 0.06 |
| 57846.3659 | 0.09 ± 0.10 | … | … | 57846.3708 | 0.09 ± 0.07 |
| 57847.3622 | 0.07 ± 0.10 | … | … | 57847.3671 | 0.09 ± 0.07 |
| 57848.2925 | 0.04 ± 0.10 | 57848.3014 | 0.13 ± 0.05 | 57848.2972 | 0.03 ± 0.06 |
| 57849.2895 | 0.04 ± 0.10 | 57849.2985 | 0.21 ± 0.05 | 57849.2942 | 0.08 ± 0.07 |
| 57850.6275 | 0.03 ± 0.10 | … | … | … | … |
| 57854.7521 | 0.07 ± 0.10 | 57854.7612 | 0.19 ± 0.05 | 57854.7569 | 0.08 ± 0.07 |
| 57855.2697 | 0.01 ± 0.10 | … | … | 57855.2746 | 0.10 ± 0.07 |
| 57856.5979 | 0.11 ± 0.10 | … | … | 57856.6029 | 0.10 ± 0.07 |
| 57857.1961 | 0.11 ± 0.10 | … | … | 57857.1986 | 0.07 ± 0.07 |
| 57860.9271 | 0.16 ± 0.10 | 57860.9363 | 0.21 ± 0.06 | 57860.9321 | 0.10 ± 0.07 |
| 57861.2517 | 0.10 ± 0.10 | … | … | 57861.2567 | 0.10 ± 0.07 |
| 57863.8484 | 0.07 ± 0.10 | … | … | 57863.8533 | 0.04 ± 0.06 |

**Note.** $F_{uvw1}$, $F_{uvm2}$, $F_{uvw2}$—flux densities within a radius of 5″ in bands $uvw1$, $uvm2$, and $uvw2$, respectively, corrected for the extinction and host galaxy contamination.

Table A5
HST-based Flux Densities of the Core and HST-1

| MJD (days) (1) | $\lambda_{\rm eff}$ (Å) (2) | $F_{\rm core}$ (mJy) (3) | $F_{\rm HST-1}$ (mJy) (4) |
|---|---|---|---|
| 57850.69 | 2714 | 0.159 ± 0.015 | 0.022 ± 0.003 |
| 57850.71 | 5779 | 0.450 ± 0.017 | 0.036 ± 0.003 |
| 57855.20 | 2714 | 0.144 ± 0.015 | 0.022 ± 0.003 |
| 57855.21 | 5779 | 0.484 ± 0.017 | 0.035 ± 0.002 |
| 57860.50 | 2714 | 0.188 ± 0.017 | 0.023 ± 0.003 |
| 57860.51 | 5779 | 0.538 ± 0.019 | 0.035 ± 0.003 |

Table A6
X-Ray Observations and Fluxes

| Observatory | Observation ID | MJD | Exposure (ks) | Total Flux[a] | Net Flux[b] |
|---|---|---|---|---|---|
| Swift-XRT | 00031105027 | 57835.2729 | 1.13 | $8.26^{+0.84}_{-0.61}$ | $3.1^{+0.31}_{-0.23}$ |
| Swift-XRT | 00031105028 | 57836.0674 | 0.83 | $7.98^{+0.78}_{-0.87}$ | $4.21^{+0.41}_{-0.46}$ |
| Swift-XRT | 00031105030 | 57838.2622 | 0.98 | $7.45^{+0.6}_{-0.48}$ | $2.33^{+0.19}_{-0.15}$ |
| Swift-XRT | 00031105032 | 57840.9248 | 1.03 | $9.43^{+0.83}_{-0.69}$ | $4.19^{+0.37}_{-0.31}$ |
| Swift-XRT | 00031105034 | 57842.0451 | 1.09 | $8.16^{+0.83}_{-1.03}$ | $2.39^{+0.24}_{-0.3}$ |
| Swift-XRT | 00031105037 | 57845.9122 | 1.86 | $8.18^{+0.42}_{-0.64}$ | $2.97^{+0.15}_{-0.23}$ |
| Swift-XRT | 00031105040 | 57848.2914 | 0.97 | $6.92^{+0.59}_{-0.65}$ | $1.85^{+0.16}_{-0.17}$ |
| Swift-XRT | 00031105041 | 57849.2885 | 0.65 | $7.61^{+0.7}_{-0.87}$ | $4.26^{+0.39}_{-0.48}$ |
| NuSTAR | 90202052002 | 57854.6848 | 24.4 | ⋯ | ⋯ |
| Swift-XRT | 00031105043 | 57854.7510 | 0.98 | $10.38^{+0.87}_{-0.93}$ | $5.97^{+0.5}_{-0.54}$ |
| Chandra | 20034 | 57854.9909 | 13.1 | ⋯ | ⋯ |
| NuSTAR | 90202052004 | 57857.0320 | 22.5 | ⋯ | ⋯ |
| Chandra | 20035 | 57857.0837 | 13.1 | ⋯ | ⋯ |
| Swift-XRT | 00031105046 | 57857.1293 | 0.89 | $9.0^{+0.92}_{-0.85}$ | $4.25^{+0.44}_{-0.4}$ |
| Swift-XRT | 00031105048 | 57860.9260 | 0.98 | $7.3^{+0.49}_{-0.72}$ | $2.2^{+0.15}_{-0.22}$ |
| Swift-XRT | 00031105051 | 57863.8473 | 0.89 | $8.22^{+0.78}_{-0.93}$ | $3.4^{+0.32}_{-0.38}$ |

**Notes.**
[a] Total flux represents the unabsorbed flux derived from the Swift-XRT spectral fits. See Table A8 for joint Chandra and NuSTAR constraints on the flux.
[b] Net flux corresponds to the unabsorbed flux from core, HST-1, outer jet only in the 2–10 keV band. The flux units are $10^{-12}$ erg cm$^{-2}$ s$^{-1}$

Table A7
VHE γ-Ray Observation Summary

| IACT | Start Time (MJD) | Stop Time (MJD) | Effective Observation Time (hr) | Zenith angle (°) | Significance (σ) | $F_{E>350\ \mathrm{GeV}}$[a] (ph cm$^{-2}$ s$^{-1}$ × 10$^{-12}$) | UL(95%)[a] |
|---|---|---|---|---|---|---|---|
| H.E.S.S. | 57840.92 | 57841.02 | 1.3 | 37–39 | 2.6 | $1.51^{+0.73}_{-0.64}$ | |
| H.E.S.S. | 57841.89 | 57842.02 | 1.8 | 36–41 | 0.4 | $0.18^{+0.52}_{-0.45}$ | 1.28 |
| H.E.S.S. | 57844.91 | 57845.03 | 2.3 | 36–45 | 1.3 | $0.57^{+0.53}_{-0.46}$ | 1.67 |
| H.E.S.S. | 57845.90 | 57846.02 | 2.3 | 36–42 | 1.7 | $0.73^{+0.50}_{-0.44}$ | 1.77 |
| MAGIC | 57834.13 | 57834.19 | 1.23 | 25–44 | −0.1 | −0.06 ± 0.67 | 1.45 |
| MAGIC | 57835.04 | 57835.10 | 0.98 | 16–20 | −0.2 | −0.23 ± 0.98 | 1.97 |
| MAGIC | 57836.01 | 57836.09 | 1.75 | 16–25 | 0.8 | 0.86 ± 0.88 | 3.34 |
| MAGIC | 57836.96 | 57837.02 | 1.47 | 22–39 | 1.3 | 1.60 ± 1.05 | 5.47 |
| MAGIC | 57838.08 | 57838.12 | 0.97 | 17–27 | 1.1 | 1.47 ± 1.21 | 5.70 |
| MAGIC | 57841.05 | 57841.09 | 1.10 | 16–21 | 1.8 | 2.22 ± 1.13 | 6.93 |
| MAGIC | 57841.96 | 57842.08 | 2.84 | 16–36 | 0.6 | 0.51 ± 0.75 | 2.74 |
| MAGIC | 57842.96 | 57843.06 | 2.16 | 16–35 | 1.1 | 1.12 ± 0.85 | 4.09 |
| MAGIC | 57844.01 | 57844.07 | 1.49 | 16–21 | 0.8 | 0.89 ± 1.00 | 4.10 |
| MAGIC | 57844.96 | 57845.02 | 1.29 | 18–33 | 0.7 | 0.86 ± 1.05 | 3.72 |
| MAGIC | 57845.98 | 57846.04 | 1.02 | 17–26 | 1.0 | 1.21 ± 1.09 | 4.97 |
| MAGIC | 57847.03 | 57847.09 | 1.46 | 16–26 | 0.5 | 0.54 ± 0.95 | 3.40 |
| MAGIC | 57848.08 | 57848.16 | 1.53 | 24–46 | 0.6 | 0.71 ± 0.98 | 3.38 |
| MAGIC | 57849.04 | 57849.16 | 2.49 | 16–45 | 1.2 | 1.27 ± 0.89 | 4.46 |
| MAGIC | 57856.91 | 57856.93 | 0.38 | 31–36 | −1.0 | −2.32 ± 1.63 | 2.31 |
| MAGIC | 57859.97 | 57860.09 | 1.78 | 16–33 | 3.0 | 2.52 ± 1.15 | |
| MAGIC | 57862.96 | 57863.04 | 1.63 | 16–21 | 0.8 | 0.75 ± 0.79 | 2.94 |
| MAGIC | 57863.95 | 57864.03 | 1.63 | 16–21 | −0.6 | −0.62 ± 0.77 | 1.45 |
| VERITAS | 57834.28 | 57834.41 | 2.64 | 20–31 | 3.0 | 1.5 ± 0.63 | |
| VERITAS | 57835.29 | 57835.33 | 0.62 | 20–21 | 1.3 | 1.1 ± 1.0 | 3.8 |
| VERITAS | 57838.26 | 57838.40 | 2.92 | 20–28 | 0.8 | 0.39 ± 0.51 | 1.5 |
| VERITAS | 57839.25 | 57839.39 | 2.97 | 20–27 | 0.2 | 0.078 ± 0.44 | 1.1 |
| VERITAS | 57846.24 | 57846.37 | 2.50 | 20–26 | 2.0 | 1.1 ± 0.61 | |
| VERITAS | 57847.32 | 57847.43 | 2.00 | 24–42 | 1.4 | 1.0 ± 0.78 | 2.7 |
| VERITAS | 57861.18 | 57861.26 | 1.32 | 20–27 | 2.0 | 1.6 ± 0.96 | |

**Note.**
[a] Only statistical 1σ errors on the fluxes are given. 95% confidence level upper limits are given for flux measurements with less than 2σ significance.

**Table A8**
SED for the M87 Core Around the EHT Observing Campaign in 2017

| (1) Observatory | (2) Band[a] | (3) $\nu$[a] (Hz) | (4) $f_\nu$[c] (Jy) | (5) $\nu f_\nu$[c] (erg s$^{-1}$ cm$^{-2}$) | (6) Angular Scale[b] (″) | (7) Observation Date (MJD) | (8) Section |
|---|---|---|---|---|---|---|---|
| EVN | 1.7 GHz | $1.7 \times 10^9$ | $0.75 \pm 0.08$ | $(1.28 \pm 0.14) \times 10^{-14}$ | $5 \times 10^{-3}$ | 2017-05-09 [57882] | Section 2.1.1 |
| HSA | 8.4 GHz | $8.4 \times 10^9$ | $1.30 \pm 0.13$ | $(1.09 \pm 0.11) \times 10^{-13}$ | $1.5 \times 10^{-3}$ | 2017-05-15 [57888] | Section 2.1.2 |
|  | 15 GHz | $1.5 \times 10^{10}$ | $1.13 \pm 0.11$ | $(1.70 \pm 0.17) \times 10^{-13}$ | $1.0 \times 10^{-3}$ | 2017-05-16 [57889] |  |
|  | 24 GHz | $2.4 \times 10^{10}$ | $1.22 \pm 0.12$ | $(2.93 \pm 0.29) \times 10^{-13}$ | $1.0 \times 10^{-3}$ | 2017-05-20 [57893] |  |
| VERA | 22 GHz | $2.2 \times 10^{10}$ | $1.32 \pm 0.13$ | $(2.90 \pm 0.29) \times 10^{-13}$ | $1 \times 10^{-3}$ | 2017-04-02 [57845] | Section 2.1.3 |
| EAVN | 22 GHz | $2.2 \times 10^{10}$ | $1.34 \pm 0.13$ | $(2.95 \pm 0.29) \times 10^{-13}$ | $1 \times 10^{-3}$ | 2017-04-03 [57846] | Section 2.1.4 |
|  | 43 GHz | $4.3 \times 10^{10}$ | $1.10 \pm 0.11$ | $(4.73 \pm 0.47) \times 10^{-13}$ | $1 \times 10^{-3}$ | 2017-04-09 [57852] |  |
| KVN | 21 GHz | $2.17 \times 10^{10}$ | $1.63 \pm 0.17$ | $(3.54 \pm 0.37) \times 10^{-13}$ | $4.3 \times 10^{-3}$ | 2017-04-19 [57862] | Section 2.1.5 |
|  | 43 GHz | $4.34 \times 10^{10}$ | $1.31 \pm 0.13$ | $(5.69 \pm 0.56) \times 10^{-13}$ | $2.2 \times 10^{-3}$ | 2017-04-19 [57862] |  |
|  | 87 GHz | $8.68 \times 10^{10}$ | $1.09 \pm 0.11$ | $(9.46 \pm 0.95) \times 10^{-13}$ | $1.1 \times 10^{-3}$ | 2017-04-19 [57862] |  |
|  | 129 GHz | $1.29 \times 10^{11}$ | $0.83 \pm 0.25$ | $(1.07 \pm 0.32) \times 10^{-12}$ | $0.8 \times 10^{-3}$ | 2017-04-19 [57862] |  |
| VLBA | 24 GHz | $2.4 \times 10^{10}$ | $1.12 \pm 0.11$ | $(2.00 \pm 0.20) \times 10^{-13}$ | $6 \times 10^{-4}$ | 2017-05-05 [57878] | Section 2.1.6 |
|  | 43 GHz | $4.3 \times 10^{10}$ | $1.09 \pm 0.11$ | $(3.62 \pm 0.36) \times 10^{-13}$ | $3 \times 10^{-4}$ | 2017-05-05 [57878] |  |
| GMVA | 86 GHz | $8.6 \times 10^{10}$ | $0.73 \pm 0.22$ | $(6.30 \pm 0.60) \times 10^{-13}$ | $\simeq 1.5 \times 10^{-4}$ | 2017-03-30 [57842] | Section 2.1.7 |
| ALMA | 220 GHz | $2.2 \times 10^{11}$ | $1.31 \pm 0.13$ | $(2.87 \pm 0.29) \times 10^{-12}$ | $\simeq 1.8$ | 2017-04-05–11 [57848–57854] | Section 2.1.8 |
| SMA | 220 GHz | $2.2 \times 10^{11}$ | $1.32 \pm 0.13$ | $(2.90 \pm 0.29) \times 10^{-12}$ | $\simeq 3.0$ | 2017-04-05–11 [57848–57854] | Section 2.1.9 |
| EHT | 230 GHz | $2.3 \times 10^{11}$ | $0.66 \pm 0.16$ | $(1.51 \pm 0.37) \times 10^{-12}$ | $2 \times 10^{-5}$ | 2017-04-05–11 [57848–57854] | Section 3.2 |
| HST | F606W | $5.2 \times 10^{14}$ | $(4.91 \pm 0.40) \times 10^{-4}$ | $(2.55 \pm 0.21) \times 10^{-12}$ | $\simeq 0.1$ | 2017-04-07–17 [57850–57860] | Section 2.2.2 |
|  | F275W | $1.1 \times 10^{15}$ | $(1.64 \pm 0.24) \times 10^{-4}$ | $(1.80 \pm 0.26) \times 10^{-12}$ |  |  |  |
| *Swift* | V | $5.47 \times 10^{14}$ | $(7.92 \pm 2.85) \times 10^{-4}$ | $(4.33 \pm 1.56) \times 10^{-12}$ | $\simeq 3$ | 2017-04-06–12 [57849–57855] | Section 2.2.1 |
|  | B | $6.82 \times 10^{14}$ | $(3.87 \pm 2.09) \times 10^{-4}$ | $(2.64 \pm 1.43) \times 10^{-12}$ |  |  |  |
|  | U | $8.65 \times 10^{14}$ | $(1.49 \pm 0.81) \times 10^{-4}$ | $(1.29 \pm 0.70) \times 10^{-12}$ |  |  |  |
|  | UVW1 | $1.15 \times 10^{15}$ | $(8.20 \pm 3.70) \times 10^{-5}$ | $(9.43 \pm 4.25) \times 10^{-13}$ |  |  |  |
|  | UVM2 | $1.33 \times 10^{15}$ | $(1.92 \pm 0.29) \times 10^{-4}$ | $(2.55 \pm 0.39) \times 10^{-12}$ |  |  |  |
|  | UVW2 | $1.55 \times 10^{15}$ | $(8.90 \pm 2.30) \times 10^{-5}$ | $(1.38 \pm 0.36) \times 10^{-12}$ |  |  |  |
|  | 2–10 keV | $1.08 \times 10^{18}$ | $< 2.08 \times 10^{-7}$ | $< 2.25 \times 10^{-12}$ | 36 | 2017-04-05–11 [57851–57857] | Section 2.3.4 |
| Chandra +NuSTAR | 0.40–0.51 keV | $1.09 \times 10^{17}$ | $(9.52 \pm 0.84) \times 10^{-7}$ | $(1.04 \pm 0.09) \times 10^{-12}$ | 0.8 | 2017-04-11–14 [57854–57857] | Section 2.3.3 |
|  | 0.51–0.65 keV | $1.40 \times 10^{17}$ | $(7.30 \pm 0.56) \times 10^{-7}$ | $(1.02 \pm 0.08) \times 10^{-12}$ |  |  |  |
|  | 0.65–0.84 keV | $1.79 \times 10^{17}$ | $(5.61 \pm 0.36) \times 10^{-7}$ | $(1.00 \pm 0.06) \times 10^{-12}$ |  |  |  |
|  | 0.84–1.07 keV | $2.29 \times 10^{17}$ | $(4.32 \pm 0.23) \times 10^{-7}$ | $(9.89 \pm 0.52) \times 10^{-13}$ |  |  |  |
|  | 1.07–1.37 keV | $2.93 \times 10^{17}$ | $(3.31 \pm 0.14) \times 10^{-7}$ | $(9.70 \pm 0.41) \times 10^{-13}$ |  |  |  |
|  | 1.37–1.75 keV | $3.75 \times 10^{17}$ | $(2.54 \pm 0.08) \times 10^{-7}$ | $(9.53 \pm 0.31) \times 10^{-13}$ |  |  |  |
|  | 1.75–2.24 keV | $4.80 \times 10^{17}$ | $(1.95 \pm 0.05) \times 10^{-7}$ | $(9.36 \pm 0.25) \times 10^{-13}$ |  |  |  |
|  | 2.24–2.87 keV | $6.13 \times 10^{17}$ | $(1.50 \pm 0.04) \times 10^{-7}$ | $(9.92 \pm 0.24) \times 10^{-13}$ |  |  |  |
|  | 2.87–3.67 keV | $7.85 \times 10^{17}$ | $(1.15 \pm 0.04) \times 10^{-7}$ | $(9.03 \pm 0.28) \times 10^{-13}$ |  |  |  |
|  | 3.67–4.69 keV | $1.00 \times 10^{18}$ | $(8.85 \pm 0.34) \times 10^{-8}$ | $(8.85 \pm 0.35) \times 10^{-13}$ |  |  |  |
|  | 4.69–6.00 keV | $1.28 \times 10^{18}$ | $(6.80 \pm 0.33) \times 10^{-8}$ | $(8.70 \pm 0.43) \times 10^{-13}$ |  |  |  |
|  | 6.00–7.67 keV | $1.64 \times 10^{18}$ | $(5.23 \pm 0.32) \times 10^{-8}$ | $(8.58 \pm 0.52) \times 10^{-13}$ |  |  |  |
|  | 7.67–9.82 keV | $2.10 \times 10^{18}$ | $(4.02 \pm 0.29) \times 10^{-8}$ | $(8.44 \pm 0.60) \times 10^{-13}$ |  |  |  |

**Table A8**
(Continued)

| (1) Observatory | (2) Band[a] | (3) $\nu$[a] (Hz) | (4) $f_\nu$[c] (Jy) | (5) $\nu f_\nu$[c] (erg s$^{-1}$ cm$^{-2}$) | (6) Angular Scale[b] (″) | (7) Observation Date (MJD) | (8) Section |
|---|---|---|---|---|---|---|---|
| | 9.82–12.6 keV | $2.69 \times 10^{18}$ | $(3.09 \pm 0.26) \times 10^{-8}$ | $(8.31 \pm 0.69) \times 10^{-13}$ | | | |
| | 12.6–16.1 keV | $3.44 \times 10^{18}$ | $(2.37 \pm 0.23) \times 10^{-8}$ | $(8.15 \pm 0.78) \times 10^{-13}$ | | | |
| | 16.1–20.5 keV | $4.40 \times 10^{18}$ | $(1.83 \pm 0.20) \times 10^{-8}$ | $(8.05 \pm 0.86) \times 10^{-13}$ | | | |
| | 20.5–26.3 keV | $5.62 \times 10^{18}$ | $(1.41 \pm 0.17) \times 10^{-8}$ | $(7.92 \pm 0.94) \times 10^{-13}$ | | | |
| | 26.3–33.6 keV | $7.19 \times 10^{18}$ | $(1.08 \pm 0.14) \times 10^{-8}$ | $(7.77 \pm 1.02) \times 10^{-13}$ | | | |
| | 33.6–43.0 keV | $9.20 \times 10^{18}$ | $(8.32 \pm 1.20) \times 10^{-9}$ | $(7.65 \pm 1.10) \times 10^{-13}$ | | | |
| | 43.0–55.0 keV | $1.18 \times 10^{19}$ | $(6.40 \pm 1.00) \times 10^{-9}$ | $(7.55 \pm 1.18) \times 10^{-13}$ | | | |
| Fermi | 0.1–1 GeV | $7.65 \times 10^{22}$ | $(4.27 \pm 1.32) \times 10^{-12}$ | $(3.27 \pm 1.01) \times 10^{-12}$ | $\simeq 7700$ | 2017-03-01–05-31 [57813–57904] | Section 2.4.1 |
| | 1–10 GeV | $7.65 \times 10^{23}$ | $(2.45 \pm 0.98) \times 10^{-13}$ | $(1.87 \pm 0.75) \times 10^{-12}$ | $\simeq 1250$ | | |
| | 10–100 GeV | $7.65 \times 10^{24}$ | $(6.49 \pm 3.87) \times 10^{-14}$ | $(4.96 \pm 2.96) \times 10^{-12}$ | $\simeq 400$ | | |
| | 100 GeV–1 TeV | $7.65 \times 10^{25}$ | $<6.88 \times 10^{-14}$ | $<5.26 \times 10^{-11}$ | $\simeq 350$ | | |
| H.E.S.S. | 240–380 GeV | $7.31 \times 10^{25}$ | $<4.40 \times 10^{-15}$ | $<3.22 \times 10^{-12}$ | $\simeq 360$ | 2017-03-28–04-03 [57840–57846] | Section 2.4.2 |
| | 380–603 GeV | $1.16 \times 10^{26}$ | $<1.24 \times 10^{-15}$ | $<1.44 \times 10^{-12}$ | | | |
| | 603–955 GeV | $1.83 \times 10^{26}$ | $<5.31 \times 10^{-16}$ | $<9.74 \times 10^{-13}$ | | | |
| | 955–1512 GeV | $2.91 \times 10^{26}$ | $<3.05 \times 10^{-16}$ | $<8.88 \times 10^{-13}$ | | | |
| | 1512–2397 GeV | $4.61 \times 10^{26}$ | $<1.67 \times 10^{-16}$ | $<7.70 \times 10^{-13}$ | | | |
| MAGIC | 126–199 GeV | $3.82 \times 10^{25}$ | $(3.68 \pm 2.14) \times 10^{-15}$ | $(1.40 \pm 0.82) \times 10^{-13}$ | $\simeq 240$ | 2017-03-22–04-21 [57834–57864] | Section 2.4.2 |
| | 199–315 GeV | $6.06 \times 10^{25}$ | $(1.10 \pm 0.77) \times 10^{-15}$ | $(6.67 \pm 4.68) \times 10^{-13}$ | $\simeq 210$ | | |
| | 315–500 GeV | $9.60 \times 10^{25}$ | $(3.99 \pm 3.69) \times 10^{-16}$ | $(3.83 \pm 3.54) \times 10^{-13}$ | $\simeq 180$ | | |
| | 500–792 GeV | $1.52 \times 10^{26}$ | $(5.08 \pm 1.94) \times 10^{-16}$ | $(7.72 \pm 2.94) \times 10^{-13}$ | $\simeq 160$ | | |
| | 792–1256 GeV | $2.41 \times 10^{26}$ | $(1.95 \pm 1.00) \times 10^{-16}$ | $(4.70 \pm 2.41) \times 10^{-13}$ | $\simeq 150$ | | |
| | 1256–1991 GeV | $3.82 \times 10^{26}$ | $(2.15 \pm 0.86) \times 10^{-16}$ | $(8.21 \pm 3.30) \times 10^{-13}$ | $\simeq 130$ | | |
| VERITAS | 200–281 GeV | $5.73 \times 10^{25}$ | $<1.74 \times 10^{-14}$ | $<9.97 \times 10^{-12}$ | $\simeq 360$ | 2017-03-22–04-18 [57834–57861] | Section 2.4.2 |
| | 281–397 GeV | $8.10 \times 10^{25}$ | $<2.10 \times 10^{-15}$ | $<1.70 \times 10^{-12}$ | | | |
| | 398–562 GeV | $1.14 \times 10^{26}$ | $(6.42 \pm 3.42) \times 10^{-16}$ | $(7.32 \pm 3.90) \times 10^{-13}$ | | | |
| | 562–793 GeV | $1.61 \times 10^{26}$ | $(4.02 \pm 2.30) \times 10^{-16}$ | $(6.47 \pm 3.70) \times 10^{-13}$ | | | |
| | 793–1120 GeV | $2.28 \times 10^{26}$ | $<4.22 \times 10^{-16}$ | $<9.62 \times 10^{-13}$ | | | |
| | 1120–1583 GeV | $3.22 \times 10^{26}$ | $<3.51 \times 10^{-16}$ | $<1.13 \times 10^{-12}$ | | | |

**Notes.**
[a] Column (2) uses nomenclature common in the given frequency or energy range, while column (3) gives the central frequency of the band.
[b] Estimate of the spatial scale of the emission: typically, FWHM of the instrumental beam or PSF. For elliptical beams, the average between axes is given here. At the adopted distance of M87, 1″ corresponds to 81 pc, or $\sim 2.5 \times 10^5\, r_g$ in projection.
[c] Data analysis yields either flux density or flux within a band; the other value is calculated by multiplication or division with the representative frequency given in column (3). See Section 3.2 for details.

4. UV flux densities from Swift-UVOT observations in Table A4,
5. HST-based flux densities of the core and HST-1 in Table A5,
6. X-ray observations and fluxes in Table A6,
7. VHE γ-ray observation summary in Table A7, and
8. SED for the M87 core around the EHT observing campaign in 2017 in Table A8.

## Appendix B
## Supplementary Material

Data products presented in this Letter are available for download through the EHT Collaboration Data Webpage (https://eventhorizontelescope.org/for-astronomers/data), or directly from the CyVerse repository accessible via the following permanent DOI:https://doi.org/10.25739/mhh2-cw46. The repository contains the following data products.

1. Broadband spectrum table with frequency, flux density, its uncertainty, and instrument index (format: CSV)
2. Description of observations and data processing (format: text)
3. Fluxes from H.E.S.S., MAGIC, and VERITAS observations (format: CSV)
4. Fluxes from Swift-XRT observations (format: CSV)
5. Flux densities from Swift-UVOT observations (format: CSV)
6. Flux densities from HST observations (format: CSV)
7. Flux densities from radio observations (format: CSV)
8. Scripts, spectral, and response files for modeling Swift-XRT data (format: standard X-ray data formats)
9. Scripts, spectral, and response files for modeling Chandra and NuSTAR data (format: standard X-ray data formats)
10. Sampled posterior distributions of X-ray spectral model based on Chandra and NuSTAR data (format: FITS)
11. Sampled posterior distributions of the single-zone broadband spectral model described in Section 3.3.2 (format: FITS).

## ORCID iDs

J. C. Algaba https://orcid.org/0000-0001-6993-1696
J. Anczarski https://orcid.org/0000-0003-4317-3385
M. Baloković https://orcid.org/0000-0003-0476-6647
S. Chandra https://orcid.org/0000-0002-8776-1835
Y.-Z. Cui https://orcid.org/0000-0001-6311-4345
A. D. Falcone https://orcid.org/0000-0002-5068-7344
M. Giroletti https://orcid.org/0000-0002-8657-8852
C. Goddi https://orcid.org/0000-0002-2542-7743
K. Hada https://orcid.org/0000-0001-6906-772X
D. Haggard https://orcid.org/0000-0001-6803-2138
S. Jorstad https://orcid.org/0000-0001-6158-1708
A. Kaur https://orcid.org/0000-0002-0878-1193
T. Kawashima https://orcid.org/0000-0001-8527-0496
G. Keating https://orcid.org/0000-0002-3490-146X
J.-Y. Kim https://orcid.org/0000-0001-8229-7183
M. Kino https://orcid.org/0000-0002-2709-7338
S. Komossa https://orcid.org/0000-0002-9214-4428
E. V. Kravchenko https://orcid.org/0000-0003-4540-4095
T. P. Krichbaum https://orcid.org/0000-0002-4892-9586
S.-S. Lee https://orcid.org/0000-0002-6269-594X
R.-S. Lu (路如森) https://orcid.org/0000-0002-7692-7967
S. Markoff https://orcid.org/0000-0001-9564-0876
J. Neilsen https://orcid.org/0000-0002-8247-786X
M. A. Nowak https://orcid.org/0000-0001-6923-1315
J. Park https://orcid.org/0000-0001-6558-9053
G. Principe https://orcid.org/0000-0003-0406-7387
V. Ramakrishnan https://orcid.org/0000-0002-9248-086X
M. T. Reynolds https://orcid.org/0000-0003-1621-9392
M. Sasada https://orcid.org/0000-0001-5946-9960
S. S. Savchenko https://orcid.org/0000-0003-4147-3851
K. E. Williamson https://orcid.org/0000-0003-1318-8535
Kazunori Akiyama https://orcid.org/0000-0002-9475-4254
Antxon Alberdi https://orcid.org/0000-0002-9371-1033
Richard Anantua https://orcid.org/0000-0003-3457-7660
Rebecca Azulay https://orcid.org/0000-0002-2200-5393
Anne-Kathrin Baczko https://orcid.org/0000-0003-3090-3975
John Barrett https://orcid.org/0000-0002-9290-0764
Bradford A. Benson https://orcid.org/0000-0002-5108-6823
Lindy Blackburn https://orcid.org/0000-0002-9030-642X
Raymond Blundell https://orcid.org/0000-0002-5929-5857
Katherine L. Bouman https://orcid.org/0000-0003-0077-4367
Geoffrey C. Bower https://orcid.org/0000-0003-4056-9982
Hope Boyce https://orcid.org/0000-0002-6530-5783
Christiaan D. Brinkerink https://orcid.org/0000-0002-2322-0749
Roger Brissenden https://orcid.org/0000-0002-2556-0894
Silke Britzen https://orcid.org/0000-0001-9240-6734
Avery E. Broderick https://orcid.org/0000-0002-3351-760X
Thomas Bronzwaer https://orcid.org/0000-0003-1151-3971
Do-Young Byun https://orcid.org/0000-0003-1157-4109
Andrew Chael https://orcid.org/0000-0003-2966-6220
Chi-kwan Chan https://orcid.org/0000-0001-6337-6126
Shami Chatterjee https://orcid.org/0000-0002-2878-1502
Koushik Chatterjee https://orcid.org/0000-0002-2825-3590
Paul M. Chesler https://orcid.org/0000-0001-6327-8462
Ilje Cho https://orcid.org/0000-0001-6083-7521
Pierre Christian https://orcid.org/0000-0001-6820-9941
John E. Conway https://orcid.org/0000-0003-2448-9181
Thomas M. Crawford https://orcid.org/0000-0001-9000-5013
Geoffrey B. Crew https://orcid.org/0000-0002-2079-3189
Alejandro Cruz-Osorio https://orcid.org/0000-0002-3945-6342
Jordy Davelaar https://orcid.org/0000-0002-2685-2434
Mariafelicia De Laurentis https://orcid.org/0000-0002-9945-682X
Roger Deane https://orcid.org/0000-0003-1027-5043
Jessica Dempsey https://orcid.org/0000-0003-1269-9667
Gregory Desvignes https://orcid.org/0000-0003-3922-4055
Jason Dexter https://orcid.org/0000-0003-3903-0373
Sheperd S. Doeleman https://orcid.org/0000-0002-9031-0904
Ralph P. Eatough https://orcid.org/0000-0001-6196-4135
Heino Falcke https://orcid.org/0000-0002-2526-6724
Joseph Farah https://orcid.org/0000-0003-4914-5625
Vincent L. Fish https://orcid.org/0000-0002-7128-9345
H. Alyson Ford https://orcid.org/0000-0002-9797-0972

Raquel Fraga-Encinas  https://orcid.org/0000-0002-5222-1361
Antonio Fuentes  https://orcid.org/0000-0002-8773-4933
Peter Galison  https://orcid.org/0000-0002-6429-3872
Charles F. Gammie  https://orcid.org/0000-0001-7451-8935
Roberto García  https://orcid.org/0000-0002-6584-7443
Boris Georgiev  https://orcid.org/0000-0002-3586-6424
Roman Gold  https://orcid.org/0000-0003-2492-1966
Arturo I. Gómez-Ruiz  https://orcid.org/0000-0001-9395-1670
Minfeng Gu (顾敏峰)  https://orcid.org/0000-0002-4455-6946
Mark Gurwell  https://orcid.org/0000-0003-0685-3621
Ronald Hesper  https://orcid.org/0000-0003-1918-6098
Luis C. Ho (何子山)  https://orcid.org/0000-0001-6947-5846
Mareki Honma  https://orcid.org/0000-0003-4058-9000
Chih-Wei L. Huang  https://orcid.org/0000-0001-5641-3953
Shiro Ikeda  https://orcid.org/0000-0002-2462-1448
Sara Issaoun  https://orcid.org/0000-0002-5297-921X
David J. James  https://orcid.org/0000-0001-5160-4486
Michael Janssen  https://orcid.org/0000-0001-8685-6544
Britton Jeter  https://orcid.org/0000-0003-2847-1712
Wu Jiang (江悟)  https://orcid.org/0000-0001-7369-3539
Alejandra Jiménez-Rosales  https://orcid.org/0000-0002-2662-3754
Michael D. Johnson  https://orcid.org/0000-0002-4120-3029
Taehyun Jung  https://orcid.org/0000-0001-7003-8643
Mansour Karami  https://orcid.org/0000-0001-7387-9333
Ramesh Karuppusamy  https://orcid.org/0000-0002-5307-2919
Mark Kettenis  https://orcid.org/0000-0002-6156-5617
Dong-Jin Kim  https://orcid.org/0000-0002-7038-2118
Junhan Kim  https://orcid.org/0000-0002-4274-9373
Jun Yi Koay  https://orcid.org/0000-0002-7029-6658
Patrick M. Koch  https://orcid.org/0000-0003-2777-5861
Shoko Koyama  https://orcid.org/0000-0002-3723-3372
Michael Kramer  https://orcid.org/0000-0002-4175-2271
Carsten Kramer  https://orcid.org/0000-0002-4908-4925
Tod R. Lauer  https://orcid.org/0000-0003-3234-7247
Aviad Levis  https://orcid.org/0000-0001-7307-632X
Yan-Rong Li (李彦荣)  https://orcid.org/0000-0001-5841-9179
Zhiyuan Li (李志远)  https://orcid.org/0000-0003-0355-6437
Michael Lindqvist  https://orcid.org/0000-0002-3669-0715
Rocco Lico  https://orcid.org/0000-0001-7361-2460
Greg Lindahl  https://orcid.org/0000-0002-6100-4772
Jun Liu (刘俊)  https://orcid.org/0000-0002-7615-7499
Kuo Liu  https://orcid.org/0000-0002-2953-7376
Elisabetta Liuzzo  https://orcid.org/0000-0003-0995-5201
Laurent Loinard  https://orcid.org/0000-0002-5635-3345
Nicholas R. MacDonald  https://orcid.org/0000-0002-6684-8691
Jirong Mao (毛基荣)  https://orcid.org/0000-0002-7077-7195
Nicola Marchili  https://orcid.org/0000-0002-5523-7588
Daniel P. Marrone  https://orcid.org/0000-0002-2367-1080
Alan P. Marscher  https://orcid.org/0000-0001-7396-3332
Iván Martí-Vidal  https://orcid.org/0000-0003-3708-9611
Satoki Matsushita  https://orcid.org/0000-0002-2127-7880
Lynn D. Matthews  https://orcid.org/0000-0002-3728-8082
Lia Medeiros  https://orcid.org/0000-0003-2342-6728
Karl M. Menten  https://orcid.org/0000-0001-6459-0669
Izumi Mizuno  https://orcid.org/0000-0002-7210-6264
Yosuke Mizuno  https://orcid.org/0000-0002-8131-6730
James M. Moran  https://orcid.org/0000-0002-3882-4414
Kotaro Moriyama  https://orcid.org/0000-0003-1364-3761
Monika Moscibrodzka  https://orcid.org/0000-0002-4661-6332
Cornelia Müller  https://orcid.org/0000-0002-2739-2994
Gibwa Musoke  https://orcid.org/0000-0003-1984-189X
Alejandro Mus Mejías  https://orcid.org/0000-0003-0329-6874
Hiroshi Nagai  https://orcid.org/0000-0003-0292-3645
Neil M. Nagar  https://orcid.org/0000-0001-6920-662X
Masanori Nakamura  https://orcid.org/0000-0001-6081-2420
Ramesh Narayan  https://orcid.org/0000-0002-1919-2730
Iniyan Natarajan  https://orcid.org/0000-0001-8242-4373
Chunchong Ni  https://orcid.org/0000-0003-1361-5699
Aristeidis Noutsos  https://orcid.org/0000-0002-4151-3860
Héctor Olivares  https://orcid.org/0000-0001-6833-7580
Gisela N. Ortiz-León  https://orcid.org/0000-0002-2863-676X
Daniel C. M. Palumbo  https://orcid.org/0000-0002-7179-3816
Ue-Li Pen  https://orcid.org/0000-0003-2155-9578
Dominic W. Pesce  https://orcid.org/0000-0002-5278-9221
Oliver Porth  https://orcid.org/0000-0002-4584-2557
Felix M. Pötzl  https://orcid.org/0000-0002-6579-8311
Ben Prather  https://orcid.org/0000-0002-0393-7734
Jorge A. Preciado-López  https://orcid.org/0000-0002-4146-0113
Hung-Yi Pu  https://orcid.org/0000-0001-9270-8812
Ramprasad Rao  https://orcid.org/0000-0002-1407-7944
Alexander W. Raymond  https://orcid.org/0000-0002-5779-4767
Luciano Rezzolla  https://orcid.org/0000-0002-1330-7103
Angelo Ricarte  https://orcid.org/0000-0001-5287-0452
Bart Ripperda  https://orcid.org/0000-0002-7301-3908
Freek Roelofs  https://orcid.org/0000-0001-5461-3687
Eduardo Ros  https://orcid.org/0000-0001-9503-4892
Mel Rose  https://orcid.org/0000-0002-2016-8746
Alan L. Roy  https://orcid.org/0000-0002-1931-0135
Chet Ruszczyk  https://orcid.org/0000-0001-7278-9707
Kazi L. J. Rygl  https://orcid.org/0000-0003-4146-9043
David Sánchez-Arguelles  https://orcid.org/0000-0002-7344-9920
Tuomas Savolainen  https://orcid.org/0000-0001-6214-1085
Lijing Shao  https://orcid.org/0000-0002-1334-8853
Zhiqiang Shen (沈志强)  https://orcid.org/0000-0003-3540-8746
Des Small  https://orcid.org/0000-0003-3723-5404
Bong Won Sohn  https://orcid.org/0000-0002-4148-8378
Jason SooHoo  https://orcid.org/0000-0003-1938-0720
He Sun (孙赫)  https://orcid.org/0000-0003-1526-6787
Fumie Tazaki  https://orcid.org/0000-0003-0236-0600
Alexandra J. Tetarenko  https://orcid.org/0000-0003-3906-4354
Paul Tiede  https://orcid.org/0000-0003-3826-5648
Remo P. J. Tilanus  https://orcid.org/0000-0002-6514-553X
Michael Titus  https://orcid.org/0000-0002-3423-4505
Kenji Toma  https://orcid.org/0000-0002-7114-6010

Pablo Torne https://orcid.org/0000-0001-8700-6058
Efthalia Traianou https://orcid.org/0000-0002-1209-6500
Sascha Trippe https://orcid.org/0000-0003-0465-1559
Ilse van Bemmel https://orcid.org/0000-0001-5473-2950
Huib Jan van Langevelde https://orcid.org/0000-0002-0230-5946
Daniel R. van Rossum https://orcid.org/0000-0001-7772-6131
Derek Ward-Thompson https://orcid.org/0000-0003-1140-2761
John Wardle https://orcid.org/0000-0002-8960-2942
Jonathan Weintroub https://orcid.org/0000-0002-4603-5204
Norbert Wex https://orcid.org/0000-0003-4058-2837
Robert Wharton https://orcid.org/0000-0002-7416-5209
Maciek Wielgus https://orcid.org/0000-0002-8635-4242
George N. Wong https://orcid.org/0000-0001-6952-2147
Qingwen Wu (吴庆文) https://orcid.org/0000-0003-4773-4987
Doosoo Yoon https://orcid.org/0000-0001-8694-8166
André Young https://orcid.org/0000-0003-0000-2682
Ken Young https://orcid.org/0000-0002-3666-4920
Ziri Younsi https://orcid.org/0000-0001-9283-1191
Feng Yuan (袁峰) https://orcid.org/0000-0003-3564-6437
J. Anton Zensus https://orcid.org/0000-0001-7470-3321
Guang-Yao Zhao https://orcid.org/0000-0002-4417-1659
Shan-Shan Zhao https://orcid.org/0000-0002-9774-3606
G. Principe https://orcid.org/0000-0003-0406-7387
M. Giroletti https://orcid.org/0000-0002-8657-8852
C. Arcaro https://orcid.org/0000-0002-1998-9707
M. Backes https://orcid.org/0000-0002-9326-6400
V. Barbosa Martins https://orcid.org/0000-0002-5085-8828
D. Berge https://orcid.org/0000-0002-2918-1824
M. Böttcher https://orcid.org/0000-0002-8434-5692
M. Breuhaus https://orcid.org/0000-0003-0268-5122
T. Bulik https://orcid.org/0000-0003-2045-4803
S. Caroff https://orcid.org/0000-0002-1103-130X
S. Casanova https://orcid.org/0000-0002-6144-9122
T. Chand https://orcid.org/0000-0002-1833-3749
S. Einecke https://orcid.org/0000-0001-9687-8237
S. Funk https://orcid.org/0000-0002-2012-0080
G. Giavitto https://orcid.org/0000-0002-7629-6499
J. A. Hinton https://orcid.org/0000-0002-1031-7760
C. Hoischen https://orcid.org/0000-0001-6661-2278
T. L. Holch https://orcid.org/0000-0001-5161-1168
D. Horns https://orcid.org/0000-0003-1945-0119
V. Joshi https://orcid.org/0000-0003-4467-3621
U. Katz https://orcid.org/0000-0002-7063-4418
S. Klepser https://orcid.org/0000-0002-8949-4275
Nu. Komin https://orcid.org/0000-0003-3280-0582
R. Konno https://orcid.org/0000-0003-1892-2356
K. Kosack https://orcid.org/0000-0001-8424-3621
J.-P. Lenain https://orcid.org/0000-0001-7284-9220
D. Malyshev https://orcid.org/0000-0001-9689-2194
E. Moulin https://orcid.org/0000-0003-4007-0145
T. Murach https://orcid.org/0000-0003-1128-5008
F. Niederwanger https://orcid.org/0000-0003-4263-2017
S. Ohm https://orcid.org/0000-0002-3474-2243
A. Reimer https://orcid.org/0000-0001-8604-7077
O. Reimer https://orcid.org/0000-0001-6953-1385
F. Rieger https://orcid.org/0000-0003-1334-2993
L. Rinchiuso https://orcid.org/0000-0003-0540-9967
C. Romoli https://orcid.org/0000-0003-2541-4499
G. Rowell https://orcid.org/0000-0002-9516-1581
E. Ruiz-Velasco https://orcid.org/0000-0001-6939-7825
S. Sailer https://orcid.org/0000-0001-8273-8495
A. Santangelo https://orcid.org/0000-0003-4187-9560
M. Sasaki https://orcid.org/0000-0001-5302-1866
H. M. Schutte https://orcid.org/0000-0002-1769-5617
C. Steppa https://orcid.org/0000-0003-0116-8836
R. Terrier https://orcid.org/0000-0002-8219-4667
C. van Eldik https://orcid.org/0000-0001-9669-645X
B. van Soelen https://orcid.org/0000-0003-1873-7855
C. Venter https://orcid.org/0000-0002-2666-4812
J. Vink https://orcid.org/0000-0002-4708-4219
M. Zacharias https://orcid.org/0000-0001-5801-3945
R. Zanin https://orcid.org/0000-0001-6320-1801
A. Zech https://orcid.org/0000-0002-4388-5625
J. Zorn https://orcid.org/0000-0001-9309-0700
S. Zouari https://orcid.org/0000-0002-5333-2004
N. Żywucka https://orcid.org/0000-0003-2644-6441
V. A. Acciari https://orcid.org/0000-0001-8307-2007
S. Ansoldi https://orcid.org/0000-0002-5613-7693
L. A. Antonelli https://orcid.org/0000-0002-5037-9034
A. Arbet Engels https://orcid.org/0000-0001-9076-9582
D. Baack https://orcid.org/0000-0002-2311-4460
A. Babić https://orcid.org/0000-0002-1444-5604
A. Baquero https://orcid.org/0000-0002-1757-5826
U. Barres de Almeida https://orcid.org/0000-0001-7909-588X
J. A. Barrio https://orcid.org/0000-0002-0965-0259
J. Becerra González https://orcid.org/0000-0002-6729-9022
W. Bednarek https://orcid.org/0000-0003-0605-108X
E. Bernardini https://orcid.org/0000-0003-3108-1141
A. Berti https://orcid.org/0000-0003-0396-4190
W. Bhattacharyya https://orcid.org/0000-0003-4751-0414
C. Bigongiari https://orcid.org/0000-0003-3293-8522
A. Biland https://orcid.org/0000-0002-1288-833X
O. Blanch https://orcid.org/0000-0002-8380-1633
G. Bonnoli https://orcid.org/0000-0003-2464-9077
Ž. Bošnjak https://orcid.org/0000-0001-6536-0320
G. Busetto https://orcid.org/0000-0002-2687-6380
R. Carosi https://orcid.org/0000-0002-4137-4370
M. Cerruti https://orcid.org/0000-0001-7891-699X
Y. Chai https://orcid.org/0000-0003-2816-2821
A. Chilingarian https://orcid.org/0000-0002-2018-9715
S. M. Colak https://orcid.org/0000-0001-7793-3106
E. Colombo https://orcid.org/0000-0002-3700-3745
J. L. Contreras https://orcid.org/0000-0001-7282-2394
J. Cortina https://orcid.org/0000-0003-4576-0452
S. Covino https://orcid.org/0000-0001-9078-5507
G. D'Amico https://orcid.org/0000-0001-6472-8381
V. D'Elia https://orcid.org/0000-0002-7320-5862
F. Dazzi https://orcid.org/0000-0001-5409-6544
A. De Angelis https://orcid.org/0000-0002-3288-2517
B. De Lotto https://orcid.org/0000-0003-3624-4480
M. Delfino https://orcid.org/0000-0002-9468-4751
J. Delgado https://orcid.org/0000-0002-7014-4101
C. Delgado Mendez https://orcid.org/0000-0002-0166-5464
D. Depaoli https://orcid.org/0000-0002-2672-4141
F. Di Pierro https://orcid.org/0000-0003-4861-432X
L. Di Venere https://orcid.org/0000-0003-0703-824X
E. Do Souto Espiñeira https://orcid.org/0000-0001-6974-2676

D. Dominis Prester https://orcid.org/0000-0002-9880-5039
A. Donini https://orcid.org/0000-0002-3066-724X
D. Dorner https://orcid.org/0000-0001-8823-479X
M. Doro https://orcid.org/0000-0001-9104-3214
D. Elsaesser https://orcid.org/0000-0001-6796-3205
V. Fallah Ramazani https://orcid.org/0000-0001-8991-7744
A. Fattorini https://orcid.org/0000-0002-1056-9167
G. Ferrara https://orcid.org/0000-0002-1137-6252
M. V. Fonseca https://orcid.org/0000-0003-2235-0725
L. Font https://orcid.org/0000-0003-2109-5961
C. Fruck https://orcid.org/0000-0001-5880-7518
R. J. García López https://orcid.org/0000-0002-8204-6832
M. Garczarczyk https://orcid.org/0000-0002-0445-4566
M. Gaug https://orcid.org/0000-0001-8442-7877
N. Giglietto https://orcid.org/0000-0002-9021-2888
F. Giordano https://orcid.org/0000-0002-8651-2394
P. Gliwny https://orcid.org/0000-0002-4183-391X
N. Godinović https://orcid.org/0000-0002-4674-9450
J. G. Green https://orcid.org/0000-0002-1130-6692
D. Green https://orcid.org/0000-0003-0768-2203
D. Hadasch https://orcid.org/0000-0001-8663-6461
A. Hahn https://orcid.org/0000-0003-0827-5642
L. Heckmann https://orcid.org/0000-0002-6653-8407
J. Herrera https://orcid.org/0000-0002-3771-4918
J. Hoang https://orcid.org/0000-0001-5591-5927
D. Hrupec https://orcid.org/0000-0002-7027-5021
M. Hütten https://orcid.org/0000-0002-2133-5251
S. Inoue https://orcid.org/0000-0003-1096-9424
L. Jouvin https://orcid.org/0000-0001-5119-8537
M. Karjalainen https://orcid.org/0000-0003-0751-3231
D. Kerszberg https://orcid.org/0000-0002-5289-1509
H. Kubo https://orcid.org/0000-0001-9159-9853
J. Kushida https://orcid.org/0000-0002-8002-8585
A. Lamastra https://orcid.org/0000-0003-2403-913X
D. Lelas https://orcid.org/0000-0002-8269-5760
F. Leone https://orcid.org/0000-0001-7626-3788
E. Lindfors https://orcid.org/0000-0002-9155-6199
S. Lombardi https://orcid.org/0000-0002-6336-865X
F. Longo https://orcid.org/0000-0003-2501-2270
R. López-Coto https://orcid.org/0000-0002-3882-9477
M. López-Moya https://orcid.org/0000-0002-8791-7908
A. López-Oramas https://orcid.org/0000-0003-4603-1884
S. Loporchio https://orcid.org/0000-0003-4457-5431
B. Machado de Oliveira Fraga https://orcid.org/0000-0002-6395-3410
C. Maggio https://orcid.org/0000-0003-0670-7771
P. Majumdar https://orcid.org/0000-0002-5481-5040
M. Makariev https://orcid.org/0000-0002-1622-3116
M. Mallamaci https://orcid.org/0000-0003-4068-0496
G. Maneva https://orcid.org/0000-0002-5959-4179
M. Manganaro https://orcid.org/0000-0003-1530-3031
K. Mannheim https://orcid.org/0000-0002-2950-6641
M. Mariotti https://orcid.org/0000-0003-3297-4128
M. Martínez https://orcid.org/0000-0002-9763-9155
D. Mazin https://orcid.org/0000-0002-2010-4005
S. Mender https://orcid.org/0000-0002-0755-0609
S. Mićanović https://orcid.org/0000-0002-0076-3134
D. Miceli https://orcid.org/0000-0002-2686-0098
J. M. Miranda https://orcid.org/0000-0002-1472-9690
R. Mirzoyan https://orcid.org/0000-0003-0163-7233
E. Molina https://orcid.org/0000-0003-1204-5516
A. Moralejo https://orcid.org/0000-0002-1344-9080
D. Morcuende https://orcid.org/0000-0001-9400-0922
V. Moreno https://orcid.org/0000-0002-8358-2098
E. Moretti https://orcid.org/0000-0001-5477-9097
V. Neustroev https://orcid.org/0000-0003-4772-595X
C. Nigro https://orcid.org/0000-0001-8375-1907
K. Nilsson https://orcid.org/0000-0002-1445-8683
K. Nishijima https://orcid.org/0000-0002-1830-4251
K. Noda https://orcid.org/0000-0003-1397-6478
S. Nozaki https://orcid.org/0000-0002-6246-2767
T. Oka https://orcid.org/0000-0002-9924-9978
J. Otero-Santos https://orcid.org/0000-0002-4241-5875
S. Paiano https://orcid.org/0000-0002-2239-3373
M. Palatiello https://orcid.org/0000-0002-4124-5747
D. Paneque https://orcid.org/0000-0002-2830-0502
R. Paoletti https://orcid.org/0000-0003-0158-2826
J. M. Paredes https://orcid.org/0000-0002-1566-9044
L. Pavletić https://orcid.org/0000-0002-9926-0405
C. Perennes https://orcid.org/0000-0002-0766-4446
M. Persic https://orcid.org/0000-0003-1853-4900
P. G. Prada Moroni https://orcid.org/0000-0001-9712-9916
E. Prandini https://orcid.org/0000-0003-4502-9053
C. Priyadarshi https://orcid.org/0000-0002-9160-9617
I. Puljak https://orcid.org/0000-0001-7387-3812
W. Rhode https://orcid.org/0000-0003-2636-5000
M. Ribó https://orcid.org/0000-0002-9931-4557
J. Rico https://orcid.org/0000-0003-4137-1134
C. Righi https://orcid.org/0000-0002-1218-9555
A. Rugliancich https://orcid.org/0000-0001-5471-4701
L. Saha https://orcid.org/0000-0002-3171-5039
N. Sahakyan https://orcid.org/0000-0003-2011-2731
K. Satalecka https://orcid.org/0000-0002-7669-266X
F. G. Saturni https://orcid.org/0000-0002-1946-7706
K. Schmidt https://orcid.org/0000-0002-9883-4454
J. Sitarek https://orcid.org/0000-0002-1659-5374
D. Sobczynska https://orcid.org/0000-0003-4973-7903
A. Spolon https://orcid.org/0000-0001-8770-9503
A. Stamerra https://orcid.org/0000-0002-9430-5264
D. Strom https://orcid.org/0000-0003-2108-3311
Y. Suda https://orcid.org/0000-0002-2692-5891
F. Tavecchio https://orcid.org/0000-0003-0256-0995
P. Temnikov https://orcid.org/0000-0002-9559-3384
T. Terzić https://orcid.org/0000-0002-4209-3407
J. van Scherpenberg https://orcid.org/0000-0002-6173-867X
G. Vanzo https://orcid.org/0000-0003-1539-3268
M. Vazquez Acosta https://orcid.org/0000-0002-2409-9792
S. Ventura https://orcid.org/0000-0001-7065-5342
V. Verguilov https://orcid.org/0000-0001-7911-1093
C. F. Vigorito https://orcid.org/0000-0002-0069-9195
V. Vitale https://orcid.org/0000-0001-8040-7852
I. Vovk https://orcid.org/0000-0003-3444-3830
M. Will https://orcid.org/0000-0002-7504-2083
D. Zarić https://orcid.org/0000-0001-5763-9487
W. Benbow https://orcid.org/0000-0003-2098-170X
J. L. Christiansen https://orcid.org/0000-0002-8035-4778
M. Errando https://orcid.org/0000-0002-1853-863X
Q. Feng https://orcid.org/0000-0001-6674-4238
L. Fortson https://orcid.org/0000-0002-1067-8558
A. Furniss https://orcid.org/0000-0003-1614-1273
O. Hervet https://orcid.org/0000-0003-3878-1677
W. Jin https://orcid.org/0000-0002-1089-1754
P. Kaaret https://orcid.org/0000-0002-3638-0637

D. Kieda https://orcid.org/0000-0003-4785-0101
G. Maier https://orcid.org/0000-0001-9868-4700
R. Mukherjee https://orcid.org/0000-0002-3223-0754
M. Pohl https://orcid.org/0000-0001-7861-1707
E. Pueschel https://orcid.org/0000-0002-0529-1973
M. Santander https://orcid.org/0000-0001-7297-8217
Tomoya Hirota https://orcid.org/0000-0003-1659-095X
Lang Cui https://orcid.org/0000-0003-0721-5509
Kotaro Niinuma https://orcid.org/0000-0002-8169-3579
Nobuyuki Sakai https://orcid.org/0000-0002-5814-0554
Satoko Sawada-Satoh https://orcid.org/0000-0001-7719-274X
Kiyoaki Wajima https://orcid.org/0000-0003-3823-7954
Yoshinori Yonekura https://orcid.org/0000-0001-5615-5464

J. C. Algaba[1], J. Anczarski[2], K. Asada[3], M. Baloković[4,5], S. Chandra[6], Y.-Z. Cui[7,8], A. D. Falcone[9], M. Giroletti[10], C. Goddi[11,12], K. Hada[7,8], D. Haggard[13,14], S. Jorstad[15,16], A. Kaur[9], T. Kawashima[17], G. Keating[18], J.-Y. Kim[19,20], M. Kino[21,22], S. Komossa[20], E. V. Kravchenko[10,23,24], T. P. Krichbaum[20], S.-S. Lee[19], R.-S. Lu (路如森)[20,25,26], M. Lucchini[27], S. Markoff[27,28], J. Neilsen[2], M. A. Nowak[29], J. Park[30,31,244], G. Principe[10,32,33], V. Ramakrishnan[34], M. T. Reynolds[35], M. Sasada[21,36], S. S. Savchenko[37,38], K. E. Williamson[15]

The Event Horizon Telescope Collaboration[246],

Kazunori Akiyama[21,39,40], Antxon Alberdi[41], Walter Alef[20], Richard Anantua[18,40,42], Rebecca Azulay[20,43,44], Anne-Kathrin Baczko[20], David Ball[45], John Barrett[39], Dan Bintley[46], Bradford A. Benson[47,48], Lindy Blackburn[18,40], Raymond Blundell[18], Wilfred Boland[49], Katherine L. Bouman[18,40,50], Geoffrey C. Bower[51], Hope Boyce[52,53], Michael Bremer[54], Christiaan D. Brinkerink[11], Roger Brissenden[18,40], Silke Britzen[20], Avery E. Broderick[55,56,57], Dominique Broguiere[54], Thomas Bronzwaer[256], Do-Young Byun[19,58], John E. Carlstrom[48,59,60,61], Andrew Chael[62,239], Chi-kwan Chan[45,63], Shami Chatterjee[64], Koushik Chatterjee[65], Ming-Tang Chen[51], Yongjun Chen (陈永军)[25,26], Paul M. Chesler[40], Ilje Cho[19,58], Pierre Christian[66], John E. Conway[67], James M. Cordes[64], Thomas M. Crawford[48,59], Geoffrey B. Crew[39], Alejandro Cruz-Osorio[68], Jordy Davelaar[11,42,69], Mariafelicia De Laurentis[68,70,71], Roger Deane[72,73,74], Jessica Dempsey[46], Gregory Desvignes[75], Jason Dexter[76], Sheperd S. Doeleman[18,40], Ralph P. Eatough[20,77], Heino Falcke[11], Joseph Farah[18,40,78], Vincent L. Fish[39], Ed Fomalont[79], H. Alyson Ford[80], Raquel Fraga-Encinas[11], Per Friberg[46], Christian M. Fromm[18,40,68], Antonio Fuentes[41], Peter Galison[40,81,82], Charles F. Gammie[83,84], Roberto García[54], Olivier Gentaz[54], Boris Georgiev[56,57], Roman Gold[55,85], José L. Gómez[41], Arturo I. Gómez-Ruiz[86,87], Minfeng Gu (顾敏峰)[25,88], Mark Gurwell[18], Michael H. Hecht[39], Ronald Hesper[89], Luis C. Ho (何子山)[90,91], Paul Ho[3], Mareki Honma[7,8,92], Chih-Wei L. Huang[3], Lei Huang (黄磊)[25,88], David H. Hughes[86], Shiro Ikeda[21,93,94,95], Makoto Inoue[3], Sara Issaoun[11], David J. James[18,40], Buell T. Jannuzi[45], Michael Janssen[20], Britton Jeter[56,57], Wu Jiang (江悟)[25], Alejandra Jiménez-Rosales[11], Michael D. Johnson[18,40], Taehyun Jung[19,58], Mansour Karami[55,56], Ramesh Karuppusamy[20], Mark Kettenis[96], Dong-Jin Kim[20], Jongsoo Kim[19], Junhan Kim[45,50], Jun Yi Koay[3], Yutaro Kofuji[8,92], Patrick M. Koch[3], Shoko Koyama[3], Michael Kramer[20], Carsten Kramer[54], Cheng-Yu Kuo[3,97], Tod R. Lauer[98], Aviad Levis[50], Yan-Rong Li (李彦荣)[99], Zhiyuan Li (李志远)[100,101], Michael Lindqvist[67], Rocco Lico[20,41], Greg Lindahl[18], Jun Liu (刘俊)[20], Kuo Liu[20], Elisabetta Liuzzo[102], Wen-Ping Lo[3,103], Andrei P. Lobanov[20], Laurent Loinard[104,105], Colin Lonsdale[39], Nicholas R. MacDonald[20], Jirong Mao (毛基荣)[106,107,108], Nicola Marchili[20,102], Daniel P. Marrone[45], Alan P. Marscher[15], Iván Martí-Vidal[43,44], Satoki Matsushita[3], Lynn D. Matthews[39], Lia Medeiros[45,109], Karl M. Menten[20], Izumi Mizuno[46], Yosuke Mizuno[68,110], James M. Moran[18,40], Kotaro Moriyama[8,39], Monika Moscibrodzka[11], Cornelia Müller[11,20], Gibwa Musoke[11,65], Alejandro Mus Mejías[43,44], Hiroshi Nagai[7,21], Neil M. Nagar[34], Masanori Nakamura[3,111], Ramesh Narayan[18,40], Gopal Narayanan[112], Iniyan Natarajan[72,74,113], Antonios Nathanail[68,114], Roberto Neri[54], Chunchong Ni[56,57], Aristeidis Noutsos[20], Hiroki Okino[8,92], Héctor Olivares[11], Gisela N. Ortiz-León[20], Tomoaki Oyama[8], Feryal Özel[45], Daniel C. M. Palumbo[18,40], Nimesh Patel[18], Ue-Li Pen[55,115,116,117], Dominic W. Pesce[18,40], Vincent Piétu[54], Richard Plambeck[118], Aleksandar PopStefanija[112], Oliver Porth[65,68], Felix M. Pötzl[20], Ben Prather[83], Jorge A. Preciado-López[55], Dimitrios Psaltis[45], Hung-Yi Pu[3,55,119], Ramprasad Rao[18], Mark G. Rawlings[46], Alexander W. Raymond[18,40], Luciano Rezzolla[120,121,122], Angelo Ricarte[18,40], Bart Ripperda[42,123], Freek Roelofs[11], Alan Rogers[39], Eduardo Ros[20], Mel Rose[45], Arash Roshanineshat[45], Helge Rottmann[20], Alan L. Roy[20], Chet Ruszczyk[39], Kazi L. J. Rygl[102], Salvador Sánchez[124], David Sánchez-Arguelles[86,87], Tuomas Savolainen[20,125,126], F. Peter Schloerb[112], Karl-Friedrich Schuster[54], Lijing Shao[20,91], Zhiqiang Shen (沈志强)[25,26], Des Small[96], Bong Won Sohn[19,58,127], Jason SooHoo[39], He Sun (孙赫)[50], Fumie Tazaki[8], Alexandra J. Tetarenko[128], Paul Tiede[56,57], Remo P. J. Tilanus[11,12,45,129], Michael Titus[39], Kenji Toma[130,131], Pablo Torne[20,124], Tyler Trent[45], Efthalia Traianou[20], Sascha Trippe[31], Ilse van Bemmel[96], Huib Jan van Langevelde[96,132], Daniel R. van Rossum[11], Jan Wagner[20], Derek Ward-Thompson[133], John Wardle[134], Jonathan Weintroub[18,40], Norbert Wex[20], Robert Wharton[20], Maciek Wielgus[18,40], George N. Wong[83], Qingwen Wu (吴庆文)[135], Doosoo Yoon[65], André Young[11], Ken Young[18], Ziri Younsi[68,136,245], Feng Yuan (袁峰)[25,88,137], Ye-Fei Yuan (袁业飞)[138], J. Anton Zensus[20], Guang-Yao Zhao[41], Shan-Shan Zhao[25]

The Fermi Large Area Telescope Collaboration,

G. Principe[10,32,33], M. Giroletti[10], F. D'Ammando[10], M. Orienti[10]

H.E.S.S. Collaboration,

H. Abdalla[6], R. Adam[139], F. Aharonian[140,141,142], F. Ait Benkhali[141], E. O. Angüner[143], C. Arcaro[6,*], C. Armand[144], T. Armstrong[145], H. Ashkar[146], M. Backes[6,147], V. Baghmanyan[148], V. Barbosa Martins[149], A. Barnacka[150], M. Barnard[6], Y. Becherini[151], D. Berge[149], K. Bernlöhr[141], B. Bi[152], M. Böttcher[6], C. Boisson[153], J. Bolmont[154], M. de Bony de Lavergne[144], M. Breuhaus[141], F. Brun[146], P. Brun[146], M. Bryan[155], M. Büchele[156], T. Bulik[157], T. Bylund[151],

S. Caroff[154], A. Carosi[144], S. Casanova[148,141], T. Chand[6], A. Chen[158], G. Cotter[145], M. Curyło[157],
J. Damascene Mbarubucyeye[149], I. D. Davids[147], J. Davies[145], C. Deil[141], J. Devin[159], P. deWilt[160], L. Dirson[161],
A. Djannati-Ataï[162], A. Dmytriiev[153], A. Donath[141], V. Doroshenko[152], C. Duffy[163], J. Dyks[164], K. Egberts[165], F. Eichhorn[156],
S. Einecke[160], G. Emery[154], J.-P. Ernenwein[143], K. Feijen[160], S. Fegan[139], A. Fiasson[144], G. Fichet de Clairfontaine[153],
G. Fontaine[139], S. Funk[156], M. Füßling[149], S. Gabici[162], Y. A. Gallant[166], G. Giavitto[149], L. Giunti[162], D. Glawion[156,*],
J. F. Glicenstein[146], D. Gottschall[152], M.-H. Grondin[159], J. Hahn[141], M. Haupt[149], G. Hermann[141], J. A. Hinton[141],
W. Hofmann[141], C. Hoischen[165], T. L. Holch[167], M. Holler[168], M. Hörbe[145], D. Horns[161], D. Huber[168], M. Jamrozy[150],
D. Jankowsky[156], F. Jankowsky[169], A. Jardin-Blicq[141], V. Joshi[156], I. Jung-Richardt[156], E. Kasai[147], M. A. Kastendieck[161],
K. Katarzyński[170], U. Katz[156], D. Khangulyan[171], B. Khélifi[162], S. Klepser[149], W. Kluźniak[164], Nu. Komin[158],
R. Konno[149], K. Kosack[146], D. Kostunin[149], M. Kreter[6], G. Lamanna[144], A. Lemière[162], M. Lemoine-Goumard[159],
J.-P. Lenain[154], C. Levy[154], T. Lohse[167], I. Lypova[149], J. Mackey[140], J. Majumdar[149], D. Malyshev[152], D. Malyshev[156],
V. Marandon[141], P. Marchegiani[158], A. Marcowith[166], A. Mares[159], G. Martí-Devesa[168], R. Marx[169,141], G. Maurin[144],
P. J. Meintjes[172], M. Meyer[156], R. Moderski[164], M. Mohamed[169], L. Mohrmann[156], A. Montanari[146], C. Moore[163], P. Morris[145],
E. Moulin[146], J. Muller[139], T. Murach[149], K. Nakashima[156], A. Nayerhoda[148], M. de Naurois[139], H. Ndiyavala[6],
F. Niederwanger[168], J. Niemiec[148], L. Oakes[167], P. O'Brien[163], H. Odaka[173], S. Ohm[149], L. Olivera-Nieto[141],
E. de Ona Wilhelmi[149], M. Ostrowski[150], M. Panter[141], S. Panny[168], R. D. Parsons[167], G. Peron[141], B. Peyaud[146], Q. Piel[144],
S. Pita[162], V. Poireau[144], A. Priyana Noel[150], D. A. Prokhorov[155], H. Prokoph[149], G. Pühlhofer[152,*], M. Punch[151,162],
A. Quirrenbach[169], R. Rauth[168], P. Reichherzer[146], A. Reimer[168], O. Reimer[168], Q. Remy[141], M. Renaud[166], F. Rieger[141],
L. Rinchiuso[146], C. Romoli[141], G. Rowell[160], B. Rudak[164], E. Ruiz-Velasco[141], V. Sahakian[174], S. Sailer[141],
D. A. Sanchez[144,*], A. Santangelo[152], M. Sasaki[156], M. Scalici[152], H. M. Schutte[6], U. Schwanke[167], S. Schwemmer[169],
M. Seglar-Arroyo[146], M. Senniappan[151], A. S. Seyffert[6], N. Shafi[158], K. Shiningayamwe[147], R. Simoni[155], A. Sinha[162], H. Sol[153],
A. Specovius[156], S. Spencer[145], M. Spir-Jacob[162], Ł. Stawarz[150], L. Sun[155], R. Steenkamp[147], C. Stegmann[149,165], S. Steinmassl[141],
C. Steppa[165], T. Takahashi[175], T. Tavernier[146], A. M. Taylor[149], R. Terrier[162], D. Tiziani[156], M. Tluczykont[161],
L. Tomankova[156], C. Trichard[139], M. Tsirou[166], R. Tuffs[141], Y. Uchiyama[171], D. J. van der Walt[6], C. van Eldik[156],
C. van Rensburg[6], B. van Soelen[172], G. Vasileiadis[166], J. Veh[156], C. Venter[6], P. Vincent[154], J. Vink[155], H. J. Völk[141],
T. Vuillaume[144], Z. Wadiasingh[6], S. J. Wagner[169], J. Watson[145], F. Werner[141], R. White[141], A. Wierzcholska[148,169],
Yu Wun Wong[156], A. Yusafzai[156], M. Zacharias[6], R. Zanin[141], D. Zargaryan[140,142], A. A. Zdziarski[164], A. Zech[153],
S. J. Zhu[149], J. Zorn[141], S. Zouari[162], N. Żywucka[6]

MAGIC Collaboration,

V. A. Acciari[176], S. Ansoldi[177], L. A. Antonelli[178], A. Arbet Engels[179], M. Artero[180], K. Asano[181], D. Baack[182],
A. Babić[183], A. Baquero[184], U. Barres de Almeida[185], J. A. Barrio[184], J. Becerra González[176], W. Bednarek[186],
L. Bellizzi[187], E. Bernardini[188], M. Bernardos[189], A. Berti[190], J. Besenrieder[191], W. Bhattacharyya[188], C. Bigongiari[178],
A. Biland[179], O. Blanch[180], G. Bonnoli[187], Ž. Bošnjak[183], G. Busetto[189], R. Carosi[192], G. Ceribella[191],
M. Cerruti[193], Y. Chai[191], A. Chilingarian[194], S. Cikota[183], S. M. Colak[180], E. Colombo[176], J. L. Contreras[184],
J. Cortina[195], S. Covino[178], G. D'Amico[191], V. D'Elia[178], P. Da Vela[192,240], F. Dazzi[178], A. De Angelis[189],
B. De Lotto[177], M. Delfino[180,241], J. Delgado[180,241], C. Delgado Mendez[195], D. Depaoli[190], F. Di Pierro[190],
L. Di Venere[196], E. Do Souto Espiñeira[180], D. Dominis Prester[197], A. Donini[177], D. Dorner[198], M. Doro[189],
D. Elsaesser[182], V. Fallah Ramazani[199], A. Fattorini[182], G. Ferrara[178], M. V. Fonseca[184], L. Font[200], C. Fruck[191],
S. Fukami[181], R. J. García López[176], M. Garczarczyk[188], S. Gasparyan[201], M. Gaug[200], N. Giglietto[196], F. Giordano[196],
P. Gliwny[186], N. Godinović[202], J. G. Green[178], D. Green[191], D. Hadasch[181], A. Hahn[191,†], L. Heckmann[191],
J. Herrera[176], J. Hoang[184], D. Hrupec[203], M. Hütten[191], T. Inada[181], S. Inoue[204], K. Ishio[191], Y. Iwamura[181],
I. Jiménez[195], J. Jormanainen[199], L. Jouvin[180], Y. Kajiwara[205], M. Karjalainen[176], D. Kerszberg[180], Y. Kobayashi[181],
H. Kubo[205], J. Kushida[206], A. Lamastra[178], D. Lelas[202], F. Leone[178], E. Lindfors[199], S. Lombardi[178],
F. Longo[177,242], R. López-Coto[189], M. López-Moya[184], A. López-Oramas[176], S. Loporchio[196],
B. Machado de Oliveira Fraga[185], C. Maggio[200], P. Majumdar[207], M. Makariev[208], M. Mallamaci[189], G. Maneva[208],
M. Manganaro[197], K. Mannheim[198], L. Maraschi[178], M. Mariotti[189], M. Martínez[180], D. Mazin[181,191,†],
S. Menchiari[187], S. Mender[182], S. Mićanović[197], D. Miceli[177], T. Miener[184], M. Minev[208], J. M. Miranda[187],
R. Mirzoyan[191], E. Molina[193], A. Moralejo[180], D. Morcuende[184], V. Moreno[200], E. Moretti[180], V. Neustroev[209],
C. Nigro[180], K. Nilsson[199], K. Nishijima[206], K. Noda[181], S. Nozaki[205], Y. Ohtani[181], T. Oka[205],
J. Otero-Santos[176], S. Paiano[178], M. Palatiello[177], D. Paneque[191], R. Paoletti[187], J. M. Paredes[193], L. Pavletić[197],
P. Peñil[184], C. Perennes[189], M. Persic[177,243], P. G. Prada Moroni[192], E. Prandini[189], C. Priyadarshi[180], I. Puljak[202],
W. Rhode[182], M. Ribó[193], J. Rico[180], C. Righi[178], A. Rugliancich[192], L. Saha[184], N. Sahakyan[201], T. Saito[181],
S. Sakurai[181], K. Satalecka[188], F. G. Saturni[178], B. Schleicher[198], K. Schmidt[182], T. Schweizer[191], J. Sitarek[186],
I. Šnidarić[210], D. Sobczynska[186], A. Spolon[189], A. Stamerra[178], D. Strom[191], M. Strzys[181], Y. Suda[191], T. Surić[210],
M. Takahashi[181], F. Tavecchio[178], P. Temnikov[208], T. Terzić[197], M. Teshima[181,191], L. Tosti[211], S. Truzzi[187], A. Tutone[178],

S. Ubach[200], J. van Scherpenberg[191], G. Vanzo[176], M. Vazquez Acosta[176], S. Ventura[187], V. Verguilov[208],
C. F. Vigorito[190], V. Vitale[212], I. Vovk[181], M. Will[191], C. Wunderlich[187], D. Zarić[202]

VERITAS Collaboration,
C.B. Adams[213], W. Benbow[18], A. Brill[214], M. Capasso[213], J. L. Christiansen[215], A. J. Chromey[216], M. K. Daniel[18],
M. Errando[217], K. A Farrell[218], Q. Feng[213], J. P. Finley[219], L. Fortson[220], A. Furniss[221], A. Gent[222], C. Giuri[223],
T. Hassan[223], O. Hervet[224], J. Holder[225], G. Hughes[18], T. B. Humensky[214], W. Jin[226,‡], P. Kaaret[227], M. Kertzman[228],
D. Kieda[229], S. Kumar[230], M. J. Lang[231], M. Lundy[230], G. Maier[223], P. Moriarty[231], R. Mukherjee[213], D. Nieto[232],
M. Nievas-Rosillo[223], S. O'Brien[230], R. A. Ong[233], A. N. Otte[222], S. Patel[227], K. Pfrang[223], M. Pohl[234], R. R. Prado[223],
E. Pueschel[223], J. Quinn[218], K. Ragan[230], P. T. Reynolds[235], D. Ribeiro[214], G. T. Richards[225], E. Roache[18], C. Rulten[220],
J. L. Ryan[233], M. Santander[226,‡], G. H. Sembroski[219], R. Shang[233], A. Weinstein[216], D. A. Williams[224], T. J Williamson[225],
and
EAVN Collaboration,
Tomoya Hirota[7,21], Lang Cui[236], Kotaro Niinuma[237], Hyunwook Ro, Nobuyuki Sakai, Satoko Sawada-Satoh[237],
Kiyoaki Wajima, Na Wang[236], Xiang Liu[236], and Yoshinori Yonekura[238]

[1] Department of Physics, Faculty of Science, University of Malaya, 50603 Kuala Lumpur, Malaysia
[2] Department of Physics, Villanova University, 800 E. Lancaster Avenue, Villanova, PA 19085, USA
[3] Institute of Astronomy and Astrophysics, Academia Sinica, 11F of Astronomy-Mathematics Building, AS/NTU No. 1, Sec. 4, Roosevelt Rd, Taipei 10617, Taiwan, R.O.C.
[4] Yale Center for Astronomy & Astrophysics, 52 Hillhouse Avenue, New Haven, CT 06511, USA
[5] Department of Physics, Yale University, P.O. Box 2018120, New Haven, CT 06520, USA
[6] Centre for Space Research, North-West University, Potchefstroom 2520, South Africa
[7] Department of Astronomical Science, The Graduate University for Advanced Studies (SOKENDAI), 2-21-1 Osawa, Mitaka, Tokyo 181-8588, Japan
[8] Mizusawa VLBI Observatory, 2-12 Hoshigaoka, Mizusawa, Oshu, Iwate 023-0861, Japan
[9] 525 Davey Laboratory, Department of Astronomy and Astrophysics, Pennsylvania State University, University Park, PA 16802, USA
[10] INAF Istituto di Radioastronomia, Via P. Gobetti, 101, I-40129 Bologna, Italy
[11] Department of Astrophysics, Institute for Mathematics, Astrophysics and Particle Physics (IMAPP), Radboud University, P.O. Box 9010, 6500 GL Nijmegen, The Netherlands
[12] Leiden Observatory—Allegro, Leiden University, P.O. Box 9513, 2300 RA Leiden, The Netherlands
[13] Department of Physics, McGill University, 3600 University Street, Montréal, QC H3A 2T8, Canada
[14] McGill Space Institute, McGill University, 3550 University Street, Montréal, QC H3A 2A7, Canada
[15] Institute for Astrophysical Research, Boston University, 725 Commonwealth Avenue, Boston, MA 02215, USA
[16] Astronomical Institute, St. Petersburg University, Universitetskij Pr. 28, Petrodvorets, 198504 St. Petersburg, Russia
[17] Institute for Cosmic Ray Research, The University of Tokyo, 5-1-5 Kashiwanoha, Kashiwa, Chiba 277-8582, Japan
[18] Center for Astrophysics, Harvard & Smithsonian, 60 Garden Street, Cambridge, MA 02138, USA
[19] Korea Astronomy and Space Science Institute, Daedeok-daero 776, Yuseong-gu, Daejeon 34055, Republic of Korea
[20] Max-Planck-Institut für Radioastronomie, Auf dem Hügel 69, D-53121 Bonn, Germany
[21] National Astronomical Observatory of Japan, 2-21-1 Osawa, Mitaka, Tokyo 181-8588, Japan
[22] Kogakuin University of Technology & Engineering, Academic Support Center, 2665-1 Nakano, Hachioji, Tokyo 192-0015, Japan
[23] Moscow Institute of Physics and Technology, Institutsky per. 9, Moscow region, Dolgoprudny, 141700, Russia
[24] Astro Space Center, Lebedev Physical Institute, Profsouznaya 84/32, Moscow 117997, Russia
[25] Shanghai Astronomical Observatory, Chinese Academy of Sciences, 80 Nandan Road, Shanghai 200030, People's Republic of China
[26] Key Laboratory of Radio Astronomy, Chinese Academy of Sciences, Nanjing 210008, People's Republic of China
[27] API—Anton Pannekoek Institute for Astronomy, University of Amsterdam, Science Park 904, 1098 XH Amsterdam, The Netherlands
[28] GRAPPA—Gravitation and AstroParticle Physics Amsterdam, University of Amsterdam, Science Park 904, 1098 XH Amsterdam, The Netherlands
[29] Physics Department, Washington University CB 1105, St Louis, MO 63130, USA
[30] Institute of Astronomy and Astrophysics, Academia Sinica, P.O. Box 23-141, Taipei 10617, Taiwan
[31] Department of Physics and Astronomy, Seoul National University, Gwanak-gu, Seoul 08826, Republic of Korea
[32] Dipartimento di Fisica, Universitá di Trieste, I-34127 Trieste, Italy
[33] Istituto Nazionale di Fisica Nucleare, Sezione di Trieste, I-34127 Trieste, Italy
[34] Astronomy Department, Universidad de Concepción, Casilla 160-C, Concepción, Chile
[35] University of Michigan, 1085 S. University Ave., Ann Arbor, MI 48109, USA
[36] Hiroshima Astrophysical Science Center, Hiroshima University, 1-3-1 Kagamiyama, Higashi-Hiroshima, Hiroshima 739-8526, Japan
[37] Saint Petersburg State University, 7/9, Universitetskaya nab., St. Petersburg, 199034, Russia
[38] Special Astrophysical Observatory, Russian Academy of Science, Nizhnii Arkhyz 369167, Russia
[39] Massachusetts Institute of Technology Haystack Observatory, 99 Millstone Road, Westford, MA 01886, USA
[40] Black Hole Initiative at Harvard University, 20 Garden Street, Cambridge, MA 02138, USA
[41] Instituto de Astrofísica de Andalucía-CSIC, Glorieta de la Astronomía s/n, E-18008 Granada, Spain
[42] Center for Computational Astrophysics, Flatiron Institute, 162 Fifth Avenue, New York, NY 10010, USA
[43] Departament d'Astronomia i Astrofísica, Universitat de València, C. Dr. Moliner 50, E-46100 Burjassot, València, Spain
[44] Observatori Astronòmic, Universitat de València, C. Catedrático José Beltrán 2, E-46980 Paterna, València, Spain
[45] Steward Observatory and Department of Astronomy, University of Arizona, 933 N. Cherry Ave., Tucson, AZ 85721, USA
[46] East Asian Observatory, 660 N. A'ohoku Place, Hilo, HI 96720, USA
[47] Fermi National Accelerator Laboratory, MS209, P.O. Box 500, Batavia, IL 60510, USA
[48] Department of Astronomy and Astrophysics, University of Chicago, 5640 South Ellis Avenue, Chicago, IL 60637, USA
[49] Nederlandse Onderzoekschool voor Astronomie (NOVA), P.O. Box 9513, 2300 RA Leiden, The Netherlands
[50] California Institute of Technology, 1200 East California Boulevard, Pasadena, CA 91125, USA
[51] Institute of Astronomy and Astrophysics, Academia Sinica, 645 N. A'ohoku Place, Hilo, HI 96720, USA
[52] Department of Physics, McGill University, 3600 rue University, Montréal, QC H3A 2T8, Canada
[53] McGill Space Institute, McGill University, 3550 rue University, Montréal, QC H3A 2A7, Canada

[54] Institut de Radioastronomie Millimétrique, 300 rue de la Piscine, F-38406 Saint Martin d'Hères, France
[55] Perimeter Institute for Theoretical Physics, 31 Caroline Street North, Waterloo, ON, N2L 2Y5, Canada
[56] Department of Physics and Astronomy, University of Waterloo, 200 University Avenue West, Waterloo, ON, N2L 3G1, Canada
[57] Waterloo Centre for Astrophysics, University of Waterloo, Waterloo, ON N2L 3G1 Canada
[58] University of Science and Technology, Gajeong-ro 217, Yuseong-gu, Daejeon 34113, Republic of Korea
[59] Kavli Institute for Cosmological Physics, University of Chicago, 5640 South Ellis Avenue, Chicago, IL 60637, USA
[60] Department of Physics, University of Chicago, 5720 South Ellis Avenue, Chicago, IL 60637, USA
[61] Enrico Fermi Institute, University of Chicago, 5640 South Ellis Avenue, Chicago, IL 60637, USA
[62] Princeton Center for Theoretical Science, Jadwin Hall, Princeton University, Princeton, NJ 08544, USA
[63] Data Science Institute, University of Arizona, 1230 N. Cherry Ave., Tucson, AZ 85721, USA
[64] Cornell Center for Astrophysics and Planetary Science, Cornell University, Ithaca, NY 14853, USA
[65] Anton Pannekoek Institute for Astronomy, University of Amsterdam, Science Park 904, 1098 XH, Amsterdam, The Netherlands
[66] Physics Department, Fairfield University, 1073 North Benson Road, Fairfield, CT 06824, USA
[67] Department of Space, Earth and Environment, Chalmers University of Technology, Onsala Space Observatory, SE-43992 Onsala, Sweden
[68] Institut für Theoretische Physik, Goethe-Universität Frankfurt, Max-von-Laue-Straße 1, D-60438 Frankfurt am Main, Germany
[69] Department of Astronomy and Columbia Astrophysics Laboratory, Columbia University, 550 W 120th Street, New York, NY 10027, USA
[70] Dipartimento di Fisica "E. Pancini", Universitá di Napoli "Federico II," Compl. Univ. di Monte S. Angelo, Edificio G, Via Cinthia, I-80126, Napoli, Italy
[71] INFN Sez. di Napoli, Compl. Univ. di Monte S. Angelo, Edificio G, Via Cinthia, I-80126, Napoli, Italy
[72] Wits Centre for Astrophysics, University of the Witwatersrand, 1 Jan Smuts Avenue, Braamfontein, Johannesburg 2050, South Africa
[73] Department of Physics, University of Pretoria, Hatfield, Pretoria 0028, South Africa
[74] Centre for Radio Astronomy Techniques and Technologies, Department of Physics and Electronics, Rhodes University, Makhanda 6140, South Africa
[75] LESIA, Observatoire de Paris, Université PSL, CNRS, Sorbonne Université, Université de Paris, 5 place Jules Janssen, F-92195 Meudon, France
[76] JILA and Department of Astrophysical and Planetary Sciences, University of Colorado, Boulder, CO 80309, USA
[77] National Astronomical Observatories, Chinese Academy of Sciences, 20A Datun Road, Chaoyang District, Beijing 100101, People's Republic of China
[78] University of Massachusetts Boston, 100 William T. Morrissey Boulevard, Boston, MA 02125, USA
[79] National Radio Astronomy Observatory, 520 Edgemont Rd, Charlottesville, VA 22903, USA
[80] Steward Observatory and Department of Astronomy, University of Arizona, 933 North Cherry Avenue, Tucson, AZ 85721, USA
[81] Department of History of Science, Harvard University, Cambridge, MA 02138, USA
[82] Department of Physics, Harvard University, Cambridge, MA 02138, USA
[83] Department of Physics, University of Illinois, 1110 West Green Street, Urbana, IL 61801, USA
[84] Department of Astronomy, University of Illinois at Urbana-Champaign, 1002 West Green Street, Urbana, IL 61801, USA
[85] CP3-Origins, University of Southern Denmark, Campusvej 55, DK-5230 Odense M, Denmark
[86] Instituto Nacional de Astrofísica, Óptica y Electrónica. Apartado Postal 51 y 216, 72000. Puebla Pue., México
[87] Consejo Nacional de Ciencia y Tecnología, Av. Insurgentes Sur 1582, 03940, Ciudad de México, México
[88] Key Laboratory for Research in Galaxies and Cosmology, Chinese Academy of Sciences, Shanghai 200030, People's Republic of China
[89] NOVA Sub-mm Instrumentation Group, Kapteyn Astronomical Institute, University of Groningen, Landleven 12, 9747 AD Groningen, The Netherlands
[90] Department of Astronomy, School of Physics, Peking University, Beijing 100871, People's Republic of China
[91] Kavli Institute for Astronomy and Astrophysics, Peking University, Beijing 100871, People's Republic of China
[92] Department of Astronomy, Graduate School of Science, The University of Tokyo, 7-3-1 Hongo, Bunkyo-ku, Tokyo 113-0033, Japan
[93] The Institute of Statistical Mathematics, 10-3 Midori-cho, Tachikawa, Tokyo, 190-8562, Japan
[94] Department of Statistical Science, The Graduate University for Advanced Studies (SOKENDAI), 10-3 Midori-cho, Tachikawa, Tokyo 190-8562, Japan
[95] Kavli Institute for the Physics and Mathematics of the Universe, The University of Tokyo, 5-1-5 Kashiwanoha, Kashiwa, 277-8583, Japan
[96] Joint Institute for VLBI ERIC (JIVE), Oude Hoogeveensedijk 4, 7991 PD Dwingeloo, The Netherlands
[97] Physics Department, National Sun Yat-Sen University, No. 70, Lien-Hai Rd, Kaosiung City 80424, Taiwan, R.O.C
[98] National Optical Astronomy Observatory, 950 North Cherry Ave., Tucson, AZ 85719, USA
[99] Key Laboratory for Particle Astrophysics, Institute of High Energy Physics, Chinese Academy of Sciences, 19B Yuquan Road, Shijingshan District, Beijing, People's Republic of China
[100] School of Astronomy and Space Science, Nanjing University, Nanjing 210023, People's Republic of China
[101] Key Laboratory of Modern Astronomy and Astrophysics, Nanjing University, Nanjing 210023, People's Republic of China
[102] Italian ALMA Regional Centre, INAF-Istituto di Radioastronomia, Via P. Gobetti 101, I-40129 Bologna, Italy
[103] Department of Physics, National Taiwan University, No.1, Sect.4, Roosevelt Rd., Taipei 10617, Taiwan, R.O.C
[104] Instituto de Radioastronomía y Astrofísica, Universidad Nacional Autónoma de México, Morelia 58089, México
[105] Instituto de Astronomía, Universidad Nacional Autónoma de México, CdMx 04510, México
[106] Yunnan Observatories, Chinese Academy of Sciences, 650011 Kunming, Yunnan Province, People's Republic of China
[107] Center for Astronomical Mega-Science, Chinese Academy of Sciences, 20A Datun Road, Chaoyang District, Beijing, 100012, People's Republic of China
[108] Key Laboratory for the Structure and Evolution of Celestial Objects, Chinese Academy of Sciences, 650011 Kunming, People's Republic of China
[109] School of Natural Sciences, Institute for Advanced Study, 1 Einstein Drive, Princeton, NJ 08540, USA
[110] Tsung-Dao Lee Institute and School of Physics and Astronomy, Shanghai Jiao Tong University, Shanghai, 200240, People's Republic of China
[111] National Institute of Technology, Hachinohe College, 16-1 Uwanotai, Tamonoki, Hachinohe City, Aomori 039-1192, Japan
[112] Department of Astronomy, University of Massachusetts, 01003, Amherst, MA, USA
[113] South African Radio Astronomy Observatory, Observatory 7925, Cape Town, South Africa
[114] Department of Physics, National and Kapodistrian University of Athens, Panepistimiopolis, GR 15783 Zografos, Greece
[115] Canadian Institute for Theoretical Astrophysics, University of Toronto, 60 St. George Street, Toronto, ON M5S 3H8, Canada
[116] Dunlap Institute for Astronomy and Astrophysics, University of Toronto, 50 St. George Street, Toronto, ON M5S 3H4, Canada
[117] Canadian Institute for Advanced Research, 180 Dundas St West, Toronto, ON M5G 1Z8, Canada
[118] Radio Astronomy Laboratory, University of California, Berkeley, CA 94720, USA
[119] Department of Physics, National Taiwan Normal University, No. 88, Sec. 4, Tingzhou Rd., Taipei 116, Taiwan, R.O.C.
[120] Institut für Theoretische Physik, Max-von-Laue-Strasse 1, D-60438 Frankfurt, Germany
[121] Frankfurt Institute for Advanced Studies, Ruth-Moufang-Strasse 1, D-60438 Frankfurt, Germany
[122] School of Mathematics, Trinity College, Dublin 2, Ireland
[123] Department of Astrophysical Sciences, Peyton Hall, Princeton University, Princeton, NJ 08544, USA
[124] Instituto de Radioastronomía Milimétrica, IRAM, Avenida Divina Pastora 7, Local 20, E-18012, Granada, Spain
[125] Aalto University Department of Electronics and Nanoengineering, PL 15500, FI-00076 Aalto, Finland
[126] Aalto University Metsähovi Radio Observatory, Metsähovintie 114, FI-02540 Kylmälä, Finland
[127] Department of Astronomy, Yonsei University, Yonsei-ro 50, Seodaemun-gu, 03722 Seoul, Republic of Korea

[128] East Asian Observatory, 660 North A'ohoku Place, Hilo, HI 96720, USA
[129] Netherlands Organisation for Scientific Research (NWO), Postbus 93138, 2509 AC Den Haag, The Netherlands
[130] Frontier Research Institute for Interdisciplinary Sciences, Tohoku University, Sendai 980-8578, Japan
[131] Astronomical Institute, Tohoku University, Sendai 980-8578, Japan
[132] Leiden Observatory, Leiden University, Postbus 2300, 9513 RA Leiden, The Netherlands
[133] Jeremiah Horrocks Institute, University of Central Lancashire, Preston PR1 2HE, UK
[134] Physics Department, Brandeis University, 415 South Street, Waltham, MA 02453, USA
[135] School of Physics, Huazhong University of Science and Technology, Wuhan, Hubei, 430074, People's Republic of China
[136] Mullard Space Science Laboratory, University College London, Holmbury St. Mary, Dorking, Surrey, RH5 6NT, UK
[137] School of Astronomy and Space Sciences, University of Chinese Academy of Sciences, No. 19A Yuquan Road, Beijing 100049, People's Republic of China
[138] Astronomy Department, University of Science and Technology of China, Hefei 230026, People's Republic of China
[139] Laboratoire Leprince-Ringuet, École Polytechnique, CNRS, Institut Polytechnique de Paris, F-91128 Palaiseau, France
[140] Dublin Institute for Advanced Studies, 31 Fitzwilliam Place, Dublin 2, Ireland
[141] Max-Planck-Institut für Kernphysik, P.O. Box 103980, D-69029 Heidelberg, Germany
[142] High Energy Astrophysics Laboratory, RAU, 123 Hovsep Emin St Yerevan 0051, Armenia
[143] Aix Marseille Université, CNRS/IN2P3, CPPM, Marseille, France
[144] Laboratoire d'Annecy de Physique des Particules, Univ. Grenoble Alpes, Univ. Savoie Mont Blanc, CNRS, LAPP, F-74000 Annecy, France
[145] University of Oxford, Department of Physics, Denys Wilkinson Building, Keble Road, Oxford OX1 3RH, UK
[146] IRFU, CEA, Université Paris-Saclay, F-91191 Gif-sur-Yvette, France
[147] University of Namibia, Department of Physics, Private Bag 13301, Windhoek 10005, Namibia
[148] Instytut Fizyki Jądrowej PAN, ul. Radzikowskiego 152, 31-342 Kraków, Poland
[149] DESY, D-15738 Zeuthen, Germany
[150] Obserwatorium Astronomiczne, Uniwersytet Jagielloński, ul. Orla 171, 30-244 Kraków, Poland
[151] Department of Physics and Electrical Engineering, Linnaeus University, SE-351 95 Växjö, Sweden
[152] Institut für Astronomie und Astrophysik, Universität Tübingen, Sand 1, D-72076 Tübingen, Germany
[153] Laboratoire Univers et Théories, Observatoire de Paris, Université PSL, CNRS, Université de Paris, F-92190 Meudon, France
[154] Sorbonne Université, Université Paris Diderot, Sorbonne Paris Cité, CNRS/IN2P3, Laboratoire de Physique Nucléaire et de Hautes Energies, LPNHE, 4 Place Jussieu, F-75252 Paris, France
[155] GRAPPA, Anton Pannekoek Institute for Astronomy, University of Amsterdam, Science Park 904, 1098 XH Amsterdam, The Netherlands
[156] Friedrich-Alexander-Universität Erlangen-Nürnberg, Erlangen Centre for Astroparticle Physics, Erwin-Rommel-Str. 1, D-91058 Erlangen, Germany
[157] Astronomical Observatory, The University of Warsaw, Al. Ujazdowskie 4, 00-478 Warsaw, Poland
[158] School of Physics, University of the Witwatersrand, 1 Jan Smuts Avenue, Braamfontein, Johannesburg, 2050 South Africa
[159] Université Bordeaux, CNRS/IN2P3, Centre d'Études Nucléaires de Bordeaux Gradignan, F-33175 Gradignan, France
[160] School of Physical Sciences, University of Adelaide, Adelaide 5005, Australia
[161] Universität Hamburg, Institut für Experimentalphysik, Luruper Chaussee 149, D-22761 Hamburg, Germany
[162] Université de Paris, CNRS, Astroparticule et Cosmologie, F-75013 Paris, France
[163] Department of Physics and Astronomy, The University of Leicester, University Road, Leicester, LE1 7RH, UK
[164] Nicolaus Copernicus Astronomical Center, Polish Academy of Sciences, ul. Bartycka 18, 00-716 Warsaw, Poland
[165] Institut für Physik und Astronomie, Universität Potsdam, Karl-Liebknecht-Strasse 24/25, D-14476 Potsdam, Germany
[166] Laboratoire Univers et Particules de Montpellier, Université Montpellier, CNRS/IN2P3, CC 72, Place Eugène Bataillon, F-34095 Montpellier Cedex 5, France
[167] Institut für Physik, Humboldt-Universität zu Berlin, Newtonstr. 15, D-12489 Berlin, Germany
[168] Institut für Astro- und Teilchenphysik, Leopold-Franzens-Universität Innsbruck, A-6020 Innsbruck, Austria
[169] Landessternwarte, Universität Heidelberg, Königstuhl, D-69117 Heidelberg, Germany
[170] Institute of Astronomy, Faculty of Physics, Astronomy and Informatics, Nicolaus Copernicus University, Grudziadzka 5, 87-100 Torun, Poland
[171] Department of Physics, Rikkyo University, 3-34-1 Nishi-Ikebukuro, Toshima-ku, Tokyo 171-8501, Japan
[172] Department of Physics, University of the Free State, P.O. Box 339, Bloemfontein 9300, South Africa
[173] Department of Physics, The University of Tokyo, 7-3-1 Hongo, Bunkyo-ku, Tokyo 113-0033, Japan
[174] Yerevan Physics Institute, 2 Alikhanian Brothers St., 375036 Yerevan, Armenia
[175] Kavli Institute for the Physics and Mathematics of the Universe (WPI), The University of Tokyo Institutes for Advanced Study (UTIAS), The University of Tokyo, 5-1-5 Kashiwa-no-Ha, Kashiwa, Chiba, 277-8583, Japan
[176] Instituto de Astrofísica de Canarias and Dpto. Astrofísica, Universidad de La Laguna, 38200, La Laguna, Tenerife, Spain
[177] Università di Udine and INFN Trieste, I-33100 Udine, Italy
[178] National Institute for Astrophysics (INAF), I-00136 Rome, Italy
[179] ETH Zürich, CH-8093 Zürich, Switzerland
[180] Institut de Física d'Altes Energies (IFAE), The Barcelona Institute of Science and Technology (BIST), E-08193 Bellaterra (Barcelona), Spain
[181] Japanese MAGIC Group: Institute for Cosmic Ray Research (ICRR), The University of Tokyo, Kashiwa, 277-8582 Chiba, Japan
[182] Technische Universität Dortmund, D-44221 Dortmund, Germany
[183] Croatian MAGIC Group: University of Zagreb, Faculty of Electrical Engineering and Computing (FER), 10000 Zagreb, Croatia
[184] IPARCOS Institute and EMFTEL Department, Universidad Complutense de Madrid, E-28040 Madrid, Spain
[185] Centro Brasileiro de Pesquisas Físicas (CBPF), 22290-180 URCA, Rio de Janeiro (RJ), Brazil
[186] University of Lodz, Faculty of Physics and Applied Informatics, Department of Astrophysics, 90-236 Lodz, Poland
[187] Università di Siena and INFN Pisa, I-53100 Siena, Italy
[188] Deutsches Elektronen-Synchrotron (DESY), D-15738 Zeuthen, Germany
[189] Università di Padova and INFN, I-35131 Padova, Italy
[190] INFN MAGIC Group: INFN Sezione di Torino and Università degli Studi di Torino, I-10125 Torino, Italy
[191] Max-Planck-Institut für Physik, D-80805 München, Germany
[192] Università di Pisa and INFN Pisa, I-56126 Pisa, Italy
[193] Universitat de Barcelona, ICCUB, IEEC-UB, E-08028 Barcelona, Spain
[194] Armenian MAGIC Group: A. Alikhanyan National Science Laboratory, Armenia
[195] Centro de Investigaciones Energéticas, Medioambientales y Tecnológicas, E-28040 Madrid, Spain
[196] INFN MAGIC Group: INFN Sezione di Bari and Dipartimento Interateneo di Fisica dell'Università e del Politecnico di Bari, I-70125 Bari, Italy
[197] Croatian MAGIC Group: University of Rijeka, Department of Physics, 51000 Rijeka, Croatia
[198] Universität Würzburg, D-97074 Würzburg, Germany
[199] Finnish MAGIC Group: Finnish Centre for Astronomy with ESO, University of Turku, FI-20014 Turku, Finland
[200] Departament de Física, and CERES-IEEC, Universitat Autònoma de Barcelona, E-08193 Bellaterra, Spain

[201] Armenian MAGIC Group: ICRANet-Armenia at NAS RA, Armenia
[202] Croatian MAGIC Group: University of Split, Faculty of Electrical Engineering, Mechanical Engineering and Naval Architecture (FESB), 21000 Split, Croatia
[203] Croatian MAGIC Group: Josip Juraj Strossmayer University of Osijek, Department of Physics, 31000 Osijek, Croatia
[204] Japanese MAGIC Group: RIKEN, Wako, Saitama 351-0198, Japan
[205] Japanese MAGIC Group: Department of Physics, Kyoto University, 606-8502 Kyoto, Japan
[206] Japanese MAGIC Group: Department of Physics, Tokai University, Hiratsuka, 259-1292 Kanagawa, Japan
[207] Saha Institute of Nuclear Physics, HBNI, 1/AF Bidhannagar, Salt Lake, Sector-1, Kolkata 700064, India
[208] Inst. for Nucl. Research and Nucl. Energy, Bulgarian Academy of Sciences, BG-1784 Sofia, Bulgaria
[209] Finnish MAGIC Group: Astronomy Research Unit, University of Oulu, FI-90014 Oulu, Finland
[210] Croatian MAGIC Group: Ruđer Bošković Institute, 10000 Zagreb, Croatia
[211] INFN MAGIC Group: INFN Sezione di Perugia, I-06123 Perugia, Italy
[212] INFN MAGIC Group: INFN Roma Tor Vergata, I-00133 Roma, Italy
[213] Department of Physics and Astronomy, Barnard College, Columbia University, NY 10027, USA
[214] Physics Department, Columbia University, New York, NY 10027, USA
[215] Physics Department, California Polytechnic State University, San Luis Obispo, CA 94307, USA
[216] Department of Physics and Astronomy, Iowa State University, Ames, IA 50011, USA
[217] Department of Physics, Washington University, St. Louis, MO 63130, USA
[218] School of Physics, University College Dublin, Belfield, Dublin 4, Ireland
[219] Department of Physics and Astronomy, Purdue University, West Lafayette, IN 47907, USA
[220] School of Physics and Astronomy, University of Minnesota, Minneapolis, MN 55455, USA
[221] Department of Physics, California State University—East Bay, Hayward, CA 94542, USA
[222] School of Physics and Center for Relativistic Astrophysics, Georgia Institute of Technology, 837 State Street NW, Atlanta, GA 30332-0430, USA
[223] DESY, Platanenallee 6, D-15738 Zeuthen, Germany
[224] Santa Cruz Institute for Particle Physics and Department of Physics, University of California, Santa Cruz, CA 95064, USA
[225] Department of Physics and Astronomy and the Bartol Research Institute, University of Delaware, Newark, DE 19716, USA
[226] Department of Physics and Astronomy, University of Alabama, Tuscaloosa, AL 35487, USA
[227] Department of Physics and Astronomy, University of Iowa, Van Allen Hall, Iowa City, IA 52242, USA
[228] Department of Physics and Astronomy, DePauw University, Greencastle, IN 46135-0037, USA
[229] Department of Physics and Astronomy, University of Utah, Salt Lake City, UT 84112, USA
[230] Physics Department, McGill University, Montreal, QC H3A 2T8, Canada
[231] School of Physics, National University of Ireland Galway, University Road, Galway, Ireland
[232] Institute of Particle and Cosmos Physics, Universidad Complutense de Madrid, E-28040 Madrid, Spain
[233] Department of Physics and Astronomy, University of California, Los Angeles, CA 90095, USA
[234] Institute of Physics and Astronomy, University of Potsdam, D-14476 Potsdam-Golm, Germany and DESY, Platanenallee 6, D-15738 Zeuthen, Germany
[235] Department of Physical Sciences, Cork Institute of Technology, Bishopstown, Cork, Ireland
[236] Xinjiang Astronomical Observatory, Chinese Academy of Sciences, Urumqi 830011, People's Republic of China
[237] Graduate School of Sciences and Technology for Innovation, Yamaguchi University, Yoshida 1677-1, Yamaguchi, Yamaguchi 753-8512, Japan
[238] Center for Astronomy, Ibaraki University, 2-1-1 Bunkyo, Mito, Ibaraki 310-8512, Japan